\documentclass[11pt]{article}
\pdfoutput=1
\usepackage{jheppub}
\usepackage{physics}
\usepackage{bbm}

\newcommand\nn{\nonumber}

\newcommand\fft[2]{\frac{#1}{#2}}

\newcommand\tf{\text{f}}

\newtheorem{definition}{Definition}

\begin{document}

\title{Decomposition of BPS Moduli Spaces and Asymptotics of Supersymmetric Partition Functions}


\author[a]{Arash Arabi Ardehali}
\affiliation[a]{C.N. Yang Institute for Theoretical Physics, Stony Brook University,\\ Stony Brook, NY 11794, USA}
\author[b]{and Junho Hong}
\affiliation[b]{Institute for Theoretical Physics, KU Leuven,\\
Celestijnenlaan 200D, B-3001 Leuven, Belgium}

\emailAdd{a.a.ardehali@gmail.com}\emailAdd{junhophysics@gmail.com}

\abstract{We present a prototype for Wilsonian analysis of asymptotics of supersymmetric partition functions of non-abelian gauge theories. Localization allows expressing such partition functions as an integral over a BPS moduli space. When the limit of interest introduces a scale hierarchy in the problem, asymptotics of the partition function is obtained in the Wilsonian approach by $i)$ decomposing (in some suitable scheme) the BPS moduli space into various patches according to the set of light fields (lighter than the scheme dependent cut-off~$\Lambda$) they support, $ii)$ localizing the partition function of the effective field theory on each patch (with cut-offs set by the scheme), and $iii)$~summing up the contributions of all patches to obtain the final asymptotic result (which is scheme-independent and accurate as $\Lambda\to\infty$). Our prototype concerns the Cardy-like asymptotics of the 4d superconformal index, which has been of interest recently for its application to black hole microstate counting in AdS$_5$/CFT$_4$. As a byproduct of our analysis we obtain the most general asymptotic expression for the index of gauge theories in the Cardy-like limit, encompassing and extending all previous results.}

\maketitle \flushbottom

\section{Introduction}

A fruitful theme in supersymmetric quantum field theory (QFT) has been the use of exact results to complement semi-classical reasoning. The idea is that semi-classical (or effective) QFT should emerge as a systematic set of prescriptions to reproduce various aspects of exact results in some---often weak-coupling and/or low-energy---limit. An outstanding landmark in this line of research has been the use of exact supersymmetric localization results by Nekrasov and Okounkov~\cite{Nekrasov:2002qd,Nekrasov:2003rj} to  shed light on the Seiberg-Witten solution of certain 4d $\mathcal{N}=2$ gauge theories, originally obtained via effective field theory (EFT) arguments~\cite{Seiberg:1994rs,Seiberg:1994aj}.

In the opposite direction, semi-classical/effective QFT can be used to shed ``physical'' light on the asymptotics of exact results. A major component of the present work is in this direction, following in the footsteps of Di~Pietro, Honda, and Komargodski \cite{DiPietro:2014bca,DiPietro:2016ond}. The exact result that we consider is the 4d superconformal index \cite{Romelsberger:2005eg,Kinney:2005ej}. We study a certain Cardy-like limit \cite{Cardy:1986ie,Choi:2018hmj} of the index, and obtain asymptotic expressions for it that we explain through \emph{perturbative} EFT techniques\footnote{Our results regarding the interplay between exact results and EFT are hence more limited than the ones in \cite{Nekrasov:2002qd,Nekrasov:2003rj} which concern non-perturbative effects. However, at least in several examples, there seems to be a curious connection between some of the perturbative effects that we study here and non-perturbative effects on $\mathbb{R}^3\times S^1$ \cite{ArabiArdehali:2019zac}, which might bridge our investigation to that of \cite{Nekrasov:2002qd,Nekrasov:2003rj}. We will not explore this connection any further in this work.} pioneered in \cite{DiPietro:2014bca,DiPietro:2016ond}.

The superconformal index is a graded partition function of 4d $\mathcal{N}=1$ QFTs with a $U(1)_R$ symmetry, encoding the protected BPS spectrum of the theory, and defined explicitly as~\cite{Kinney:2005ej,Romelsberger:2005eg}
\begin{equation}
\mathcal{I}(p,q;\boldsymbol{v})=\mathrm{Tr}\, (-1)^F\, p^{ J_1+\frac{r}{2}}q^{J_2+\frac{r}{2}} \boldsymbol{v}^{q_F},\label{eq:indexDef}
\end{equation}
with the trace in the Hilbert space of radial quantization. Here $F$ is the fermion number, $J_{1,2}$~are the rotation quantum numbers, and $r$ is the $U(1)_R$ charge.
The symbol $\boldsymbol{v}^{q_F}$ stands for $v_1^{q_1}\times\dots\times v_{r_F}^{q_{r_F}}$, where $v_a$ are fugacities associated to the Cartan of the flavor symmetry, while $q_a$ are the corresponding charges. We assume $|v_a|=1$ and $0<|p|,|q|<1$ throughout this paper. We will refer to $v_a$ as flavor fugacities, and refer to $p,q$ as spacetime fugacities.

Closed-form expressions are available for the index of large classes of 4d $\mathcal{N}=1$ QFTs (see \emph{e.g.} \cite{Dolan:2008qi,Spiridonov:2009za,Spiridonov:2011hf,Gaiotto:2012xa}). The closed formulas are rather complicated however, and hence often not particularly illuminating. Substantial simplification occurs in the Cardy-like limit \cite{Cardy:1986ie} where the spacetime fugacities approach 1. More precisely, defining $\beta>0$ and $\omega_{1,2}\in\mathbb{H}$ (the upper half-plane) via $p=e^{i\beta\omega_1}$, $q=e^{i\beta\omega_2}$, and defining $\xi_a\in\mathbb{R}$ via $v_a=e^{2\pi i \xi_a}$, we consider the limit
\begin{equation}
\beta\to 0,\ \text{with\
}\, \omega_{1,2},\, \xi_a \ \,
\text{fixed}.\label{eq:CKKNlimitIntro}
\end{equation}
This limit of the index is of physical interest for its application to testing supersymmetric dualities~\cite{Ardehali:2015bla}, to studying supersymmetric gauge dynamics on $\mathbb{R}^3\times S^1$ \cite{ArabiArdehali:2019zac}, and to microstate counting of BPS (possibly multi-center) black holes via AdS$_5$/CFT$_4$ \cite{Choi:2018hmj,Honda:2019cio,ArabiArdehali:2019tdm,Cabo-Bizet:2019osg,Kim:2019yrz,ArabiArdehali:2019orz}.

\subsection*{Asymptotics from the exact result}

Cardy-like asymptotics of the index has been worked out in various special cases (in special theories, or for special choices of parameters; see below for more details on relation of our results to those of previous work). In Section~\ref{sec:asymptoticAnalysis}, using direct asymptotic analysis we present the most general expression, up to exponentially small error, for arbitrary $\mathcal{N}=1$ gauge theories with a $U(1)_R$ symmetry and a semi-simple compact gauge group.\footnote{Relaxing the semi-simplicity constraint appears to be straightforward, but we do not attempt that here.} We do this by employing the exact matrix-integral expression for the index of supersymmetric gauge theories~\cite{Dolan:2008qi,Spiridonov:2009za,Spiridonov:2011hf}.

The matrix-integral is over a moduli space of matrix eigenvalues. It turns out that in general on the one hand the simplest asymptotic estimate for the integrand of the index is \emph{not} uniformly valid over all of the moduli space, and on the other hand the strongest asymptotic estimate contains too much irrelevant (exponentially suppressed) information on most of the moduli space. This necessitates \emph{decomposing the moduli space} into various patches, and using appropriate uniform estimates on each patch. The efficient uniform grip  over individual patches then allows us to further simplify the asymptotic contribution of each patch using some additional input from the structure of the integrand. Summing up the contributions of all patches we obtain
the asymptotics of the matrix integral as in~\eqref{eq:ItotGenSimp}. While the exponential piece in~\eqref{eq:ItotGenSimp} had been demonstrated earlier in special cases where the matrix-integral has isolated saddles in the Cardy-like limit (see \emph{e.g.} \cite{Ardehali:2015bla,ArabiArdehali:2021nsx}), our result that the most general asymptotic in presence of extended saddles is that exponential dressed by a polynomial in $1/\beta$ is new.

We demonstrate the power of this method of asymptotic analysis in our case study of the index of SU($N$) $\mathcal{N}=4$ theory, where we obtain the following significant improvements over previous results:
\begin{itemize}
    \item For $\xi_a=0$, the Cardy-like asymptotics of the $\mathcal{N}=4$ index was found in \cite{Ardehali:2015bla} to be of the form $\frac{f(\omega_{1,2})}{\beta^{N-1}}(1+o(\beta^0))$ for the limited range of parameters $\omega_{1,2}\in i\mathbb{R}_{>0}$ (\emph{i.e.}~$p,q\in\mathbb{R}$). In Section~\ref{sec:asymptoticAnalysis} we show that the $\xi_a=0$ asymptotics is more generally for $\omega_{1,2}\in\mathbb{H}$ given, up to exponentially small error, by a degree $N-1$ polynomial in $1/\beta$ whose leading monomial coincides with the result of \cite{Ardehali:2015bla} when $\omega_{1,2}\in i\mathbb{R}_{>0}$. 

    \item The leading Cardy-like asymptotics of the $\mathcal{N}=4$ index in the fully-deconfined phase (see Section~\ref{sec:asymptoticAnalysis}) was obtained in \cite{Choi:2018hmj,Honda:2019cio,ArabiArdehali:2019tdm} using non-uniform estimates in the integrand of the matrix integral. But there is no guarantee in general that non-uniform estimates of the integrand give correct asymptotics for an integral. Moreover, the papers \cite{Choi:2018hmj,Honda:2019cio,ArabiArdehali:2019tdm} assumed but did not demonstrate absence of leading-order cancellations (completely destructive interference) due to phase oscillations in the integrand. These amount to two significant gaps in those derivations, and the analysis in Section~\ref{sec:asymptoticAnalysis} fills both.
    
    \item In the special case with $p=q$, the Cardy-like asymptotics of the $\mathcal{N}=4$ index in the fully-deconfined phase was obtained up to exponentially small error in \cite{GonzalezLezcano:2020yeb}.\footnote{See \cite{Amariti:2020jyx,Amariti:2021ubd} for related work studying different gauge groups as well as various $\mathcal{N}=1$ SCFTs.} The analysis in Section~\ref{sec:asymptoticAnalysis} yields the generalization of that result to $p\neq q$.
    
    \item The strongest results on the Cardy-like asymptotics of the $\mathcal{N}=4$ index in partially-deconfined or confined phases were obtained in \cite{ArabiArdehali:2019orz,Lezcano:2021qbj}. In particular, in the special case with $p=q$ an asymptotic expression of the form \eqref{eq:ItotGenSimp} valid with exponentially small error was obtained in \cite{Lezcano:2021qbj} for SU(2) gauge group, and a similarly accurate expression was suggested for SU(3). The analysis in Section~\ref{sec:asymptoticAnalysis} generalizes the SU(2) result of \cite{Lezcano:2021qbj} to $p\neq q$, and demonstrates validity of \eqref{eq:ItotGenSimp} not only for SU(3) but for arbitrary rank.
\end{itemize}

\subsection*{Asymptotics from effective field theory}

A more physical approach to finding the Cardy-like asymptotics of the index was initiated by Di~Pietro and Komargodski~\cite{DiPietro:2014bca} within a Lagrangian, path-integral formulation of the index (rather than the Hamiltonian, trace formulation as in Eq.~\eqref{eq:indexDef}).

Before explaining how the Cardy-like asymptotics of the index is obtained in this approach, let us discuss how the exact matrix-integral expression for it is obtained via path-integration in gauge theories. The 4d $\mathcal{N}=1$ gauge theory with $U(1)_R$ symmetry is placed on a Hopf surface that is topologically $S^3\times S^1$. Turning on appropriate background fields and deforming the Lagrangian as necessary for preserving supersymmetry \cite{Festuccia:2011ws}, one uses supersymmetric localization \cite{Nekrasov:2002qd,Pestun:2007rz,Assel:2014paa} to compute the path-integral. The result is the matrix integral expression\footnote{The localization computation is done in \cite{Assel:2014paa} for real-valued $p,q$, but the result can presumably be analytically continued---at least to a neighborhood of the real line(s).} for the index, up to a supersymmetric Casimir energy factor (see Section~\ref{sec:EFT} for details). The integral over the moduli space of matrix eigenvalues arises in the 4d localization calculation as the integral over the moduli space of BPS field configurations (or the BPS moduli space, for short).

The Cardy-like limit $p,q\to1$ corresponds to shrinking the circle (see Section~\ref{sec:EFT}). To study the asymptotics of the index in the small-circle limit, we write the Hopf surface metric in a Kaluza-Klein (KK) form as an $S^1$ fiber over a three-manifold $\mathcal{M}_3$. The inverse of the circle size $\beta$ sets a large scale $\Lambda\propto 1/\beta$ in the problem, and one can use a 3d effective field theory on $\mathcal{M}_3$ with only the KK fields lighter than $\Lambda$ as the dynamical degrees of freedom, and with the effect of the heavy KK fields encoded in the effective action of the light 3d fields. The partition function of the 3d EFT should reproduce---essentially by definition of an EFT---the large-$\Lambda$ (hence small-$\beta$, Cardy-like) asymptotics of the 4d partition function. Therefore the Cardy-like asymptotics of the index should be governed by the effective action of the 3d~EFT on $\mathcal{M}_3$.

The general considerations of the previous paragraph are sharpened and made computationally efficient when complemented with the following ideas.
\begin{enumerate}
  \setcounter{enumi}{0}
    \item The effective action contains a one-loop piece, which can be computed by first fixing the coefficients of certain Chern-Simons (CS) terms that are easily under control, and then supersymmetrizing those CS terms.
\end{enumerate}
This was the breakthrough idea of the pioneering work of Di~Pietro and Komargodski \cite{DiPietro:2014bca}. Using CS actions involving only background fields, the exponential asymptotics of the index was found in \cite{DiPietro:2014bca} in some special cases with $\xi_a=0$.\footnote{See \cite{Choi:2018hmj,Kim:2019yrz,Cassani:2021fyv} for related work pertaining to $\xi_a\neq 0$.}

Even restricting to $\xi_a=0$, however, the asymptotic obtained in \cite{DiPietro:2014bca} is not valid for general theories \cite{Ardehali:2015bla,DiPietro:2016ond,Hwang:2018riu} or for general choices of the parameters $\omega_{1,2}$ (\emph{cf.} \cite{ArabiArdehali:2021nsx}). To fix that, the following crucial idea was provided by Di~Pietro and Honda in~\cite{DiPietro:2016ond}.
\begin{enumerate}
  \setcounter{enumi}{1}
    \item Besides the CS terms studied in \cite{DiPietro:2014bca} involving only the background fields, one must include in the one-loop part of the effective action also the CS terms involving the dynamical (\emph{i.e.} light) gauge fields of the 3d EFT.
\end{enumerate}
With this added ingredient, the work \cite{DiPietro:2016ond} found expressions for the exponential asymptotics of the index that are valid for arbitrary gauge theories with a $U(1)_R$ symmetry and a compact semi-simple gauge group. More precisely, the derivation of \cite{DiPietro:2016ond} was valid only for the limited range of parameters $\omega_{1,2}\in i\mathbb{R}_{>0}$ (\emph{cf.} \cite{Ardehali:2015bla} where $\xi_a$ are set to zero but can be straightforwardly restored). But the limitation $\omega_{1,2}\in i\mathbb{R}_{>0}$ was purely technical, and due to the Hopf surface background of \cite{DiPietro:2016ond} having real-valued $p,q$. This limitation can be overcome using the more general background with $\omega_{1,2}\in\mathbb{H}$ discussed in an elegant recent paper by Cassani and Komargodski \cite{Cassani:2021fyv}. We will heavily rely on the background of \cite{Cassani:2021fyv} in Section~\ref{sec:EFT}.

Once dynamical fields are included in the effective action as in \cite{DiPietro:2016ond}, they should be integrated over. In \cite{DiPietro:2016ond}, instead of performing the path-integral the dynamical fields were set to their BPS values and then an integral over the BPS moduli space was introduced. While this is natural and yields correctly the exponential asymptotics of the index, it begs for a sharp justification and is in fact not sufficient for obtaining the complete subleading asymptotics with exponentially small error. A more systematic approach was proposed in \cite{ArabiArdehali:2021nsx}.

\begin{enumerate}
  \setcounter{enumi}{2}
    \item The partition function of the 3d EFT is obtained by supersymmetric localization of its path-integral on $\mathcal{M}_3$, with the effective action consisting of the one-loop piece involving supersymmetrized CS terms, as well as a tree-level piece coming from KK expansion of the (UV) 4d action.
\end{enumerate}
This idea was implemented in \cite{ArabiArdehali:2021nsx} for a special choice\footnote{The index studied in \cite{ArabiArdehali:2021nsx} (and also \cite{Cassani:2021fyv,Cabo-Bizet:2019osg,Kim:2019yrz}) depending on a parameter $n_0$, is recovered in our setting by formally associating a $U(1)$ flavor fugacity $v_\chi=e^{2\pi i(-n_0 r_\chi/2)}$ to every chiral multiplet $\chi$ and taking $q_\chi=1$.} of $\xi_a$, but the generalization to arbitrary $\xi_a\in\mathbb{R}$ is straightforward. The restriction of the one-loop supersymmetrized CS terms to the BPS locus is justified by (and follows from) supersymmetric localization in this approach. Also, with the inclusion of the tree-level action, the subleading asymptotics of the index is reproduced in \cite{ArabiArdehali:2021nsx} correctly\footnote{We emphasize that the EFT derivation in \cite{ArabiArdehali:2021nsx} relied on a certain assumption regarding the supersymmetrized gravitational CS term. We propose in Section~\ref{sec:EFT} that actually a modified version of that assumption is needed (see Eq.~\eqref{eq:modifiedG}).} up to exponentially small error\footnote{In particular, the hyperbolic sines (resp. the exponential factor signalling the CS level) in the integrand of the Chern-Simons partition function found in the subleading  asymptotics of the SU$(N)$ $\mathcal{N}=4$ index in \cite{GonzalezLezcano:2020yeb} arise from localization of the tree-level (resp. one-loop) piece of the 3d effective action.} in cases of interest in that work, although not in general as we explain shortly.

In this work we add one more idea to the mix, which in our view completes the conceptual framework---though important puzzles remain as discussed in Section~\ref{sec:discussion}. We make the EFT cut-offs explicit, and emphasize that the set of light 3d fields is different on different patches of the BPS moduli space.

\begin{enumerate}
  \setcounter{enumi}{3}
    \item The BPS moduli space should be decomposed in some suitable scheme into various patches, each supporting its own set of light 3d fields, and hence each having its own 3d EFT. The asymptotics of the index is obtained (up to exponentially small error) by summing the (perturbative) EFT partition functions of all patches.
\end{enumerate}
The earlier work \cite{ArabiArdehali:2021nsx} had considered only two of these patches (the outer patch and in$_0$  in the language of Section~\ref{sec:asymptoticAnalysis}). Those were sufficient for the cases of interest in that work, but when other patches dominate the index (as may happen in partially-deconfined phases for example) the EFT results of \cite{ArabiArdehali:2021nsx} are not enough to reproduce the correct asymptotics.\\

Our EFT analysis in Section~\ref{sec:EFT} parallels the general asymptotic analysis of Section~\ref{sec:asymptoticAnalysis} with exponential accuracy. It gives a crisp physical understanding of the Cardy-like asymptotics of the index, and in particular through the analysis of light 3d fields on different parts of the BPS moduli space clarifies why the decomposition procedure of Section~\ref{sec:asymptoticAnalysis} is so powerful: it is---with some stretch of imagination---what Wilson would do.

\subsection{Notation and terminology}

Throughout this work the symbol $\simeq$ means that the ratio of the two sides is 1, up to exponentially suppressed error of the form $\mathcal O(e^{-1/\beta})$. More precisely
\begin{equation}
    \text{$A(\beta)\simeq B(\beta)$ \ \ if \ \ $\frac{A(\beta)}{ B(\beta)}=1+\mathcal{O}(e^{-c/\beta})$ \ as $\beta\to0$,  with some fixed $c>0$.}
\end{equation}

If there is dependence on extra parameters $x_j$, we say $A(\beta,x_j)\simeq B(\beta,x_j)$ \emph{uniformly} over a certain domain, if there is a $c>0$ that works as above (except possibly for removable singularities in $\frac{A(\beta,x_j)}{ B(\beta,x_j)}$) for all $x_j$ in that domain.

For $x\in\mathbb{R}$, the symbol $x^{}_\mathbb{Z}$ will denote $x-\mathrm{nint}(x)$, with nint the nearest integer function.\\

\noindent\textbf{Note added:} while finalizing this manuscript we learned of work by A.~Cabo-Bizet \cite{Cabo-Bizet:2021plf} also using a decomposition method to study Cardy-like asymptotics of 4d superconformal indices.




\section{Cardy-like asymptotics of the 4d superconformal index}\label{sec:asymptoticAnalysis}

Consider a 4d $\mathcal{N}=1$ gauge theory with a $U(1)_R$ symmetry, a semi-simple (compact Lie) gauge group $G$ of finite rank $r_G$, a finite number of chiral multiplets $\chi$ in various representations of $G$, and a global flavor symmetry (compact Lie) group of rank $r_F$.

The index can be evaluated in closed form as an elliptic hypergeometric integral \cite{Dolan:2008qi,Spiridonov:2009za,Spiridonov:2011hf}:
\begin{equation}
	\mathcal{I}(p,q;\boldsymbol{v})=(p;p)^{r_G}(q;q)^{r_G}\int_{\mathfrak{h}_{\text{cl}}} D\boldsymbol{x}\ \frac{\prod_\chi\prod_{\rho^\chi}\Gamma_e\big(r_\chi (\frac{\tau+\sigma}{2})+\rho^\chi\cdot \boldsymbol{x}+q^\chi\cdot\boldsymbol{\xi}\big)}{\prod_{\alpha_+}\Gamma_e(\alpha_+\cdot\boldsymbol{x})\, \Gamma_e(-\alpha_+\cdot\boldsymbol{x})}.\label{eq:EHIgen}
\end{equation}
The measure is defined as
\begin{equation}
    D\boldsymbol{x}:=\frac{1}{|W|} \prod_{j=1}^{r_G} \mathrm{d} x_j,\label{eq:defDx}
\end{equation} with $|W|$ the order of the Weyl group of $G$. The parameters $x_j$ will be referred to as the \emph{holonomies}, and their moduli space (parametrized by $x_1,\dots,x_{r_G}$ mod 1) is denoted $\mathfrak{h}_{\text{cl}}$, which we write explicitly as
\begin{equation}
\mathfrak{h}_\text{cl}=(-\frac{1}{2},\frac{1}{2}]^{r_G}.    
\end{equation}
The $r_G$-tuple $x_1,\dots,x_{r_G}$ is denoted $\boldsymbol{x}$, and the $r_F$-tuple $\xi_1,\dots,\xi_{r_F}$ (with $\xi_a$ related to $v_a$ via $v_a=e^{2\pi i \xi_a}$) is denoted $\boldsymbol{\xi}$. The positive roots of the gauge group are denoted $\alpha_+$, and the weights of the gauge group representation of the chiral multiplet $\chi$ are denoted $\rho^\chi$. The $r_F$-tuple flavor charge of $\chi$ is denoted $q^\chi$, and its $U(1)_R$ charge is denoted $r_\chi$. We assume $r_\chi\in(0,2)$ for all $\chi$. The special function $(\cdot;\cdot)$ is the \emph{Pochhammer symbol}
\begin{equation}
    (z;q):=\prod_{k=0}^{\infty}(1-zq^k),\label{eq:PochDef}
\end{equation}
and the function
$\Gamma_e(\cdot):=\Gamma(e^{2\pi i(\cdot)};p,q)$ is
the \emph{elliptic gamma function} \cite{Ruijsenaars:1997}:
\begin{align}
    \Gamma(z;p,q)&:=\prod_{j,k\ge 0}\frac{1-z^{-1}p^{j+1}q^{k+1}}{1-z
    p^{j}q^{k}}.\label{eq:GammaDef}
\end{align}
Finally, the parameters  $\sigma$, $\tau$ are defined through
\begin{equation}
    p=e^{2\pi i \sigma},\qquad q=e^{2\pi i\tau}.\label{eq:p&qDef}
\end{equation}

For simplicity throughout this paper we restrict our attention to the special case where the flavor fugacities $v_a$ are on the unit circle. Alternatively, we take $\xi_{a}\in\mathbb{R}$.

Defining the four real parameters $b,\beta\in\mathbb{R}_{>0}$ and $k_{1,2}\in\mathbb{R}$ through (\emph{cf.} \cite{Cassani:2021fyv})
\begin{equation}
    \sigma=\frac{i\beta }{2\pi}(b+ik_1),\qquad \tau=\frac{i\beta }{2\pi}(b^{-1}+ik_2),\label{eq:csmSec2}
\end{equation}
the Cardy-like \cite{Cardy:1986ie} limit of our interest in this work corresponds to (\emph{cf.} \cite{Choi:2018hmj})
\begin{equation}
\beta\to 0^+,\ \text{with fixed\
}\, b\in\mathbb{R}_{>0},\  k_{1,2},
\xi_{a}\in\mathbb{R}.\label{eq:CKKNlimit}
\end{equation}
%

The parameters $\omega_{1,2}$ defined as
\begin{equation}
    \omega_{1}:=\frac{2\pi\sigma}{\beta}=i(b+ik_1),\qquad \omega_{2}:=\frac{2\pi\tau}{\beta}=i(b^{-1}+ik_2),\label{eq:thfModuli}
\end{equation}
will be useful below and find a natural interpretation in the 3d EFT describing the Cardy-like limit of the index in Section~\ref{sec:EFT}. Note that they are finite in the limit \eqref{eq:CKKNlimit} and inside the upper-half plane. Below we will assume 
\begin{equation}
\mathrm{Re}\big(\frac{i}{\omega_1\omega_2}\big)\neq0\qquad\text{and}\qquad \mathrm{Re}\big(\frac{i(\omega_1+\omega_2)}{\omega_1\omega_2}\big)\neq0.\label{eq:restrictOmegas}
\end{equation}
This will streamline the analysis of generic theories.\footnote{That is because otherwise $\mathrm{Re}V^{\text{out}}_2$ or $\mathrm{Re}V^{\text{out}}_1$, defined in Eqs.~\eqref{eq:V2out}--\eqref{eq:V1out} and featuring in Definition~\ref{def:hqu}, would vanish. In non-generic situations ($\mathcal{N}=4$ theory with $\xi_a=0$ for example) it may be that $\mathrm{Re}V^{\text{out}}_2$ or $\mathrm{Re}V^{\text{out}}_1$ vanish regardless of such restrictions on $\omega_{1,2}$, in which case those restrictions are not needed/useful, and our analysis applies without demanding~\eqref{eq:restrictOmegas}. Note in particular that in \cite{Ardehali:2015bla}, since the focus was on non-chiral theories with $\xi_a=0$ where $\mathrm{Re}V^{\text{out}}_2$ (\emph{cf.} the function $Q_h$ there) vanishes, $\mathrm{Re}\big(\frac{i}{\omega_1\omega_2}\big)\neq0$ was not assumed. In fact for concreteness (though it was not necessary) it was assumed that $p,q\in\mathbb{R}$, so that $\mathrm{Re}\big(\frac{i}{\omega_1\omega_2}\big)=0$.}

We now proceed to the asymptotic analysis of the index \eqref{eq:EHIgen} in the limit \eqref{eq:CKKNlimit}. Asymptotics of the Pochhammer symbols are elementary (see \emph{e.g.} \cite{Ardehali:2015hya}):
\begin{equation}
    (p;p)(q;q)\simeq e^{-2\pi i\frac{\sigma+\tau}{24\sigma\tau}}\cdot\frac{1}{\sqrt{-\sigma\tau}}\cdot e^{-2\pi i\frac{\sigma+\tau}{24}}\qquad(\text{as }\beta\to0).\label{eq:PochEst}
\end{equation}
Here the choice of branch for the square root is through analytic continuation from $\sigma,\tau\in i\mathbb{R}_{>0}$ where the positive sign is picked. (So in particular for $\sigma=\tau$ we have $1/\sqrt{-\tau^2}=i/\tau$.)

The challenging part is the integral in \eqref{eq:EHIgen}. Asymptotic analysis of an integral requires uniform estimates for its integrand.\footnote{Alternatively, one may be able to integrate (using residue calculus in particular) and then asymptotically analyze. See \cite{GonzalezLezcano:2020yeb,Lezcano:2021qbj,Goldstein:2020yvj,Jejjala:2021hlt} for such analyses in the present context.} Below we divide $\mathfrak{h}_{\text{cl}}$ into an ``outer patch'' and its complement---consisting of various ``inner patches''. We then give separate asymptotic estimates for the elliptic gammas to be used with uniform validity on each patch. Asymptotics of the index is determined by the sum of the contributions from the individual patches. As the reader might anticipate, the competition between the various patches can lead to interesting phase transitions \cite{ArabiArdehali:2019orz}.

\subsection{The outer patch}\label{subsec:outer}

The simplest all-order estimate that we can use for the elliptic gammas in the integrand of \eqref{eq:EHIgen} is (\emph{cf.} \cite{Rains:2006dfy,Ardehali:2015bla,ArabiArdehali:2021nsx})
\begin{equation}
\begin{split}
    \Gamma_e\big((pq)^{\frac{r}{2}}e^{2\pi i x}\big)&\simeq \exp \biggl(-2\pi i \, \biggl(\,\frac{1}{\sigma\tau}\, \frac{\overline{B}_3(x)}{3!}
    + \frac{1}{\sigma\tau}\big(\frac{\sigma+\tau}{2}\big)\, (r -1) \frac{\overline{B}_2(x)}{2!}\\
    &\quad\qquad\qquad\qquad+\frac{3(r-1)^2 (\sigma+\tau)^2-(\sigma^2+\tau^2)}{24\sigma\tau} \, \overline{B}_1(x)\\ 
    &\quad\qquad\qquad\qquad+\big((r-1)^3-(r-1)\big)\frac{(\sigma+\tau)^3}{48\sigma\tau}+(r-1)\frac{\sigma+\tau}{24}\biggr) \biggr) ,\label{eq:outerEst}
    \end{split}
\end{equation}
valid for any $r\in\mathbb{R}$, and point-wise for $x\in\mathbb R\setminus\mathbb Z$. The functions $\overline{B}_{1,2,3}$ above are the \emph{periodic Bernoulli polynomials}, explicitly given by
\begin{equation}
\begin{split}
    \overline{B}_3(x)&:=B_3(\{x\})=\frac{1}{2}\{x\}(1-\{x\})(1-2\{x\}),\\
    \overline{B}_2(x)&:=B_2(\{x\})=-\{x\}(1-\{x\})+\frac{1}{6},\\
    \overline{B}_1(x)&:=\begin{cases}
    B_1(\{x\})=\{x\}-\frac{1}{2}\qquad\, \text{for $x\notin\mathbb{Z}$},\\
    0 \qquad\qquad\qquad\qquad\qquad\text{for $x\in\mathbb{Z}$},
    \end{cases}
    \end{split}
\end{equation}
with $\{\cdot\}:=\cdot-\lfloor\cdot\rfloor$ the \emph{fractional part} function (SawtoothWave[$\cdot$] in Mathematica). Note that $\overline{B}_{3},\overline{B}_{1}$ are odd, while $\overline{B}_{2}$ is even.

We call \eqref{eq:outerEst} the ``outer estimate'', because it applies uniformly for $x$ in compact subsets of $\mathbb{R}\setminus\mathbb{Z}$. That is, it applies \emph{outside} an open neighborhood of $\mathbb{Z}$. To be more precise, we can pick a small positive number $\epsilon$, and indicate the said open neighborhood of $\mathbb{Z}$ by $\mathrm{min}_{n\in\mathbb{Z}}(|x-n|)<\epsilon$. The \emph{outer patch} of the elliptic gamma function in \eqref{eq:outerEst} is then specified by
\begin{equation}
\mathrm{min}_{n\in\mathbb{Z}}(|x-n|)\ge\epsilon.    
\end{equation}
The estimate \eqref{eq:outerEst} is uniformly valid over this outer patch. On the other hand, the open $\epsilon$-neighborhood of an integer can be called an \emph{inner patch}, and the estimate \eqref{eq:outerEst} is not uniformly valid there.

To adapt the preceding discussion to the product of gamma functions in the integrand of the index in \eqref{eq:EHIgen}, we begin by defining the ``singular set'' $\mathcal{S}$ as
\begin{equation}
\begin{split}
\mathcal{S}_g:=\bigcup_{\alpha_+}\{\boldsymbol{x}\in
\mathfrak{h}_{cl}|\, \alpha_+\cdot
\boldsymbol{x}\in&\, \mathbb{Z}\},\quad\quad\mathcal{S}_\chi:=\bigcup_{\rho^\chi\neq0}\{\boldsymbol{x}\in\mathfrak{h}_{cl}|\, \rho^{\chi}\cdot
\boldsymbol{x}+q^\chi\cdot\boldsymbol{\xi}\in\mathbb{Z}\},\\
&\mathcal{S}:=\bigcup_{\chi}\mathcal{S}_\chi\cup
\mathcal{S}_g.\label{eq:SSsDef}
\end{split}
\end{equation}
In the next section the singular set will be interpreted as the subset of the 3d $\mathcal{N}=2$ EFT Coulomb branch where charged massless fields arise. In the present section the significance of $\mathcal{S}$ is that in its neighborhood the outer estimate \eqref{eq:outerEst} loses its uniform validity.

Denote an open $\epsilon$-neighborhood of the singular set by $\mathcal{S}_\epsilon$. The outer patch of the index \eqref{eq:EHIgen}, which we denote by $\mathcal{S}'_\epsilon$, is defined as the complement of $\mathcal{S}_\epsilon$ in $\mathfrak h_{cl}$. More explicitly we define it as follows.
\begin{definition}
The \emph{outer patch} $\mathcal{S}'_\epsilon$ is
the subset of $\mathfrak{h}_{\text{cl}}$ in which for all nonzero $\rho^{\chi}$ we have  $\mathrm{min}_{n\in\mathbb{Z}}(|\rho^{\chi}\cdot
\boldsymbol{x}+q^\chi\cdot\boldsymbol{\xi}-n|)\ge\epsilon$, and also for all $\alpha_+$ we have $\mathrm{min}_{n\in\mathbb{Z}}(|\alpha_+\cdot
\boldsymbol{x}-n|)\ge\epsilon$.\label{def:outer}
\end{definition}

The actual numerical value of $\epsilon$ does not matter at this stage, although for convenience we assume it to be small enough.\footnote{For any given 4d gauge theory with a fixed set of $\xi_a$, there exists a positive number $\epsilon_c$ such that the qualitative structure of the outer patch $\mathcal{S}'_\epsilon$---as well as that of the inner patches discussed below---is similar for all $\epsilon<\epsilon_c.$ (For example, take SU(2) $\mathcal{N}=4$ theory where $\mathfrak{h}_{\text{cl}}$ consists of $x_1\in(-\frac{1}{2},\frac{1}{2}]$, and consider the simple case where $\xi_{1,2,3}=0$, so that $\mathcal{S}=\{0,\frac{1}{2}\}$. Then the outer patch is empty for $\epsilon>\frac{1}{2},$ while it is non-empty with two connected components for $\epsilon<\frac{1}{2}.$ See Figure~\ref{fig:su2decomp}. Therefore in this simple case $\epsilon_c=\frac{1}{2}$.) To avoid undue complications we always take $\epsilon$ to be less than $\epsilon_c$.} Fixing a particular value for $\epsilon$ can be thought of as a choice of ``scheme'', and we will frequently see that unnecessary technical difficulties can be avoided in schemes with small enough $\epsilon$.

With $\epsilon$ suitably small, the outer patch is where the outer estimate \eqref{eq:outerEst} can be applied to all the gamma functions in the index \eqref{eq:EHIgen} (except those with $\rho^\chi=0$ and $q^\chi\cdot\boldsymbol{\xi}\in\mathbb{Z}$, if present, on which we comment below~\eqref{eq:Vin}).

To compute the asymptotic contribution of the outer patch to the index \eqref{eq:EHIgen}---that is the contribution from integrating over $\mathcal{S}'_{\epsilon}\subset\mathfrak{h}_\text{cl}$---we first consider generic $q^\chi\cdot\boldsymbol{\xi}$. Then the outer estimate \eqref{eq:outerEst} can be applied to the gamma functions with $\rho^\chi=0$, if present, as well. (We will relax this genericity assumption below.) Then using \eqref{eq:PochEst} and \eqref{eq:outerEst}
we find 
\begin{equation}
    \mathcal{I}_{\text{out}}(p,q;\boldsymbol{v})\simeq e^{\,\beta E_{\text{susy}}}\int_{\mathcal{S}'_\epsilon}\frac{D\boldsymbol{x}}{(\sqrt{-\omega_1 \omega_2})^{r_G}}\, \big(\frac{2\pi}{\beta}\big)^{r_G} \, e^{-V^{\text{out}}(\boldsymbol{x})}\qquad(\text{for generic }q^\chi\cdot\boldsymbol{\xi}).\label{eq:Iout}
\end{equation}
The potential $V^{\text{out}}$ is given by
\begin{equation}
   V^{\text{out}}(\boldsymbol{x})=\frac{V^{\text{out}}_2(\boldsymbol{x})}{\beta^2}+\frac{V^{\text{out}}_1(\boldsymbol{x})}{\beta}+V^{\text{out}}_0(\boldsymbol{x}),\label{V:out}
\end{equation}
with
\begin{subequations}
\begin{align}
    V^{\text{out}}_2(\boldsymbol{x})&=\frac{(2\pi)^3 i}{\omega_1\omega_2}\, \sum_{\chi}\sum_{\rho^\chi}\frac{\overline{B}_3 \bigl(\rho^\chi \cdot \boldsymbol{x}  +q^\chi\cdot\boldsymbol{\xi} \bigr)}{3!},
\label{eq:V2out}\\
    V^{\text{out}}_1(\boldsymbol{x})&=\frac{(2\pi)^2 i }{\omega_1\omega_2}\big(\frac{\omega_1+\omega_2}{2})\, \bigg( \sum_{\chi}\sum_{\rho^\chi}(r_\chi-1)\frac{\overline{B}_2 \bigl(\rho^\chi \cdot \boldsymbol{x}  +q^\chi\cdot\boldsymbol{\xi} \bigr)}{2!}+ \sum_{\alpha}\frac{\overline{B}_2 \bigl(\alpha \cdot \boldsymbol{x}\bigr)}{2!}\bigg),
\label{eq:V1out}\\
    V^{\text{out}}_0(\boldsymbol{x})&=\frac{2\pi i}{\omega_1\omega_2}\, \sum_{\chi}\sum_{\rho^\chi}\bigg(\frac{\big(\omega_1+\omega_2\big)^2}{8} (r_\chi-1)^2 -\frac{\omega_1^2+\omega_2^2}{24}\bigg)\overline{B}_1 \bigl(\rho^\chi \cdot \boldsymbol{x}  +q^\chi\cdot\boldsymbol{\xi} \bigr).
    \label{eq:V0out}
\end{align}\label{eq:Vout}%
\end{subequations}
Finally, the supersymmetric Casimir energy in \eqref{eq:Iout} is given by (\emph{cf.} \cite{Assel:2015nca})
\begin{equation}
\begin{split}
     E_\text{susy}&=-i\frac{(\omega_1+\omega_2)^3}{48\omega_1\omega_2}\big(\mathrm{Tr}R^3-\mathrm{Tr}R\big)-i\frac{\omega_1+\omega_2}{24}\mathrm{Tr}R.
     \end{split}\label{eq:Esusy}
\end{equation}

Some general remarks are in order.

First, the outer patch $\mathcal{S}'_\epsilon$ is a union of finitely many, disjoint, convex polytopes $\mathcal{P}^\epsilon_n$.\footnote{The polytopes $\mathcal{P}_n$ in \cite{Ardehali:2015bla} are related to our $\mathcal{P}^\epsilon_n$ as $\lim_{\epsilon\to0}\mathcal{P}^\epsilon_n=\mathcal{P}_n $.} On each $\mathcal{P}^\epsilon_n$ the functions $V^{\text{out}}_{2,1,0}$ are analytic. Their non-analyticity occurs across the singular set $\mathcal{S}$ which is not part of $\mathcal{S}'_\epsilon$.

Next, although $\overline{B}_3$ is piecewise cubic, thanks to the (gauge)$^3$ anomaly cancellation the potential $V^{\text{out}}_2$ is piecewise quadratic---and hence quadratic on each $\mathcal{P}^\epsilon_n$. Similarly, because of the $U(1)_R$-(gauge)$^2$ anomaly cancellation $V^{\text{out}}_1$ is piecewise linear. Also, since we are considering semi-simple gauge groups, for a chiral multiplet $\chi$ in any representation we have $\sum_\chi \rho^\chi=0$, and therefore $V^{\text{out}}_0$ is piecewise constant.

Finally, note that $\mathcal{I}_{\text{out}}$ has in general a complicated dependence on $\epsilon$ (\emph{i.e.} ``scheme dependence'') through the range of integration in \eqref{eq:Iout}. However, it is not difficult to see that when $\mathrm{Re}V^{\text{out}}_2$ is minimized strictly within the outer patch, the sensitivity of $\mathcal{I}_{\text{out}}$ to $\epsilon$ is exponentially small. This is because
the integral in \eqref{eq:Iout} is determined, up to exponentially small error, by the contribution from a small neighborhood of the locus of minima of $\mathrm{Re}V^{\text{out}}_2.$\footnote{As we will discuss momentarily, when this locus is extended, the dominant contribution to the asymptotics of the index comes from near the subset of it where $\mathrm{Re}V^{\text{out}}_1$ is minimized.}\\

Let us now consider a simple such scenario, where $\mathrm{Re}V^{\text{out}}_2$ is minimized on a single point $\boldsymbol{x}^\ast\in\mathcal{P}^\epsilon_{n_\ast} \subset\mathcal{S}'_\epsilon$, and evaluate $\mathcal{I}_{\text{out}}$. We can of course restrict the integration domain in \eqref{eq:Iout} to $\mathcal{P}^\epsilon_{n_\ast}$ (or in fact any compact subset of it containing $\boldsymbol{x}^\ast$) because our assumption that $\mathrm{Re}V^{\text{out}}_2$ is minimized at $\boldsymbol{x}^\ast$ implies that the integrand is exponentially suppressed away from $\boldsymbol{x}^\ast$. Moreover, on $\mathcal{P}^\epsilon_{n_\ast}\subset\mathcal{S}'_\epsilon$ the Taylor expansions of $V^{\text{out}}_{2,1,0}$ are exact. Therefore using $\partial_{x_j}V^{\text{out}}_2(\boldsymbol{x}^\ast)=0$ and $\overline{B}'_j(x)=j\overline{B}_{j-1}(x)$ we can estimate the integral \eqref{eq:Iout} as
\begin{equation}
    \mathcal{I}_{\text{out}}(p,q;\boldsymbol{v})\simeq e^{\,\beta E_{\text{susy}}}\, e^{-V^{\text{out}}_\ast}\, Z_{n_\ast}^{3\mathrm{d}}(\omega_1,\omega_2;\frac{2\pi\epsilon}{\beta}) ,\label{eq:IoutIsoCut}
\end{equation}
where $V^{\text{out}}_\ast:=V^{\text{out}}(\boldsymbol{x}^\ast)$, and\footnote{In the latest version of this manuscript, we have adopted a convention related by a parity transformation $k_{ij},k_{jR}\to -k_{ij},-k_{jR}$ to the earlier versions. As a result, there are sign changes in Eqs.~\eqref{eq:3dZout}, \eqref{eq:CSandFIasy}, \eqref{eq:3dZoutInfty}, \eqref{eq:3dZin}, \eqref{eq:CSandFIasy}, and \eqref{eq:3dZinInfty} below. Note that the new convention (unlike the previous one) is also consistent with Eqs.~\eqref{eq:3dZcut} and \eqref{eq:CScoupling}.}
\begin{equation}
    Z_{n_\ast}^{3\mathrm{d}}(\omega_1,\omega_2;\frac{2\pi\epsilon}{\beta}):=\int_{\frac{2\pi}{\beta}\mathcal{P}^\epsilon_{n_\ast}}\frac{D\tilde{\boldsymbol{x}}}{(\sqrt{-\omega_1 \omega_2})^{r_G}}\, e^{\frac{2\pi i}{\omega_1\omega_2}k^\text{out}_{ij}\frac{\tilde{x}_i\tilde{x}_j}{2}\, +\frac{2\pi i}{\omega_1\omega_2}(\frac{\omega_1+\omega_2}{2})k^\text{out}_{jR}\tilde{x}_j}.\label{eq:3dZout}
\end{equation}
Here $\tilde{x}_j:=\frac{2\pi}{\beta}(x_j-x^\ast_j)$, and we have denoted the rescaled polytope by $\frac{2\pi}{\beta}\mathcal{P}^\epsilon_{n_\ast}$. The coefficients $k^\text{out}_{ij}$ and $k^\text{out}_{jR}$ are given by
\begin{equation}
\begin{split}
    k^\text{out}_{ij}&=-\sum_{\chi}\sum_{\rho^\chi}\overline{B}_1 \bigl(\rho^\chi \cdot \boldsymbol{x}^\ast  +q^\chi\cdot\boldsymbol{\xi} \bigr)\, \rho^{\chi}_i\, \rho^{\chi}_j,\\
    k^\text{out}_{j R}&=-\sum_{\chi}\sum_{\rho^\chi}\overline{B}_1 \bigl(\rho^\chi \cdot \boldsymbol{x}^\ast  +q^\chi\cdot\boldsymbol{\xi} \bigr)\, \rho^{\chi}_j\,  (r_\chi-1)-\sum_{\alpha}\overline{B}_1 \bigl(\alpha \cdot \boldsymbol{x}^\ast \bigr)\, \alpha_j.
    \end{split}\label{eq:CSandFIasy}
\end{equation}
Note that our assumption that $\mathrm{Re}V^{\text{out}}_2$ is minimized at $\boldsymbol{x}^\ast$ implies that the second derivative of $\mathrm{Re}V^{\text{out}}_2$ is positive definite at $\boldsymbol{x}^\ast$, and hence
\begin{equation}
    \mathrm{Re}\big(\frac{i}{\omega_1\omega_2}\big)k_{ij}^\text{out}
\end{equation}
is a negative definite matrix. Therefore by replacing the integration domain in \eqref{eq:3dZout} with $x_j\in(-\infty,\infty)$ we would only introduce an exponentially small error. This removes the $\epsilon$-dependence of $Z_{n_\ast}^{3\mathrm{d}}$, and demonstrates the exponentially small sensitivity of $\mathcal{I}_{\text{out}}$ to $\epsilon$ in the present scenario as mentioned above. The end result is the simplification of \eqref{eq:IoutIsoCut} to
\begin{equation}
    \mathcal{I}_{\text{out}}(p,q;\boldsymbol{v})\simeq e^{\,\beta E_{\text{susy}}}\, e^{-V^{\text{out}}_\ast}\, Z_{n_\ast}^{3\mathrm{d}}(\omega_1,\omega_2), \label{eq:IoutIsoInfty}
\end{equation}
where
\begin{equation}
    Z_{n_\ast}^{3\mathrm{d}}(\omega_1,\omega_2):=Z_{n_\ast}^{3\mathrm{d}}(\omega_1,\omega_2;\infty)=\int_{-\infty}^{\infty}\frac{D\tilde{\boldsymbol{x}}}{(\sqrt{-\omega_1 \omega_2})^{r_G}}\, e^{\frac{2\pi i}{\omega_1\omega_2}k^\text{out}_{ij}\frac{\tilde{x}_i\tilde{x}_j}{2}\, +\frac{2\pi i}{\omega_1\omega_2}(\frac{\omega_1+\omega_2}{2})k^\text{out}_{jR}\tilde{x}_j}.\label{eq:3dZoutInfty}
\end{equation}
We emphasize that \eqref{eq:IoutIsoInfty} is valid for isolated $\boldsymbol{x}^\ast$ in the outer patch (and for $q^\chi\cdot\boldsymbol{\xi}\notin\mathbb{Z}$, or more generally in absence of light zero weights for chiral multiplets with $q^\chi\cdot\boldsymbol{\xi}\in\mathbb{Z}$).

If $\mathrm{Re}V^{\text{out}}_2$ is minimized on a finite number of isolated points within the outer patch, each would contribute to $\mathcal{I}_{\text{out}}$ as in \eqref{eq:IoutIsoInfty}, therefore to get the total contribution we should replace $Z_{n_\ast}^{3\mathrm{d}}(\omega_1,\omega_2)$ on the RHS of \eqref{eq:IoutIsoInfty} with $\sum_{n_\ast}Z_{n_\ast}^{3\mathrm{d}}(\omega_1,\omega_2)$.

If $\mathrm{Re}V^{\text{out}}_2$ is minimized not at isolated points but on an extended locus within the outer patch, then $\mathcal{I}_{\text{out}}$ is determined, up to exponentially small error, by the contribution from near the subset of that extended locus where $\mathrm{Re}V^{\text{out}}_1$ is minimized. We denote this subset by $\mathfrak{h}_{\text{qu}}$. If the latter consists of an isolated point, then the derivation of \eqref{eq:IoutIsoInfty} still applies (and if there is a degeneracy, $Z_{n_\ast}^{3\mathrm{d}}$ on the RHS of \eqref{eq:IoutIsoInfty} should be replaced with $\sum_{n_\ast}Z_{n_\ast}^{3\mathrm{d}}$ as in the previous paragraph). Now $\mathrm{Re}\big(\frac{i}{\omega_1\omega_2}\big)k_{ij}^\text{out}$ becomes negative \emph{semi}-definite, with those entries associated with the flat directions of $\mathrm{Re}V^{\text{out}}_2$ being zero. Still, $Z_{n_\ast}^{3\mathrm{d}}(\omega_1,\omega_2)$ remains well-defined, as $k_{jR}^\text{out}$ guarantee exponential decay of the integrand along the flat direction of $\mathrm{Re}V^{\text{out}}_2$ in the present scenario.

If the locus of minima of $\mathrm{Re}V^{\text{out}}_2$ is extended within the outer patch, and the subset of it where $\mathrm{Re}V^{\text{out}}_1$ is minimized---denoted $\mathfrak{h}_{\text{qu}}$---is also extended, then the derivation of \eqref{eq:IoutIsoInfty} fails! This is because the integrand of \eqref{eq:3dZoutInfty} does not decay along the flat directions of both $\mathrm{Re}V^{\text{out}}_2$ and $\mathrm{Re}V^{\text{out}}_1$, so the integration domain in \eqref{eq:3dZoutInfty} cannot be replaced with $x_j\in(-\infty,\infty)$. In other words, now the asymptotics of $\mathcal{I}_{\text{out}}$ is sensitive to $\epsilon.$ As we will see, this sensitivity turns out to be in an overall $\mathcal{O}(\beta^0)$ multiplicative factor in \eqref{eq:IoutIsoCut}, so let us for the moment consider only the singular asymptotics of $\log\mathcal{I}_{\text{out}}$ which remains insensitive to $\epsilon.$ We denote the number of flat directions by $\mathrm{dim}\mathfrak{h}_{\text{qu}}$. (We will give a more precise definition of $\mathrm{dim}\mathfrak{h}_{\text{qu}}$ below.) Then $Z_{n_\ast}^{3\mathrm{d}}(\omega_1,\omega_2;\frac{2\pi\epsilon}{\beta})$ would have a
\begin{equation}
    \big(\frac{1}{\beta}\big)^{\mathrm{dim}\mathfrak{h}_{\text{qu}}}
\end{equation}
divergence from integrating along the flat directions. Denoting the value of $V_{1,2}^{\text{out}}$ on $\mathfrak{h}_{\text{qu}}$ by $V_{1,2\ast}^{\text{out}}$, from \eqref{V:out} and \eqref{eq:IoutIsoCut} we obtain 
\begin{equation}
    \mathcal{I}_{\text{out}}(p,q;\boldsymbol{v})\approx  e^{-\frac{V_{2\ast}^{\text{out}}}{\beta^2}}\, e^{-\frac{V_{1\ast}^{\text{out}}}{\beta}}\,  \big(\frac{1}{\beta}\big)^{\mathrm{dim}\mathfrak{h}_{\text{qu}}}, \label{eq:IoutSingular}
\end{equation}
up to an $\mathcal{O}(\beta^0)$ error upon taking logarithms of the two sides.

To have the result \eqref{eq:IoutSingular} apply equally well to all the scenarios in the previous four paragraphs, we define $\mathfrak{h}_{\text{qu}}$ (the ``quantum moduli space'') and $\mathrm{dim}\mathfrak{h}_{\text{qu}}$ in general as follows.
\begin{definition}
Fix $\omega_{1,2},\xi_a$, and consider the locus of minima of $\mathrm{Re}V^{\text{out}}_2$ (as in \eqref{eq:V2out}) inside $\mathfrak{h}_\text{cl}$. The set $\mathfrak{h}_{\text{qu}}$ is defined as the subset of that locus where $\mathrm{Re}V^{\text{out}}_1$ (as in \eqref{eq:V1out}) is minimized. If $\mathfrak{h}_{\text{qu}}$ is a finite set, $\mathrm{dim}\mathfrak{h}_{\text{qu}}$ is defined to be zero. If $\mathfrak{h}_{\text{qu}}$ is extended, it would consist of multiple, possibly intersecting, flat elements inside $\mathfrak{h}_{\text{cl}}$, and $\mathrm{dim}\mathfrak{h}_{\text{qu}}$ is defined to be the dimension of the flat element(s) with largest dimension.\label{def:hqu}
\end{definition}

The claim in the above definition regarding the structure of $\mathfrak{h}_{\text{qu}}$ when it is extended can be demonstrated as follows. Recall that on each component of $\mathfrak{h}_{\text{cl}}\setminus\mathcal{S}$ the potentials $V^{\text{out}}_{1,2}$ are analytic. Let us denote connected components of $\mathfrak{h}_{\text{cl}}\setminus\mathcal{S}$ by  $\mathcal{P}_n:=\lim_{\epsilon\to0}\mathcal{P}^\epsilon_n$. So $V^{\text{out}}_2$ is quadratic on each $\mathcal{P}_n$, and $V^{\text{out}}_1$ is linear. The locus of minima of $\mathrm{Re}V^{\text{out}}_2$ inside each component is hence determined by linear relations arising from differentiating the quadratic expression for $\mathrm{Re}V^{\text{out}}_2$ on $\mathcal{P}_n$, and the subset of this locus where $\mathrm{Re}V^{\text{out}}_1$ is minimized is similarly a linear subset. Therefore $\mathfrak{h}_{\text{qu}}$ consists of flat (or linear) subsets of $\mathfrak{h}_{\text{cl}}\setminus\mathcal{S}$, as well as their extension to $\mathcal{S}$ (because of the continuity of $V^{\text{out}}_{1,2}$). There may be other subsets of $\mathcal{S}$ that are contained in $\mathfrak{h}_{\text{qu}}$ too (besides the subsets arising from extension of $\mathfrak{h}_{\text{qu}}\cap(\mathfrak{h}_{\text{cl}}\setminus\mathcal{S})$ to $\mathcal{S}$), and those will similarly have a flat (linear) structure, as $\mathcal{S}$ is also a union of flat hyperplanes. We therefore conclude that $\mathfrak{h}_{\text{qu}}$ consists of finitely-many, possibly intersecting, flat elements inside $\mathfrak{h}_{\text{cl}}$, as claimed.\footnote{What we call a flat (or linear) subset of $\mathfrak{h}_\text{cl}$, can be alternatively thought of as a convex polytope inside $\mathfrak{h}_\text{cl}$, thanks to the \emph{half-space representation} (or \emph{H-description}) of convex polytopes. See the Wikipedia article on Convex Polytope.}

Equipped with this refined knowledge of the structure of $\mathfrak{h}_{\text{qu}}$, we can now proceed to improve \eqref{eq:IoutSingular} to exponential accuracy. Let us denote by $\mathcal{P}^\epsilon_{n_\ast}$ those outer-patch polytopes that intersect $\mathfrak{h}_{\text{qu}}$. Similarly to how we derived \eqref{eq:IoutIsoCut} we get
\begin{equation}
    \mathcal{I}_{\text{out}}(p,q;\boldsymbol{v})\simeq e^{\,\beta E_{\text{susy}}}\, e^{-V_{\ast}^{\text{out}}}\, \sum_{n_\ast}Z_{n_\ast}^{3\mathrm{d}}(\omega_1,\omega_2;\frac{2\pi\epsilon}{\beta}) .\label{eq:IoutCutGen}
\end{equation}
Here $Z_{n_\ast}^{3\mathrm{d}}(\omega_1,\omega_2;\frac{2\pi\epsilon}{\beta})$ is precisely as before (see \eqref{eq:3dZout}), and $V_{\ast}^{\text{out}}$ stands for $V^{\text{out}}$ evaluated on $\mathfrak{h}_{\text{qu}}$. Now we first note that for each $n_\ast$ the integral in \eqref{eq:3dZout} in the directions perpendicular to $\mathfrak{h}_{\text{qu}}$ gives a multiplicative $\beta$-independent contribution.\footnote{This is implied by an argument similar to the one leading to \eqref{eq:IoutIsoInfty}: the only $\beta$-dependence of $Z^{3\mathrm{d}}_{n_\ast}(\omega_{1},\omega_2;\frac{2\pi\epsilon}{\beta})$ is through the cut-off in \eqref{eq:3dZout}, which can be removed (with exponentially small error) in the directions perpendicular to $\mathfrak{h}_\text{qu}$ where the integrand in \eqref{eq:3dZout} is exponentially suppressed.} To write down the contribution from the directions along $\mathfrak{h}_{\text{qu}}$, let us denote the dimension of $\mathcal{P}^\epsilon_{n_\ast}\cap \mathfrak{h}_{\text{qu}}$ by $\mathrm{dim}\mathfrak{h}^{n_\ast}_{\text{qu}}$. The directions along $\mathfrak{h}_{\text{qu}}$ contribute to $Z_{n_\ast}^{3\mathrm{d}}(\omega_1,\omega_2;\frac{2\pi\epsilon}{\beta})$ in \eqref{eq:3dZout} as
\begin{equation}
    \frac{\big|\frac{2\pi}{\beta}\big(\mathcal{P}^\epsilon_{n_\ast}\cap \mathfrak{h}_{\text{qu}}\big)\big|}{(\sqrt{-\omega_1 \omega_2})^{\mathrm{dim}\mathfrak{h}^{n_\ast}_{\text{qu}}}},\label{eq:flatContOut}
\end{equation}
because $ k^\text{out}_{ij}$ and $k^\text{out}_{jR}$ are zero along $\mathfrak{h}_{\text{qu}}$ (thanks respectively to $V^{\text{out}}_2$ and $V^{\text{out}}_1$ being flat along it). The symbol $\big|\frac{2\pi}{\beta}(\mathcal{P}^\epsilon_{n_\ast}\cap \mathfrak{h}_{\text{qu}})\big|$ stands for the volume of the re-scaled set; it can be alternatively written as $\big(\frac{2\pi}{\beta}\big)^{\mathrm{dim}\mathfrak{h}^{n_\ast}_{\text{qu}}}\big|\mathcal{P}^\epsilon_{n_\ast}\cap \mathfrak{h}_{\text{qu}}\big|$. We can hence simplify \eqref{eq:IoutCutGen} to 
\begin{equation}
    \mathcal{I}_{\text{out}}(p,q;\boldsymbol{v})\simeq e^{\,\beta E_{\text{susy}}}\, e^{-V_\ast^{\text{out}}}\, \big(\sum_{j=0}^{\mathrm{dim}\mathfrak{h}_{\text{qu}}}\frac{C^{\text{out}}_j(\epsilon)}{\beta^j}\big) .\label{eq:IoutCutGenSimp}
\end{equation}
The coefficients $C^{\text{out}}_j(\epsilon)$ above depend on $\epsilon$, as well as on $\omega_{1,2}$ and $\xi_a$. The term $\frac{C^{\text{out}}_j(\epsilon)}{\beta^j}$ in the sum arises\footnote{Here we are assuming that all these \emph{allowed} coefficients are nonzero. That is, they do not vanish due to ``unnatural'' cancellations. This should of course be checked in specific examples.} from those $n_\ast$ for which $\mathrm{dim}\mathfrak{h}^{n_\ast}_{\text{qu}}=j$. If there is no such $n_\ast$, then $C^{\text{out}}_j(\epsilon)$ is zero of course. The highest term in the sum corresponding to $j=\mathrm{dim}\mathfrak{h}_{\text{qu}}$ recovers \eqref{eq:IoutSingular}.

We emphasize that \eqref{eq:IoutCutGenSimp} is valid only if $\mathfrak{h}_{\text{qu}}$ intersects the outer patch. Otherwise, as we will see in Subsection~\ref{subsec:competition}, the contribution of the outer patch is exponentially suppressed compared to the patches (discussed in the next subsection) that do intersect $\mathfrak{h}_{\text{qu}}$, so $\mathcal{I}_\text{out}$ would be negligible with exponentially small error.

\subsection{Inner patches}\label{subsec:inner}

A stronger all-order estimate for the elliptic gamma function is
(\emph{cf.} \cite{Rains:2006dfy,Ardehali:2015bla})
\begin{equation}
\begin{split}
    \Gamma_e\big((pq)^{\frac{r}{2}}e^{2\pi i x}\big)&\simeq \exp \biggl(-2\pi i \, \biggl(\,\frac{1}{\sigma\tau}\, \frac{\overline{K}_3(x)}{3!}
    + \frac{1}{\sigma\tau}\big(\frac{\sigma+\tau}{2}\big)\, (r -1) \frac{\overline{K}_2(x)}{2!}\\
    &\quad\qquad\qquad\qquad+\frac{3(r-1)^2 (\sigma+\tau)^2-(\sigma^2+\tau^2)}{24\sigma\tau} \, \overline{K}_1(x)\\ 
    &\quad\qquad\qquad\qquad+\big((r-1)^3-(r-1)\big)\frac{(\sigma+\tau)^3}{48\sigma\tau}+(r-1)\frac{\sigma+\tau}{24}\biggr) \biggr)\\
    &\quad\times\Gamma_h\big(\frac{2\pi}{\beta}  x^{}_\mathbb{Z}+\big(\frac{\omega_1+\omega_2}{2}\big)\, r\, ;\omega_1,\omega_2\big),\label{eq:innerEst}
    \end{split}
\end{equation}
valid for any $r\in\mathbb{R}$, and point-wise for any $x\in\mathbb{R}$. Here $\Gamma_h(\cdot ;\omega_1,\omega_2)$ is the \emph{hyperbolic gamma function} \cite{Rains:2006dfy}, while $\overline{K}_j(x)$, which we call \emph{modified periodic Bernoulli polynomials}, are defined as
\begin{equation}
    \overline{K}_j(x):=\overline{B}_j(x)+\frac{j}{2}\mathrm{sign}(x^{}_{\mathbb{Z}}) (x^{}_\mathbb{Z} )^{j-1}.\label{eq:KbarDef}
\end{equation}
Here $x^{}_\mathbb{Z} :=x-\mathrm{nint}(x)$, with nint$(\cdot)$ the \emph{nearest integer} function (Round$[\cdot]$ in Mathematica).\footnote{In principle, we can resolve the ambiguity in $\overline{K}_j(x)$ at $x\in\mathbb{Z}+1/2$ either like Mathematica by picking the even number for nint, or by defining $\overline{K}_j(x)$ to be the average of its left and right limits there. This will not be necessary for our purposes though, as outside a small ($\epsilon$-) neighborhood of $\mathbb{Z}$ we stop using \eqref{eq:innerEst} and use the outer estimate \eqref{eq:outerEst} instead.}

The functions $\overline{K}_j$ have three notable properties. First, unlike $\overline{B}_j$, the functions $\overline{K}_j$ are smooth across $\mathbb{Z}$. In other words, the second term on the RHS of \eqref{eq:KbarDef} kills the part of $\overline{B}_j$ that is  non-analytic across $\mathbb{Z}$. This can be seen explicitly in the domain $-\frac{1}{2}<x<\frac{1}{2}$ where
\begin{alignat}{3}
    \overline{B}_3(x)&=x^3-\frac{3}{2}x|x|+\frac{x}{2},\qquad\qquad\qquad &\overline{K}_3(x)&=x^3+\frac{x}{2},\label{eq:Bbar3Kbar3explicit}\\
    \overline{B}_2(x)&=x^2-|x|+\frac{1}{6}, &\overline{K}_2(x)&=x^2+\frac{1}{6},\\
    \overline{B}_1(x)&=x-\frac{\mathrm{sign}(x)}{2}, &\overline{K}_1(x)&=x.
\end{alignat}
The non-analyticity of $\overline{K}_j$ occurs instead across $\mathbb{Z}+\frac{1}{2}$. Second, since for integer $x$ we have $x^{}_\mathbb{Z}=0$, the two functions $\overline{K}_j$ and $\overline{B}_j$ coincide on $\mathbb{Z}$. Third, as can be confirmed from the above explicit expressions in the domain $-\frac{1}{2}<x<\frac{1}{2}$, we have
\begin{equation}
    \big(\frac{\mathrm{d}}{\mathrm{d}x}\big)^{j-1}\, \overline{K}_j(x)\big|^{}_\mathbb{Z}=0.\label{eq:d(j-1)Kbar}\\
\end{equation}

We call \eqref{eq:innerEst} the ``inner estimate'', because it applies uniformly for $x$ \emph{inside} the open neighborhoods of $\mathbb{Z}$ where the outer estimate is not uniformly valid. In fact, the inner estimate \eqref{eq:innerEst} applies uniformly over \emph{all} $x\in\mathbb{R}$. But for $x$ outside an $\epsilon$-neighborhood of $\mathbb{Z}$ the argument of the hyperbolic gamma function diverges, and one can use the asymptotic formula (see \cite{Rains:2006dfy})
\begin{equation}
\begin{split}
    \Gamma_h\big(\frac{2\pi}{\beta}  x+\big(\frac{\omega_1+\omega_2}{2}\big)\, r\, ;\omega_1,\omega_2\big)&\simeq \exp \biggl(2\pi i \, \biggl(\,\frac{1}{\sigma\tau}\, \frac{\frac{3}{2}x|x|}{3!}
    + \frac{1}{\sigma\tau}\big(\frac{\sigma+\tau}{2}\big)\, (r -1) \frac{|x|}{2!}\\
    &\quad\qquad\qquad\qquad+\frac{3(r-1)^2 (\sigma+\tau)^2-(\sigma^2+\tau^2)}{24\sigma\tau} \, \frac{\mathrm{sign}(x)}{2}\biggr) \biggr),\label{eq:GammaHasy}
    \end{split}
\end{equation}
which together with \eqref{eq:KbarDef}  simplifies the inner estimate \eqref{eq:innerEst} back to the outer one \eqref{eq:outerEst}. In other words, the inner estimate is the master estimate: it reproduces the outer estimate \eqref{eq:outerEst} for $x$ in compact subsets of $\mathbb{R}\setminus\mathbb{Z}$, but for $x$ inside open neighborhoods of $\mathbb{Z}$ of the form
\begin{equation}
\mathrm{min}_{n\in\mathbb{Z}}(|x-n|)<\epsilon,
\end{equation}
and in particular inside $\mathcal{O}(\beta)$ neighborhoods of $\mathbb{Z}$ (where the argument of the hyperbolic gamma function in \eqref{eq:innerEst} is finite), it contains more information than the outer estimate.

Now, when do we use the inner estimate \eqref{eq:innerEst} for the gamma functions in the integrand of the index \eqref{eq:EHIgen}? The following definition addresses this question.

\begin{definition}
The set $\mathcal{S}_\epsilon$ is the subset of $\mathfrak{h}_{\text{cl}}$ in which either for some nonzero $\rho^{\chi}$ we have  $\mathrm{min}_{n\in\mathbb{Z}}(|\rho^{\chi}\cdot
\boldsymbol{x}+q^\chi\cdot\boldsymbol{\xi}-n|)<\epsilon$, or for some $\alpha_+$ we have $\mathrm{min}_{n\in\mathbb{Z}}(|\alpha_+\cdot
\boldsymbol{x}-n|)<\epsilon$. We refer to such weights and roots as \emph{light}. Let us include among the light weights also all the zero weights of any chiral multiplet $\chi$ for which $\mathrm{min}_{n\in\mathbb{Z}}(|q^\chi\cdot\boldsymbol{\xi}-n|)<\epsilon$.\footnote{Note that by taking $\epsilon$ small enough we can ensure that only chiral multiplets with $q^\chi\cdot\boldsymbol{\xi}\in\mathbb{Z}$ have such light zero weights. To avoid undue complications, we always take $\epsilon$ as small as needed to simplify the analysis. We hence assume from now on that only the chiral multiplets with $q^\chi\cdot\boldsymbol{\xi}\in\mathbb{Z}$ have light zero weights.} Denoting the corresponding minimizing integer of a light weight $\rho^\chi$ (or a light root $\alpha_+$) by $n^{}_{\rho^\chi}$ (or $n^{}_{\alpha_+}$), we refer to the pair $(\rho^\chi,n^{}_{\rho^\chi})$ (or $(\alpha_+,n^{}_{\alpha_+})$) as a \emph{light mode}. We decompose $\mathcal{S}_\epsilon$ into finitely-many, non-intersecting patches in$_0$, in$_1$, $\dots$, to be referred to as the \emph{inner patches}, distinguished by their differing set of light modes. The inner patch in$_0$ is the one containing the origin $\boldsymbol{x}=0$. \label{def:inner}
\end{definition}

Note that all roots are light within in$_0$. In general, all weights of chiral multiplets $\chi$ satisfying $q^\chi\cdot\boldsymbol{\xi}\in\mathbb Z$ are also light inside in$_0$. An interesting special case is when all $q^\chi\cdot\boldsymbol{\xi}$ are generic $q^\chi\cdot\boldsymbol{\xi}\not\in\mathbb Z$; then in$_0$ would have no light weights.

%
%

Consider the $n$th inner patch in$_n$, and denote its set of light roots and weights by $L_n$. The other roots and weights of the index \eqref{eq:EHIgen} comprise a set that we denote by $H_n$. We apply the inner 
estimate \eqref{eq:innerEst} to the weights and roots in $L_n$, and use the outer estimate \eqref{eq:outerEst} for those in $H_n$. This way we obtain the asymptotic contribution of the patch in$_n$ to the index given in \eqref{eq:EHIgen} as 
\begin{equation}
    \boxed{\begin{split}
    \mathcal{I}_{\text{in}^{}_n}(p,q;\boldsymbol{v})&\simeq e^{\,\beta E_{\text{susy}}}\int_{\text{in}^{}_n}\frac{D\boldsymbol{x}}{(\sqrt{-\omega_1 \omega_2})^{r_G}}\, \big(\frac{2\pi}{\beta}\big)^{r_G} \, e^{-V^{\text{in}_n}(\boldsymbol{x})}\\
    &\qquad\qquad\qquad\qquad\times \frac{\prod_\chi\prod_{\rho^\chi\in L_n}\Gamma_h\big(r_\chi (\frac{\omega_1+\omega_2}{2})+\frac{2\pi}{\beta}(\rho^\chi\cdot \boldsymbol{x}+q^\chi\cdot\boldsymbol{\xi})^{}_\mathbb{Z}\big)}{\prod_{\alpha_+\in L_n}\Gamma_h\big(\frac{2\pi}{\beta}(\alpha_+\cdot\boldsymbol{x})^{}_\mathbb{Z}\big)\Gamma_h\big(\frac{2\pi}{\beta}(-\alpha_+\cdot\boldsymbol{x})^{}_\mathbb{Z}\big)}
    ,
    \end{split}}\label{eq:Iin}
\end{equation}
with $E_\mathrm{susy}$ as before (see \eqref{eq:Esusy}). The potential $V^{\text{in}_n}$ is given by
\begin{equation}
   V^{\text{in}_n}(\boldsymbol{x})=\frac{V^{\text{in}_n}_2(\boldsymbol{x})}{\beta^2}+\frac{V^{\text{in}_n}_1(\boldsymbol{x})}{\beta}+V^{\text{in}_n}_0(\boldsymbol{x}),\label{V:in}
\end{equation}
with
\begin{subequations}
\begin{align}
    V^{\text{in}_n}_2(\boldsymbol{x})&=\frac{(2\pi)^3 i}{\omega_1\omega_2}\, \sum_{\chi}  \bigg[\sum_{\rho^\chi\in H_n}\frac{\overline{B}_3 \bigl(\rho^\chi \cdot \boldsymbol{x}  +q^\chi\cdot\boldsymbol{\xi} \bigr)}{3!}+\sum_{\rho^\chi\in L_n}\frac{\overline{K}_3 \bigl(\rho^\chi \cdot \boldsymbol{x}  +q^\chi\cdot\boldsymbol{\xi} \bigr)}{3!}\bigg],
\label{eq:V2in}\\[.3cm]
    V^{\text{in}_n}_1(\boldsymbol{x})&=\frac{(2\pi)^2 i }{\omega_1\omega_2}\bigg(\frac{\omega_1+\omega_2}{2}\bigg)\, \bigg( \sum_{\chi}(r_\chi-1)\bigg[\sum_{\rho^\chi\in H_n}\frac{\overline{B}_2 \bigl(\rho^\chi \cdot \boldsymbol{x}  +q^\chi\cdot\boldsymbol{\xi} \bigr)}{2!}+\sum_{\rho^\chi\in L_n}\frac{\overline{K}_2 \bigl(\rho^\chi \cdot \boldsymbol{x}  +q^\chi\cdot\boldsymbol{\xi} \bigr)}{2!}\bigg]\nn\\
    &\qquad\qquad\qquad\qquad\qquad+ \sum_{\alpha\in H_n}\frac{\overline{B}_2 \bigl(\alpha \cdot \boldsymbol{x}\bigr)}{2!}+\sum_{\alpha\in L_n}\frac{\overline{K}_2 \bigl(\alpha \cdot \boldsymbol{x}\bigr)}{2!}\bigg),
\label{eq:V1in}\\[.3cm]
    V^{\text{in}_n}_0(\boldsymbol{x})&=\frac{2\pi i}{\omega_1\omega_2}\, \sum_{\chi}\bigg(\frac{\big(\omega_1+\omega_2\big)^2}{8} (r_\chi-1)^2 -\frac{\omega_1^2+\omega_2^2}{24}\bigg)\nn\\
    &\qquad\qquad\qquad~\times\bigg[\sum_{\rho^\chi\in H_n}\overline{B}_1 \bigl(\rho^\chi \cdot \boldsymbol{x}  +q^\chi\cdot\boldsymbol{\xi} \bigr)+\sum_{\rho^\chi\in L_n}\overline{K}_1 \bigl(\rho^\chi \cdot \boldsymbol{x}  +q^\chi\cdot\boldsymbol{\xi} \bigr)\bigg].
    \label{eq:V0in}
\end{align}\label{eq:Vin}%
\end{subequations}

We have boxed Eq.~\eqref{eq:Iin} because everything else regarding the Cardy-like asymptotics of the index follows from it, either as a special case or through elementary manipulations of asymptotic analysis. In particular, the expression \eqref{eq:Iout} for $\mathcal{I}_{\text{out}}$, valid for generic $q^\chi\cdot\boldsymbol{\xi}$, follows from it as the ``special case'' with $L_n=\emptyset$. More generally, for $q^\chi\cdot\boldsymbol{\xi}$ not necessarily generic, we can obtain the correct expression for $\mathcal{I}_{\text{out}}$ from \eqref{eq:Iin} by letting $L_n$ consist of the zero weights ($\rho^\chi=0$) of chiral multiplets $\chi$ for which $q^\chi\cdot\boldsymbol{\xi}\in\mathbb{Z}$ (as well as replacing the integration domain in$_n$ with the outer patch $\mathcal{S}'_\epsilon$ of course).\\

Some general remarks are now in order.

First, note that $V^{\text{in}_n}_{2,1,0}$ in (\ref{eq:Vin}) are obtained from $V^{\text{out}}_{2,1,0}$ in (\ref{eq:Vout}) by replacing the $\overline{B}_{j}$ functions associated to the light weights and roots with $\overline{K}_{j}$ functions. An important consequence of this replacements is that even though the inner patch in$_n$ intersects the singular set $\mathcal{S}$, the functions $V^{\text{in}_n}_{2,1,0}$ are in fact \emph{smooth} on in$_n$. This is because in$_n$, by construction, intersects $\mathcal{S}$ where the arguments of $\overline{K}_{3,2,1}$ in $V^{\text{in}_n}_{2,1,0}$ become integers while those of $\overline{B}_{3,2,1}$ are away from integers; as mentioned below \eqref{eq:KbarDef} the functions $\overline{K}_{j}(x)$ are analytic across $x\in\mathbb{Z}$, while $\overline{B}_{3,2,1}$ are analytic away from integers.

Next, according to their definition in \eqref{eq:KbarDef}, the functions $\overline{K}_{j}(x)$ do not differ from $\overline{B}_{j}(x)$ in their highest power, that is $x^j$. Consequently, the highest powers in $V^{\text{in}_n}_{2,1,0}$ cancel similarly to what we had for $V^{\text{out}}_{2,1,0}$. To be specific, $V^{\text{in}_n}_{2}$ and $V^{\text{in}_n}_{1}$ are respectively quadratic and linear on in$_n$ thanks to the (gauge)$^3$ and $U(1)_R$-(gauge)$^2$ anomaly cancellations, while $V^{\text{in}_n}_{0}$ is constant because of the gauge group being semi-simple.

Finally, in general $\mathcal{I}_{\text{in}^{}_n}$ has a complicated dependence on $\epsilon$ through the range of integration in \eqref{eq:Iin}. However, when $\mathrm{Re}V^{\text{out}}_2$ is minimized on a single point $\boldsymbol{x}^\ast\in \mathrm{in}^{}_{n}$, the sensitivity to $\epsilon$ turns out to be exponentially small and we can evaluate $\mathcal{I}_{\text{in}^{}_n}$ as follows.

We can assume $\boldsymbol{x}^\ast$ is on the singular set $\mathcal{S}$, and furthermore that the arguments of $\overline{K}_{3,2,1}$ in $V^{\text{in}_n}_{2,1,0}$ are all integers at $\boldsymbol{x}^\ast$:
\begin{equation}
    \rho^\chi \cdot \boldsymbol{x}^\ast  +q^\chi\cdot\boldsymbol{\xi}\, ,~\alpha\cdot \boldsymbol{x}^\ast\in\mathbb{Z}\qquad\text{for all }\rho^\chi,\alpha\in L_n.
\end{equation}
This is because otherwise by taking $\epsilon$ small enough we can shrink $\mathrm{in}^{}_{n}$ so that $\boldsymbol{x}^\ast$ falls out of it. (Recall that while $\epsilon$ is a finite number, we frequently take it as small as needed to simplify the analysis.) 

Noting that $\overline{K}_{j}$ coincide with $\overline{B}_{j}$ on integers, and that the substitution $\overline{K}_{j}\to\overline{B}_{j}$ leads to $V^{\text{in}_n}\to V^{\text{out}}$, we deduce that
\begin{equation}
    V^{\text{in}_n}(\boldsymbol{x}^\ast)=V^{\text{out}}(\boldsymbol{x}^\ast).
\end{equation}
Taylor expanding the potentials $V^{\text{in}_n}_{2,1,0}$ around $\boldsymbol{x}^\ast$, and using $\partial_{x_j}V^{\text{out}}_2(\boldsymbol{x}^\ast)=0$ as well as \eqref{eq:d(j-1)Kbar}, we can estimate the integral \eqref{eq:Iin} as
\begin{equation}
\begin{split}
    \mathcal{I}_{\text{in}^{}_n}(p,q;\boldsymbol{v})&\simeq e^{\,\beta E_{\text{susy}}}\, e^{-V^{\text{out}}(\boldsymbol{x}^\ast)}\, Z^{\text{in}_n}_{3d}(\omega_1,\omega_2;\frac{2\pi\epsilon}{\beta}), \label{eq:IinIsoCut}
\end{split}
\end{equation}
with
\begin{equation}
    \begin{split}
    Z^{\text{in}_n}_{3d}(\omega_1,\omega_2;\frac{2\pi\epsilon}{\beta})&:=\int_{\frac{2\pi}{\beta}\text{in}_n}\frac{D\tilde{\boldsymbol{x}}}{(\sqrt{-\omega_1 \omega_2})^{r_G}}\, \, e^{\frac{2\pi i}{\omega_1\omega_2}\, k^{\text{in}_n}_{ij}\, \frac{\tilde{x}_i\tilde{x}_j}{2}\, +\frac{2\pi i}{\omega_1\omega_2}(\frac{\omega_1+\omega_2}{2})\, k^{\text{in}_n}_{jR}\, \tilde{x}_j}\\
    &\qquad\qquad\qquad\qquad\times \frac{\prod_\chi\prod_{\rho^\chi\in L_n}\Gamma_h\big(r_\chi (\frac{\omega_1+\omega_2}{2})+\rho^\chi\cdot \tilde{\boldsymbol{x}}\big)}{\prod_{\alpha_+\in L_n}\Gamma_h\big(\alpha_+\cdot\tilde{\boldsymbol{x}}\big)\Gamma_h\big(-\alpha_+\cdot\tilde{\boldsymbol{x}}\big)}. \label{eq:3dZin} \end{split}
\end{equation}
Here $\tilde{x}_j:=\frac{2\pi}{\beta}(x_j-x^\ast_j)$, and we have denoted the rescaled patch by $\frac{2\pi}{\beta}\text{in}^{}_n$. The coefficients $k^{\text{in}_n}_{ij}$ and $k^{\text{in}_n}_{jR}$ are given by
\begin{equation}
\begin{split}
    k^{\text{in}_n}_{ij}&=-\sum_{\chi}\sum_{\rho^\chi\in H_n}\overline{B}_1 \bigl(\rho^\chi \cdot \boldsymbol{x}^\ast  +q^\chi\cdot\boldsymbol{\xi} \bigr)\, \rho^{\chi}_i\, \rho^{\chi}_j,\\
    k^{\text{in}_n}_{j R}&=-\sum_{\chi}\sum_{\rho^\chi\in H_n}\overline{B}_1 \bigl(\rho^\chi \cdot \boldsymbol{x}^\ast  +q^\chi\cdot\boldsymbol{\xi} \bigr)\, \rho^{\chi}_j\,  (r_\chi-1)-\sum_{\alpha\in H_n}\overline{B}_1 \bigl(\alpha \cdot \boldsymbol{x}^\ast \bigr)\, \alpha_j.
    \end{split}\label{eq:CSandFIasy}
\end{equation}
For large $\tilde{x}_j$, we can use the asymptotics of the hyperbolic gamma function in \eqref{eq:GammaHasy} to see that the integrand of \eqref{eq:3dZin} simplifies back to that of \eqref{eq:3dZout}. (There might be additional contributions from the hyperbolic gamma functions with light zero weights $\rho^\chi=0$ which do not simplify at large $\tilde{x}_j$, but those are independent of $\tilde{x}_j$ anyway and can be taken outside the integral in \eqref{eq:3dZin}.)
So our assumption that $\mathrm{Re}V^{\text{out}}_2$ is minimized at $\boldsymbol{x}^\ast$ implies again that by replacing the integration domain in \eqref{eq:3dZin} with $\tilde x_j\in(-\infty,\infty)$ we would only introduce exponentially small error. This removes the $\epsilon$-dependence of $Z^{\text{in}_n}_{3d}$, and demonstrates the exponentially small sensitivity of $\mathcal{I}_{\text{in}_n}$ to $\epsilon$ in the present scenario as mentioned above. The end result is the simplification of \eqref{eq:IinIsoCut} to
\begin{equation}
\begin{split}
    \mathcal{I}_{\text{in}^{}_n}(p,q;\boldsymbol{v})&\simeq e^{\,\beta E_{\text{susy}}}\, e^{-V^{\text{out}}(\boldsymbol{x}^\ast)}\, Z^{\text{in}_n}_{3d}(\omega_1,\omega_2), \label{eq:IinIsoInfty}
\end{split}
\end{equation}
where
\begin{equation}
    \begin{split}
    Z^{\text{in}_n}_{3d}(\omega_1,\omega_2):=Z^{\text{in}_n}_{3d}(\omega_1,\omega_2;\infty)&=\int_{-\infty}^{\infty}\frac{D\tilde{\boldsymbol{x}}}{(\sqrt{-\omega_1 \omega_2})^{r_G}}\, \, e^{\frac{2\pi i}{\omega_1\omega_2}\, k^{\text{in}_n}_{ij}\, \frac{\tilde{x}_i\tilde{x}_j}{2}\, +\frac{2\pi i}{\omega_1\omega_2}(\frac{\omega_1+\omega_2}{2})\, k^{\text{in}_n}_{jR}\, \tilde{x}_j}\\
    &\qquad\qquad\qquad\qquad\times \frac{\prod_\chi\prod_{\rho^\chi\in L_n}\Gamma_h\big(r_\chi (\frac{\omega_1+\omega_2}{2})+\rho^\chi\cdot \tilde{\boldsymbol{x}}\big)}{\prod_{\alpha_+\in L_n}\Gamma_h\big(\alpha_+\cdot\tilde{\boldsymbol{x}}\big)\Gamma_h\big(-\alpha_+\cdot\tilde{\boldsymbol{x}}\big)}. \label{eq:3dZinInfty} \end{split}
\end{equation}

If there are flat directions, the derivation of \eqref{eq:IinIsoInfty} fails, because the integrand of \eqref{eq:3dZin} would not decay along the flat direction. Similarly to how we derived \eqref{eq:IoutCutGenSimp} however, we now argue that a similar expression applies here.

\subsubsection*{Without singular intersections}

First assume that in$_n$ contains parts of a number of hyperplanes $\mathcal S_1,\dots,\mathcal S_m,\dots$ belonging to the singular set $\mathcal S$, but does not contain any intersections of them. That is, $\mathrm{in}_n\cap\mathcal{S}$ consists of $\mathrm{in}_n\cap\mathcal{S}_1,\dots,\mathrm{in}_n\cap\mathcal{S}_m,\dots$ that do not intersect, and are hence at $\mathcal{O}(\epsilon)$ distances from each other. 

Write the asymptotic contribution of in$_n$ to the index as $\int_{\text{in}_n} I_{\text{in}_n}$, where the integrand can be read from \eqref{eq:Iin}, and write the asymptotic contribution of the outer patch to the index as $\int_{\text{out}} I_{\text{out}}$, where the integrand can be read for generic $q^\chi\cdot\boldsymbol{\xi}$ from \eqref{eq:Iout} or more generally from \eqref{eq:Iin} as explained below \eqref{eq:Vin}. Now write
\begin{equation}
  \int_{\text{in}_n} I_{\text{in}_n}=\int_{\text{in}_n} I_{\text{out}}+\int_{\text{in}_n} (I_{\text{in}_n}-I_{\text{out}}).\label{eq:out+(in-out)}
\end{equation}
It should be clear from our earlier discussion of the outer patch contribution that the first integral on the RHS localizes with exponential accuracy onto $\text{in}_n\cap \mathfrak{h}_{\text{qu}}$. It then contributes to the index an expression similar to \eqref{eq:IoutCutGenSimp}.

To estimate the second integral on the RHS of \eqref{eq:out+(in-out)}, first note that $I_{\text{in}_n}$ approaches  $I_{\text{out}}$ exponentially fast as we move away from $\mathcal{S}$. (Because the inner estimate recovers the outer estimate exponentially accurately away from $\mathbb{Z}$.) Therefore the second integral localizes with exponential accuracy to $\text{in}_n\cap\mathcal{S}$, which we can write alternatively as $\text{in}_n\cap(\mathcal{S}_1\cup\dots\cup\mathcal{S}_{m}\dots)$.

Next, consider the contribution from a small neighborhood of one of the components of $\text{in}_n\cap\mathcal{S}$, say the component $\text{in}_n\cap\mathcal{S}_m$. We denote the small neighborhood by $\text{in}_n\cap\mathcal{S}^{\epsilon_1}_m$, with $\epsilon_1$ chosen small enough such that $\text{in}_n\cap\mathcal{S}^{\epsilon_1}_m$ does not intersect any other component of $\text{in}_n\cap\mathcal{S}$. In this region $I_{\text{in}_n}$ is approximated exponentially accurately by what we denote by $I_{m}$. The latter is obtained from \eqref{eq:Iin} by replacing the set $L_n$ with the set of the light weights/roots associated to $\mathcal{S}_m$, which we denote by $L_m$. This replacement entails in particular a replacement $V^{\mathrm{in}_n}\to V^m$, where $V^m$ is obtained from \eqref{eq:Vin} by the replacement $L_n\to L_m$, as well as $H_n\to H_m$, where $H_m$ consists of all the other weight/roots of the index that are not in $L_m$. The second integral on the RHS of \eqref{eq:out+(in-out)} can then be expressed up to exponentially small error as
\begin{equation}
    \begin{split}
\int_{\text{in}_n} (I_{\text{in}_n}-I_{\text{out}})&\simeq\sum_m \int_{\text{in}_n\cap\mathcal{S}^{\epsilon_1}_m} (I_m-I_{\text{out}})=\sum_m \int_{\text{in}_n\cap\mathcal{S}^{\epsilon_1}_m}\, I_{\text{out}}\,  \big(\frac{I_m}{I_{\text{out}}}-1\big)\\
&=\sum_m e^{\,\beta E_{\text{susy}}}\int_{\text{in}_n\cap\mathcal{S}^{\epsilon_1}_m}\frac{D\boldsymbol{x}}{(\sqrt{-\omega_1 \omega_2})^{r_G}}\, \big(\frac{2\pi}{\beta}\big)^{r_G} \, e^{-V^{\text{out}}(\boldsymbol{x})}\, g_m\big(\frac{2\pi x^\perp_m}{\beta}\big),
    \end{split}\label{eq:Iin-IoutNoIntersection}
\end{equation}
where
\begin{equation}
   g_m\big(\frac{2\pi x^\perp_m}{\beta}\big):=\frac{I_m}{I_{\text{out}}}-1=e^{-V^m(\boldsymbol{x})+V^{\text{out}}(\boldsymbol{x})}\,  \frac{\prod_\chi\prod_{\rho^\chi\in L_m}\Gamma_h\big(r_\chi (\frac{\omega_1+\omega_2}{2})+\frac{2\pi}{\beta}(\rho^\chi\cdot \boldsymbol{x}+q^\chi\cdot\boldsymbol{\xi})^{}_\mathbb{Z}\big)}{\prod_{\alpha_+\in L_m}\Gamma_h\big(\frac{2\pi}{\beta}(\alpha_+\cdot\boldsymbol{x})^{}_\mathbb{Z}\big)\Gamma_h\big(\frac{2\pi}{\beta}(-\alpha_+\cdot\boldsymbol{x})^{}_\mathbb{Z}\big)}-1.\label{eq:gDef}
\end{equation}
Above we have used the coordinate $x^\perp_m$ which measures the distance from $\mathcal{S}_m$.\footnote{Implicit in writing \eqref{eq:gDef} is the claim that the RHS is only a function of $x^\perp_m/\beta$. While this is obvious for the hyperbolic gammas because of the form of their arguments, for the combination $-V^m(\boldsymbol{x})+V^{\text{out}}(\boldsymbol{x})$ it is not so obvious. To see why that is the case, let us consider $(-V_2^m+V_2^{\text{out}})/\beta^2.$ (Similar considerations apply to the other pieces.) Using \eqref{eq:V2out} and \eqref{eq:V2in}
we have $-V_2^m+V_2^{\text{out}}\propto\sum_{\rho^\chi\in L_m}[-\overline{K}_3 \bigl(\rho^\chi \cdot \boldsymbol{x}  +q^\chi\cdot\boldsymbol{\xi} \bigr)+\overline{B}_3 \bigl(\rho^\chi \cdot \boldsymbol{x}  +q^\chi\cdot\boldsymbol{\xi} \bigr)]$; noting from \eqref{eq:Bbar3Kbar3explicit} that the cubic and linear parts of $\bar{B}_3$ and $\bar{K}_3$ are equal, we deduce that only the quadratic part of $\bar{B}_3$ survives in the expression for $-V_2^m+V_2^{\text{out}}$. Since for $\rho^\chi\in L_m$ the combination $\rho^\chi \cdot \boldsymbol{x}  +q^\chi\cdot\boldsymbol{\xi}$ is proportional to $x^\perp_m$, we conclude that $(-V_2^m+V_2^{\text{out}})/\beta^2$ is a homogeneous function of $x^\perp_m/\beta$ of degree two.} Note that $g_m\big(\frac{2\pi x^\perp_m}{\beta}\big)$ decays to zero exponentially fast for any finite $x^\perp_m$, because $I_m$ approaches $I_\text{out}$ exponentially fast as we move away from $\mathcal S_m$.

Let us assume $\text{in}_n\cap\mathcal{S}_m$ intersects $\mathfrak{h}_{\text{qu}}$; otherwise as the argument below \eqref{eq:Isum} shows the contribution from $\text{in}_n\cap\mathcal{S}^{\epsilon_1}_m$ to the index is exponentially smaller than those from the patches that do intersect $\mathfrak{h}_{\text{qu}}$, and hence negligible. This assumption requires in particular that $\text{in}_n$ intersects $\mathfrak{h}_{\text{qu}}$. Let us denote the inner patches intersecting $\mathfrak{h}_{\text{qu}}$ by $\text{in}_{n_\ast}$.

The integrand of the $m$th term in \eqref{eq:Iin-IoutNoIntersection} is the product of two functions: $e^{-V^{\text{out}}}$ and $g_m$. Intuitively, we expect the integral to localize around $\mathrm{in}_n\cap\mathcal{S}_m\cap\mathfrak{h}_{\text{qu}}$, because $e^{-V^{\text{out}}}$ is  exponentially suppressed away from $\mathfrak{h}_{\text{qu}}$, while $g_m$ is exponentially suppressed away from $\mathcal{S}_m$. To be able to control the details however, we proceed as follows. We first integrate over the constant-$x^\perp_m$ slices parallel to $\mathcal{S}_m$. At this stage $g_m$ can be treated as a constant, and the analysis is quite similar to the one we performed for the outer patch around \eqref{eq:IoutCutGen}. The integral localizes onto $\mathrm{in}_n\cap\mathfrak{h}_{\text{qu}}$, with an integrand that is proportional to $e^{\beta E_\text{susy}}e^{-V^\text{out}_\ast}g_m\big(\frac{2\pi x^\perp_m}{\beta}\big)$. The remaining integral over $x^\perp_m$ clearly localizes onto $x^\perp_m=0$, that is onto $\mathrm{in}_n\cap\mathcal{S}_m\cap\mathfrak{h}_{\text{qu}}$. Thus the $m$th term in \eqref{eq:Iin-IoutNoIntersection} would have asymptotics similar to \eqref{eq:IoutCutGenSimp}, namely
\begin{equation}
    m\text{th term in \eqref{eq:Iin-IoutNoIntersection}}\simeq \sum_{j_m} e^{\beta E_\text{susy}}e^{-V^\text{out}_*}\,\fft{C_{j_m}(\epsilon)}{\beta^{j_m}},\label{I:mth}
\end{equation}
where $C_{j_m}(\epsilon)$ is nonzero only if $\text{in}_n\cap\mathcal{S}_m\cap\mathfrak{h}_{\text{qu}}$ has a component of dimension $j_m$.
Summing the asymptotics (\ref{I:mth}) over $m$ as required in \eqref{eq:Iin-IoutNoIntersection}, we conclude that $\mathcal I_{\text{in}_n}=\int_{\text{in}_n}I_{\text{in}_n}$ has the same asymptotic form as in \eqref{eq:IoutCutGenSimp}, where the polynomial factor here would only contain those powers of $\frac{1}{\beta}$ that correspond to the dimensions of the various components of $\text{in}_n\cap\mathcal{S}\cap\mathfrak{h}_{\text{qu}}$.

To summarize, the first integral on the RHS of  \eqref{eq:out+(in-out)} localizes to $\text{in}_n\cap\mathfrak{h}_{\text{qu}}$, while the second integral localizes to $\text{in}_n\cap\mathcal{S}\cap\mathfrak{h}_{\text{qu}}$. Both yield asymptotics of the form \eqref{eq:IoutCutGenSimp}, implying a similar asymptotic for the left-hand side, as claimed.

\subsubsection*{With singular intersections}

Now assume that in$_n$ contains intersections of several hyperplanes in $\mathcal{S}$. (We can ensure no inner patch contains two or more singular intersection sets away from each other by, again, taking $\epsilon$ to be small enough.) More explicitly, say in$_n$ intersects a number of singular hyperplanes $\mathcal{S}_1,\dots, \mathcal{S}_m,\dots$ as well as  (part or all of) some intersections of those hyperplanes. Then the following refined version of \eqref{eq:out+(in-out)} becomes useful:
\begin{equation}
  \int_{\text{in}_n} I_{\text{in}_n}=\int_{\text{in}_n} I_{\text{out}}+\int_{\text{in}_n}\sum_m (I_m-I_{\text{out}})+\int_{\text{in}_n}\big((I_{\text{in}_n}-I_{\text{out}})-\sum_m(I_m-I_{\text{out}})\big).\label{eq:out+(in-out)Refined}
\end{equation}
An argument similar to the one we applied to \eqref{eq:Iin-IoutNoIntersection} shows that the first integral on the RHS of \eqref{eq:out+(in-out)Refined} localizes to $\text{in}_n\cap \mathfrak{h}_{\text{qu}}$, while the second integral localizes to $\cup_m(\mathrm{in}_n\cap\mathcal{S}_m\cap\mathfrak{h}_{\text{qu}})$, and the last integral localizes\footnote{To shows that the third integral on the RHS of \eqref{eq:out+(in-out)Refined} receives non-negligible contributions only from near intersections of singular hyperplanes, let us demonstrate exponential suppression of its integrand away from all such singular intersections: if we are also away from all singular hyperplanes, then $I_{\text{in}_n}-I_\text{out}$ and $I_j-I_\text{out}$ (for all $j$) are separately exponentially suppressed and we are done; if we are near a hyperplane $\mathcal{S}_m$, then $I_{\text{in}_n}-I_\text{out}$ is not exponentially suppressed, and neither is $I_m-I_\text{out}$, but $(I_{\text{in}_n}-I_\text{out})-(I_m-I_\text{out})=I_{\text{in}_n}-I_m$ is exponentially suppressed (because $I_m$ approximates $I_{\text{in}_n}$ exponentially accurately if we are away from the other singular hyperplanes) and we are done.} to the various singular intersections inside $\text{in}_n\cap \mathfrak{h}_{\text{qu}}$. Moreover, for the two latter cases the integrand of the localized integrals coincides with $I_{\text{out}}$ up to $\beta$-independent factors (see the paragraph of \eqref{I:mth}). Therefore all the three integrals on the right-hand side of \eqref{eq:out+(in-out)Refined} give asymptotics of the form \eqref{eq:IoutCutGenSimp}, implying a similar asymptotic for the left-hand side, as claimed:
\begin{equation}
    \mathcal{I}_{\text{in}_{n_\ast}}(p,q;\boldsymbol{v})\simeq e^{\,\beta E_{\text{susy}}}\, e^{-V_\ast^{\text{out}}}\, \big(\sum_{j=0}^{\mathrm{dim}\mathfrak{h}_{\text{qu}}}\frac{C^{\text{in}_{n_\ast}}_j(\epsilon)}{\beta^j}\big) .\label{eq:IinCutGenSimp}
\end{equation}
We emphasize again that this asymptotic relation applies only to those inner patches $\text{in}_{n_\ast}$ that intersect $\mathfrak{h}_{\text{qu}}$. An inner patch in$_n$ that does not intersect $\mathfrak{h}_{\text{qu}}$ has asymptotics that is exponentially smaller than \eqref{eq:IinCutGenSimp} as we argue in the next subsection

An important aspect of the result \eqref{eq:IinCutGenSimp} is that the singular asymptotics of $\log \mathcal{I}_{\text{in}^{}_{n_\ast}}$ is determined entirely by the \emph{outer} potentials $V_{2,1}^{\text{out}}$.


\subsection{Competition between the patches}\label{subsec:competition}

Asymptotics of the index \eqref{eq:EHIgen} is obtained by combining the contributions of the outer patch and the inner patches given in
\eqref{eq:Iout} and \eqref{eq:Iin} respectively:
\begin{equation}
     \mathcal{I}(p,q;\boldsymbol{v})= \mathcal{I}_{\text{out}}(p,q;\boldsymbol{v})+\sum_n  \mathcal{I}_{\text{in}^{}_n}(p,q;\boldsymbol{v}).\label{eq:Isum}
\end{equation}
One or several of the patches could give the dominant contribution to the index through \eqref{eq:Isum} in the Cardy-like limit. We now argue that these are the patches that intersect $\mathfrak{h}_{\text{qu}}$.

We do this in two steps. In the first step we argue that if the outer patch does not intersect $\mathfrak{h}_{\text{qu}}$, its contribution is exponentially suppressed compared to those inner patches that do intersect $\mathfrak{h}_{\text{qu}}$. This is quite straightforward in fact, starting from \eqref{eq:Iout} (or its generalization for $q^\chi\cdot\boldsymbol{\xi}$ not necessarily generic, that follows from \eqref{eq:Iin}). We simply note that the minimum of $\mathrm{Re}V^{\text{out}}_2$ over the outer patch is greater than $\mathrm{Re}V^{\text{out}}_{2\ast}$ by a strictly positive number of order $\epsilon$ or $\epsilon^2$. That is because $V^{\text{out}}_{2}(\boldsymbol{x})$ is piecewise quadratic in $x_j$, and the outer patch is a distance $\epsilon$ away from the singular set which we are assuming contains $\mathfrak{h}_\text{qu}$. Therefore the integrand of $\mathcal{I}_{\text{out}}$ is uniformly exponentially ($e^{-1/\beta^2}$ type) suppressed compared to \eqref{eq:IinCutGenSimp}, and we conclude that $\mathcal{I}_{\text{out}}$ is exponentially suppressed relative to $\mathcal{I}_{\text{in}_{n_\ast}}$.\footnote{More precisely, in this paragraph we have assumed that the locus of minima of $\mathrm{Re}V^{\text{out}}_{2}$ does not interect the outer patch. It may be that $\mathrm{Re}V^{\text{out}}_{2}$ is minimized on a locus intersecting the outer patch, but the subset of that locus where $\mathrm{Re}V^{\text{out}}_{1}$ is minimized does not intersect the outer patch. (If it did, $\mathfrak{h}_{\text{qu}}$ would intersect the outer patch, while we are assuming it does not.) A similar argument applies in that case too, only this time with an $e^{-1/\beta}$ type, rather than $e^{-1/\beta^2}$ type, exponential suppression.}

In the second step we argue that if an inner patch in$_n$ does not intersect $\mathfrak{h}_{\text{qu}}$, its contribution is exponentially smaller than the patches that intersect $\mathfrak{h}_{\text{qu}}$. To demonstrate this it would be sufficient to argue that over in$_n$ we have $I_{\text{in}_n}$ uniformly bounded above by a constant multiple of $I_{\text{out}}$,
because then the argument of the previous paragraph applies again and we are done. (Note that we previously discussed uniformly valid exponentially accurate estimates, while here we are appealing to---much weaker---bounds that are uniformly valid.) To establish the uniform boundedness, consider the ratio
\begin{equation}
   \frac{I_{\text{in}_n}}{I_\text{out}}=e^{-V^{\text{in}_n}(\boldsymbol{x})+V^{\text{out}}(\boldsymbol{x})}\,  \frac{\prod_\chi\prod_{\rho^\chi\in L_n}\Gamma_h\big(r_\chi (\frac{\omega_1+\omega_2}{2})+\frac{2\pi}{\beta}(\rho^\chi\cdot \boldsymbol{x}+q^\chi\cdot\boldsymbol{\xi})^{}_\mathbb{Z}\big)}{\prod_{\alpha_+\in L_n}\Gamma_h\big(\frac{2\pi}{\beta}(\alpha_+\cdot\boldsymbol{x})^{}_\mathbb{Z}\big)\Gamma_h\big(\frac{2\pi}{\beta}(-\alpha_+\cdot\boldsymbol{x})^{}_\mathbb{Z}\big)}=:g_{\text{in}_n}\big(\frac{2\pi \boldsymbol{x}^\perp_{\text{in}_n}}{\beta}\big)+1.\label{eq:InOutRatio}
\end{equation}
Here $\boldsymbol{x}^\perp_{\text{in}_n}$ stands for a set of new coordinates measuring distance from the singular hyperplanes intersecting in$_n$, and we have defined $g_{\text{in}_n}$ analogously to \eqref{eq:gDef}. Now define $\mathcal{S}^{(\beta)}$ as the $\mathcal{O}(\beta)$ neighborhood of $\mathcal{S}$ obtained by replacing $\epsilon\to\frac{\Lambda\beta}{2\pi}$ in  $\mathcal{S}^{\epsilon}$, for some $\Lambda>0$. Fixing a particular positive number for $\Lambda$ amounts to a choice of scheme. Now decompose in$_n$ as the union of $\mathrm{in}_n\cap\mathcal{S}^{(\beta)}$ and $\mathrm{in}_n\cap(\mathcal{S}^{\epsilon}\setminus\mathcal{S}^{(\beta)})$. On the latter set, the asymptotics of the hyperbolic gamma function \eqref{eq:GammaHasy} implies that by taking $\Lambda$ large enough we can ensure that $g_{\text{in}_n}$ is arbitrarily close to zero, and therefore we get our desired uniform bound on $I_{\text{in}_n}/I_\text{out}$. It remains to establish a uniform bound on $\mathrm{in}_n\cap\mathcal{S}^{(\beta)}$. For that, we define the re-scaled variables $\tilde{\boldsymbol{x}}^\perp_{\text{in}_n}:=\frac{2\pi}{\beta}\boldsymbol{x}^\perp_{\text{in}_n}$. Then since the function $g_{\text{in}_n}(\tilde{\boldsymbol{x}}^\perp_{\text{in}_n})$ is continuous, it is uniformly bounded on the compact domain $|\tilde{x}^\perp_j|\le\Lambda$, and we are done. 
The said continuity relies on the fact that the poles of the chiral-multiplet gamma functions and the zeros of the vector-multiplet gamma functions in \eqref{eq:InOutRatio} are avoided for $|\tilde{x}^\perp_j|\le\Lambda$, which can be verified assuming $r_\chi>0$.\footnote{To summarize, the logic of the second step is that $|\int_{\text{in}_n}I_{\text{in}_n}|<C\times |\int_{\text{in}_n}I_\text{out}|\ll |\int_{\text{in}_{n_\ast}} I_{\text{in}_{n_\ast}}|$, with some finite positive number $C$, where in$_{n_\ast}$ (but not in$_n$) intersects $\mathfrak h_\text{qu}$, and `$\ll$' represents that the LHS is exponentially suppressed compared to the RHS. The argument establishing a uniform bound is a minor variation of the one in the next-to-last paragraph of Appendix~A of \cite{Ardehali:2015bla}.}

We have thus established the claim that the patches dominating the index are those that intersect the quantum moduli space $\mathfrak{h}_{\text{qu}}$.

We see from \eqref{eq:V2out} and \eqref{eq:V1out} that $\mathfrak{h}_{\text{qu}}$ can depend on the continuous parameters $\xi_a$, as well as the discrete parameters $\mathrm{sign}(\mathrm{Re}(\frac{i}{\omega_1 \omega_2}))$ and $\mathrm{sign}(\mathrm{Re}(\frac{i(\omega_1+\omega_2)}{\omega_1 \omega_2}))$. These \emph{control parameters} provide the arena for the competition between the various patches in \eqref{eq:Isum}. By changing the control parameters the dominant patches might change, and this can be thought of as a phase transition in the index. Since $\beta$ is analogous to inverse-temperature in thermal quantum physics, the transitions that we are discussing in the $\beta\to0$ asymptotics are analogous to infinite-temperature phase transitions.

Note that the structure of the various patches also changes by changing $\xi_a$, because $\mathcal{S}_\chi$ depend on $\xi_a$.

\subsection{The general formula}

We can now restrict the sum in \eqref{eq:Isum} to the patches intersecting $\mathfrak{h}_{\text{qu}}$ to obtain the asymptotics of the index. From \eqref{eq:IoutCutGenSimp} and \eqref{eq:IinCutGenSimp} we obtain
\begin{equation}
    \mathcal{I}(p,q;\boldsymbol{v})\simeq e^{\,\beta E_{\text{susy}}}\, e^{-\frac{V_{2\ast}^{\text{out}}}{\beta^2}-\frac{V_{1\ast}^{\text{out}}}{\beta}-V_{0\ast}^{\text{out}}}\, \big(\sum_{j=0}^{\mathrm{dim}\mathfrak{h}_{\text{qu}}}\frac{C_j}{\beta^j}\big) .\label{eq:ItotGenSimp}
\end{equation}
Here $C_j$ is defined as $C^{\text{out}}_j(\epsilon)+\sum_{n_\ast}C^{\text{in}_{n_\ast}}_j(\epsilon)$ if the outer patch intersects $\mathfrak{h}_{\text{qu}}$, and as $\sum_{n_\ast}C^{\text{in}_{n_\ast}}_j(\epsilon)$ otherwise. Note that while $C^{\text{out}}_j(\epsilon)$ and $C^{\text{in}_{n_\ast}}_j(\epsilon)$ can be $\epsilon$-dependent, the coefficients $C_j$ should be independent of the decomposition scheme. In our way of decomposing $\mathfrak{h}_{\text{cl}}$ a choice of scheme corresponds to picking a specific value for $\epsilon$. Therefore $C_j$ are independent of $\epsilon$, and functions of $\omega_{1,2}$ and $\xi_a$ only.

While most previous papers have focused on cases with $\mathrm{dim}\mathfrak{h}_\text{qu}=0$ where the polynomial factor in \eqref{eq:ItotGenSimp} is simply a constant, non-trivial polynomial factors have been observed in the indices of SO($3$) SQCD with two flavors \cite{Ardehali:2015bla} and SU(2) $\mathcal{N}=4$ theory \cite{Lezcano:2021qbj}.\footnote{As well as in the Schur index \cite{Gadde:2011ik} of SU($N$) $\mathcal{N}=4$ theory \cite{Ardehali:2015bla}.} The non-trivial polynomials were discovered in \cite{Ardehali:2015bla,Lezcano:2021qbj} using alternative representations of the index, which are available in those special theories but not in general: the finding in \cite{Ardehali:2015bla} relied on the free-fermion representation of \cite{Bourdier:2015wda} (available for a limit of the index in a particular class of theories at present \cite{Bourdier:2015sga}), and the one in \cite{Lezcano:2021qbj} used the Bethe-Ansatz approach \cite{Closset:2017bse,Benini:2018mlo} which is not fully reliable at present for rank greater than one \cite{ArabiArdehali:2019orz}. Our demonstration of \eqref{eq:ItotGenSimp} for $\mathrm{dim}\mathfrak{h}_\text{qu}>0$ using the decomposition method applies much more generally to any index of the form \eqref{eq:EHIgen}.

If we are interested only in the singular asymptotics of $\log\mathcal{I}$, from \eqref{eq:ItotGenSimp} it follows that, irrespective of which patches dominate, we have
\begin{equation}
    \log\mathcal{I}(p,q;\boldsymbol{v})= -\frac{V_{2\ast}^{\text{out}}}{\beta^2}-\frac{V_{1\ast}^{\text{out}}}{\beta}+\mathrm{dim}\mathfrak{h}_{\text{qu}}\log  \big(\frac{1}{\beta}\big)+\mathcal{O}(\beta^0), \label{eq:ItotSingular}
\end{equation}
with $V_{2,1\ast}^{\text{out}}$ the value of $V_{2,1}^{\text{out}}(\boldsymbol{x})$ on $\mathfrak{h}_{\text{qu}}$. In particular, the singular asymptotics of $\log\mathcal{I}$ is entirely determined by the functions $V_{2,1}^{\text{out}}(\boldsymbol{x})$.\\

We can summarize our approach and findings as follows.
\begin{itemize}
    \item We start by decomposing $\mathfrak{h}_\text{cl}$ to an outer patch and finitely-many inner patches. See Definitions~\ref{def:outer} and \ref{def:inner}.
    \item On each patch we use uniform small-$\beta$ estimates (see \eqref{eq:outerEst} and \eqref{eq:innerEst}) to simplify the integrand of the index. This amounts to an efficient representation of the asymptotic contribution of each patch as in \eqref{eq:Iin}. (The contribution of the outer patch follows from \eqref{eq:Iin} as explained below \eqref{eq:Vin}.)
    \item We find the dominant patches and simplify their contribution by determining $\mathfrak{h}_\text{qu}$. See Definition~\ref{def:hqu}. Adding up those contributions we get the final asymptotic expression in~\eqref{eq:ItotGenSimp}.
\end{itemize}

Most of the analysis is rather universal.
The parts that are model-dependent and should be carried out on a case by case basis are: $i)$ the specific decomposition of $\mathfrak{h}_\text{cl}$, and $ii)$ the determination of $\mathfrak{h}_{\text{qu}}$. We will illustrate these model-dependent aspects of the analysis in the case study of the SU($N$) $\mathcal{N}=4$ theory index in the next subsection.

Note that more than the specific decomposition of $\mathfrak{h}_\text{cl}$ and the resulting fine structure (\emph{i.e.} the number and shapes) of the various patches, it is $\mathfrak{h}_{\text{qu}}$ that is important. This is because according to \eqref{eq:ItotSingular} the singular asymptotics of $\log\mathcal{I}$ can be obtained without any knowledge of the fine structure of the patches.

\subsection*{What about non-Lagrangian theories?}

Our derivation of the asymptotic formula \eqref{eq:ItotGenSimp} relied on the elliptic hypergeometric integral representation of the index. Such a representation is not available for general non-Lagrangian theories. But it seems reasonable to expect that the basic structure of the asymptotics would remain the same for any QFT for which the index can be defined, including non-Lagrangian theories. More precisely, we suspect that in the limit
\begin{equation}
\beta\to 0^+,\ \text{with\
}\omega_{1,2}\in \mathbb{H},\,
\xi_{a}\in\mathbb{R} \
\text{fixed},\label{eq:CKKNlimitGen}
\end{equation}
general 4d $\mathcal{N}=1$ QFTs with a $U(1)_R$ symmetry would display asymptotics of the form
\begin{equation}
    \mathcal{I}(p,q;\boldsymbol{v})\simeq \exp\left(\frac{A}{\beta^2}+\frac{B}{\beta}+\beta E_\mathrm{susy}\right)\, P(\frac{1}{\beta}),\label{eq:generalAsymptotics}
\end{equation}
with $P(1/\beta)$ a (finite-degree) polynomial in $1/\beta$.

In Lagrangian cases the degree of $P$ is bounded above by the rank of the gauge group, because $\mathrm{dim}\mathfrak{h}_\text{qu}<\mathrm{dim}\mathfrak{h}_\text{cl}=r_G$. It would be interesting if an upper bound can be found for the degree of $P$ in non-Lagrangian cases as well.


\subsection{Example: the $\mathcal{N}=4$ index}\label{sec:N=4asy}

We now proceed to illustrate various model-dependent aspects of the analysis in the specific case of SU($N$) $\mathcal{N}=4$ theory.

%
%
%
%


The elliptic hypergeometric integral expression for the index of SU$(N)$ $\mathcal{N}=4$ theory reads \cite{Spiridonov:2010qv}:
\begin{equation}
	\mathcal{I}(p,q;v_{1,2})=\frac{\big((p;p)(q;q)\big)^{N-1}}{N!}\prod_{a=1}^{3}\Gamma_e^{N-1}\big((pq)^{\frac{1}{3}}v_a\big)\oint\prod_{j=1}^{N-1}\frac{\mathrm{d}z_j}{2\pi i z_j}\prod_{1\le i,j\le N}^{i\neq j}\frac{\prod_{a=1}^{3}\Gamma_e\big((pq)^{\frac{1}{3}}v_a \frac{z_i}{z_j}\big)}{\Gamma_e\big(\frac{z_i}{z_j}\big)},\label{eq:N=4EHI}
\end{equation}
with the unit-circle contour for the $z_j=e^{2\pi i
x_j}$, while $\prod_{j=1}^N z_j=1$. In an $\mathcal{N}=1$ language the $\mathcal{N}=4$ theory has an SU(3) flavor symmetry, and $v_{1,2,3}$ satisfying $v_1v_2v_3=1$ are the fugacities associated to this flavor symmetry. Note from the formula that each chiral multiplet of the $\mathcal{N}=4$ theory is charged (with unit charge) under only one of the $v_a$.

\subsubsection{Decomposition of the classical moduli space}

We refer to $\mathfrak{h}_{\text{cl}}$ as the \emph{classical moduli space}. In the present case
\begin{equation}
\mathfrak{h}_\text{cl}=(-\frac{1}{2},\frac{1}{2}]^{N-1}.
\end{equation}

The starting point of the decomposition scheme of the previous subsection is the singular set $\mathcal{S}$ introduced in \eqref{eq:SSsDef}, which for the SU$(N)$ $\mathcal{N}=4$ theory becomes 
\begin{equation}
    \mathcal{S}=\bigcup_{1\le i<j\le N}\{\boldsymbol{x}\in
\mathfrak{h}_{cl}|\, x_i-x_j\in\, \mathbb{Z}\} \bigcup_{a=1,2,3}\, \bigcup_{1\le i\neq j\le N}\{\boldsymbol{x}\in
\mathfrak{h}_{cl}|\, x_i-x_j+\xi_a\in\, \mathbb{Z}\}.
\end{equation}
Let us consider the SU(2) case as an example. Then
\begin{equation}
\mathfrak{h}_\text{cl}=(-\frac{1}{2},\frac{1}{2}],\qquad\qquad\mathcal{S}=\{0\,,\,\frac{1}{2}\}\,\bigcup_{a=1,2,3}\{\mp\frac{\xi_a}{2}\,\mathrm{mod\,}1\,,\,\frac{1}{2}\mp\frac{\xi_a}{2}\,\mathrm{mod\,}1\},  
\end{equation}
where by mod~1 we mean the representative in $\mathfrak{h}_\text{cl}$. In particular, in the simple special case with $\xi_{1,2,3}=0$, we have $\mathcal{S}=\{0\,,\,\frac{1}{2}\}$.

The outer patch $\mathcal{S}'_\epsilon$ is
the subset of $\mathfrak{h}_{\text{cl}}$ in which for all $a$ and all $i\neq j$ we have  $\mathrm{min}_{n\in\mathbb{Z}}(|x_i-x_j+\xi_a-n|)\ge\epsilon$, and also $\mathrm{min}_{n\in\mathbb{Z}}(|x_i-x_j-n|)\ge\epsilon$. In the SU(2) case with $\xi_{1,2,3}=0$ for example, we have $\mathcal{S}'_\epsilon=[-\frac{1}{2}+\frac{\epsilon}{2}\,,\,-\frac{\epsilon}{2}]\cup[\frac{\epsilon}{2}\,,\,\frac{1}{2}-\frac{\epsilon}{2}]$.

\begin{figure}[t]
\centering
    \includegraphics[scale=.45]{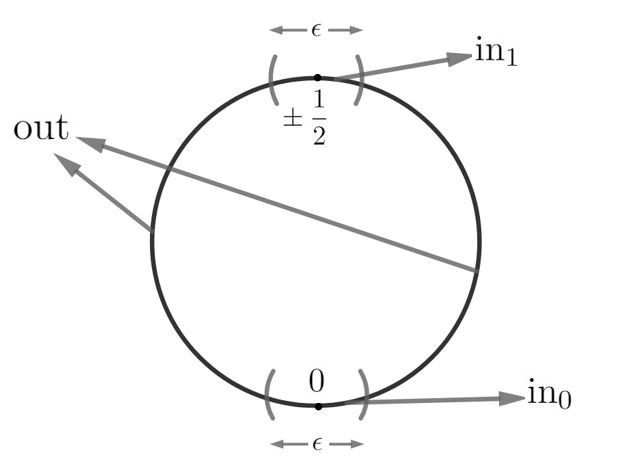}
\caption{The decomposition of the classical moduli space $\mathfrak{h}_{\text{cl}}$ for SU(2) $\mathcal{N}=4$ theory with $\xi_{1,2,3}=0$. In the SU(2) case $\mathfrak{h}_{\text{cl}}$ is parametrized by $x_1\in[-\frac{1}{2},+\frac{1}{2}]$ with $-\frac{1}{2}$ and $+\frac{1}{2}$ identified. For $\epsilon<\epsilon_c=\frac{1}{2}$, there are two inner patches, and the outer patch consists of two connected components. 
\label{fig:su2decomp}}
\end{figure}

The inner patches are the subsets of $\mathcal{S}_\epsilon(=\mathfrak{h}_\text{cl}\setminus \mathcal{S}'_\epsilon)$ distinguished from each other by their differing sets of light modes. In the SU(2) example with $\xi_{1,2,3}=0$, we have two inner patches, each an open interval of length $\epsilon$, one around 0 and the other around $\frac{1}{2}$ (note that $\frac{1}{2}$ and $-\frac{1}{2}$ are identified). See Figure~\ref{fig:su2decomp}.

\subsubsection{The quantum moduli space and the asymptotics of the index}


We refer to $\mathfrak{h}_{\text{qu}}$ as the \emph{quantum moduli space}. As discussed above, the patches that intersect $\mathfrak{h}_{\text{qu}}$ give the dominant contributions to the index in the limit \eqref{eq:CKKNlimit}. 

The quantum moduli space coincides with the locus of minima of $\mathrm{Re}V_2^{\text{out}}$, or when this locus is extended, with the subset of that locus where $\mathrm{Re}V_1^{\text{out}}$ is minimized.
For SU($N$) $\mathcal{N}=4$ theory the potentials can be found from \eqref{eq:Vout} to be
\begin{subequations}
\begin{align}
    V^{\text{out}}_2(\boldsymbol{x})
    &=\frac{(2\pi)^3 i}{6\omega_1\omega_2}\, \sum_{a=1}^3\big[\sum_{i<j}\big(\overline{B}_3(\xi_a+x_{ij})+\overline{B}_3(\xi_a-x_{ij})\big)+(N-1)\overline{B}_3(\xi_a)\big],\label{eq:V2outN=4}\\\nn
    V^{\text{out}}_1(\boldsymbol{x})
    &=\frac{(2\pi)^2 i \, (\omega_1+\omega_2)}{4\omega_1\omega_2}\, \bigg(-\frac{1}{3}\sum_{a=1}^3\bigg[\sum_{i<j}\big(\overline{B}_2(\xi_a+x_{ij})+\overline{B}_2(\xi_a-x_{ij})\big)+(N-1)\overline{B}_2(\xi_a)\bigg]\\
    &\quad\qquad\qquad\qquad\qquad+2\sum_{i<j}\overline{B}_2(x_{ij})+\frac{N-1}{6}\bigg),
    \label{eq:V1outN=4}\\
    V^{\text{out}}_0(\boldsymbol{x})&=\frac{2\pi i}{\omega_1\omega_2}\bigg(\frac{\big(\omega_1+\omega_2\big)^2}{72} -\frac{\omega_1^2+\omega_2^2}{24}\bigg)\sum_{a=1}^3\Bigl(\sum_{i\neq j}\overline{B}_1 \bigl(x_{ij} +\xi_a \bigr)+(N-1)\overline{B}_1 \bigl(\xi_a \bigr)\Bigr).
    \label{eq:V0outN=4}
\end{align}\label{eq:VoutN=4}%
\end{subequations}

The $x$-dependent piece of $\mathrm{Re}V_2^{\text{out}}$, in turn, is of the form $\sum_{i<j}V_h(x_{ij})$, with $V_h(x)$ the \emph{pairwise holonomy potential}, which up to a positive overall factor reads\footnote{To compare with \cite{ArabiArdehali:2019tdm} note that the complex parameters $b,\beta$ in that work are related to the real parameters here as $b_{\mathrm{there}}=\sqrt{\frac{b+ik_1}{b^{-1}+ik_2}}$, $\beta_{\mathrm{there}}=\beta\sqrt{(b+ik_1)(b^{-1}+ik_2)}$.}
\begin{equation}
    V_h(x)=\mathrm{sign}\big(\mathrm{Re}\big(\frac{-i}{(b+ik_1)(b^{-1}+ik_2)}\big)\big)\sum_{a=1}^3\big(\overline{B}_3(\xi_a+x)+\overline{B}_3(\xi_a-x)\big).\label{eq:Vh}
\end{equation}
We emphasize that although $\overline{B}_3$ is piecewise cubic, due to the (gauge)$^3$ anomaly cancellation $V_h$ ends up being piecewise quadratic.

\begin{figure}[t]
\centering
    \includegraphics[scale=.55]{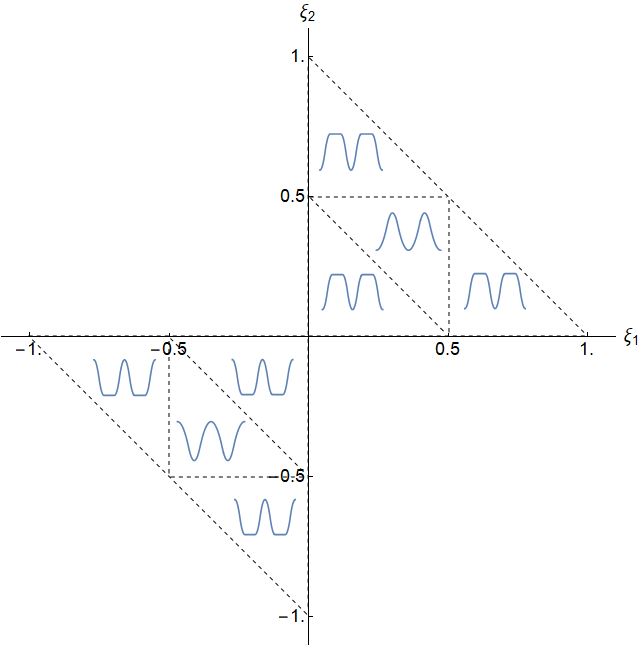}
\caption{The qualitative---or catastrophic---profile of the pairwise potential for the
holonomies, $V_h(x)$, drawn over the range $x\in(-1,1)$ of the pair's separation, for fixed
$\xi_{1,2}$ and fixed $k_{1,2}>0$, in the two
complementary wings of the
space of the control-parameters $\xi_{1,2}$. (Note that $\xi_{1,2}$ are 1-periodic: $\xi_{1,2}\sim \xi_{1,2}+1$, so the two wings cover a fundamental domain.) The dashed lines indicate the bifurcation sets across which the qualitative behavior changes. In particular, on the upper-right wing the potential is M-shaped and the holonomies attract each other, while on the lower-left wing it is W-shaped and the holonomies repel \cite{ArabiArdehali:2019tdm}. (Note that the holonomies are also 1-periodic.) The qualitative change on a single wing is more fine-grained---less catastrophic---and was not emphasized in~\cite{ArabiArdehali:2019tdm}: it only involves appearance or disappearance of plateaux in the potential. The plateaux on the W wing will be associated to the breakdown of the non-renormalization of $\log\mathcal{I}$ in Section~\ref{sec:discussion}. The M and W wings switch places if
$k_{1,2}$ are taken to be negative
instead.
\label{fig:CatastSimp}}
\end{figure}

Let us assume $k_{1,2}>0$ for the moment. Since $\xi_3$ is redundant
thanks to the balancing condition $v_1v_2v_3=1$, we can study the pairwise potential $V_h(x)$ as a function of only two \emph{control-parameters} $\xi_{1,2}\in\mathbb{R}$. The behavior of $V_h(x)$, and consequently the precise asymptotic behavior of the index, depend on where on this space of control-parameters we are \cite{ArabiArdehali:2019tdm,ArabiArdehali:2019orz}. As shown in Figure~\ref{fig:CatastSimp}, in essentially half of the $\xi_1-\xi_2$ plane the pairwise potential is M-shaped, while in the mirror half it is W-shaped. There is also a measure-zero \emph{bifurcation set} (indicated by the dashed lines in the figure) on which the potential undergoes catastrophic changes \cite{poston2014catastrophe}: it either vanishes (on the boundary of the butterfly) or develops plateaux (on the boundaries of the middle triangles inside the wings). When instead $k_{1,2}<0$, the M and W wings of Figure~\ref{fig:CatastSimp} switch places.

%
Finally, the supersymmetric Casimir energy \eqref{eq:Esusy} for SU($N$) $\mathcal{N}=4$ theory becomes
\begin{equation}
\begin{split}
     E_\text{susy}
     &=\frac{-i(\omega_1+\omega_2)^3}{54\omega_1\omega_2}\big(N^2-1\big).
     \end{split}\label{eq:EsusyN=4}
\end{equation}


\subsubsection*{M wing}

We now assume $\xi_{1,2}$ are chosen such that we land strictly inside the M~wing of the $\xi_1-\xi_2$ plane. In this case, the pairwise potential $V_h(x)$ \eqref{eq:Vh} has a minimum at $x=0$. Then, from Definition~\ref{def:hqu} for the quantum moduli space, and from the fact that the $x$-dependent piece of $\mathrm{Re}V_2^{\text{out}}$ is of the form $\sum_{i<j}V_h(x_{ij})$,  we conclude that the quantum moduli space $\mathfrak h_\text{qu}$ of the SU($N$) $\mathcal{N}=4$ theory on the M wing corresponds to the set of isolated points satisfying $x_{ij}=0$ for all $i,j\in\{1,\cdots,N\}$. In other words, the holonomies attract and $\mathfrak{h}_{\text{qu}}$ corresponds to all $x_j$ on top of each other, and equal to a given multiple of $1/N$ due to the SU($N$) constraint $\sum_{i=1}^Nx_i\in\mathbb Z$. 

Importantly, the quantum moduli space intersects the inner patch in$_0$ and its $N-1$ images under the $\mathbb{Z}_N$ center symmetry $x_j\to x_j+\frac{1}{N}$ \cite{ArabiArdehali:2019tdm}. We denote these $N$ inner patches by in$_{n_\ast}$.

The inner patch in$_0$ contributes to the index as
\begin{equation}
\begin{split}
    \mathcal{I}_{\text{in}_0}(p,q;v_{1,2})&\simeq \exp\left[-\frac{i\pi}{\sigma\tau}(N^2-1)\left(\prod_{a=1}^3\left(\{\xi_a\}+\frac{\sigma+\tau}{3}-\frac{1+\eta}{2}\right)+\frac{\eta(\sigma^2+3\sigma\tau+\tau^2)}{12}\right)\right]\\
    &\times\frac{(\frac{2\pi}{\beta})^{N-1}}{N!}\int_{\text{in}^{}_0}\prod_{\ell=1}^{N-1}\frac{\mathrm{d}x_\ell}{\sqrt{-\omega_1\omega_2}}\ e^{-\frac{i\pi\eta N}{\sigma\tau}\sum_{j=1}^N x_j^2} \prod_{1\le i<j\le N} \frac{1}{\Gamma_h(\frac{2\pi(x_i-x_j)}{\beta})\Gamma_h(\frac{2\pi(x_j-x_i)}{\beta})},\label{eq:almostThere}
    \end{split}
\end{equation}
inside the M wing. This is obtained by specializing the formula \eqref{eq:Iin} to the $\mathcal N=4$ theory. Here we have defined $\eta\in\{-1,+1\}$ following \cite{GonzalezLezcano:2020yeb} via
\begin{equation}
    \eta:=2\sum_a\{\xi_a\}-3.\label{eq:etaDef}
\end{equation}

Defining the rescaled variables $\sigma_\ell:=2\pi x_\ell/\beta$, we can rewrite (\ref{eq:almostThere}) as
\begin{equation}
\begin{split}
    \mathcal{I}_{\text{in}_0}(p,q;v_{1,2})&\simeq \exp\left[-\frac{i\pi}{\sigma\tau}(N^2-1)\left(\prod_{a=1}^3\big(\{\xi_a\}+\frac{\sigma+\tau}{3}-\frac{1+\eta}{2}\big)+\frac{\eta(\sigma^2+3\sigma\tau+\tau^2)}{12}\right)\right]\\
    &\quad\times Z_{3d}(\omega_1,\omega_2;\frac{2\pi\epsilon}{\beta}),\label{eq:N=4indexAsyCut}
    \end{split}
\end{equation}
where
\begin{equation}
    Z_{3d}(\omega_1,\omega_2;\Lambda)=\frac{1}{N!}\int_{|\sigma_i-\sigma_j|<\Lambda}\prod_{\ell=1}^{N-1}\frac{\mathrm{d}\sigma_\ell}{\sqrt{-\omega_1\omega_2}}\ e^{-\frac{i\pi\eta N}{\omega_1\omega_2}\sum_{j=1}^N \sigma_j^2} \prod_{1\le i<j\le N} \frac{1}{\Gamma_h(\sigma_i-\sigma_j)\Gamma_h(\sigma_j-\sigma_i)},\label{eq:3dZcut}
\end{equation}
and with $\sigma_j$ subject to $\sum_{j=1}^N \sigma_j=0$.

We observe that $Z_{3d}(\omega_1,\omega_2;\Lambda)$ coincides with the partition function\footnote{Compare with the expressions in Section~5 of \cite{Aharony:2013dha}, noting that the three-manifold considered there has $\omega_1=ib$, $\omega_2=ib^{-1}$. To compare with the transversely holomorphic foliation (THF) moduli in \cite{ArabiArdehali:2021nsx} note that the 3d metric there differs by an overall factor from the one in \cite{Cassani:2021fyv} that we are adopting here.} of 3d $\mathcal{N}=2$ SYM with gauge group SU($N$) and Chern-Simons coupling
\begin{equation}
    k_{ij}=-\eta N\delta_{ij},\label{eq:CScoupling}
\end{equation}
with Coulomb branch cut-off $\Lambda$, on a 3-manifold with $\omega_1,\omega_2$ as its moduli of the transversely holomorphic foliation~\cite{Closset:2013vra}.

Since $\mathfrak h_{\text{qu}}$ consists of a set of isolated points, as argued around \eqref{eq:IinIsoInfty}
sending $\epsilon\to\infty$ on the RHS of \eqref{eq:N=4indexAsyCut} introduces exponentially small error:
\begin{equation}
    \qquad\qquad\qquad\qquad\qquad Z_{3d}(\omega_1,\omega_2;\frac{2\pi\epsilon}{\beta})\simeq Z_{3d}(\omega_1,\omega_2;\infty)\qquad(\text{on the M wings}).\label{eq:epsilonInftyM}
\end{equation}

Moreover, since the index is dominated by the patches in$_{n_\ast}$ intersecting $\mathfrak{h}_\text{qu}$, we have
\begin{equation}
    \qquad\qquad\qquad\qquad\mathcal{I}(p,q;v_{1,2})\simeq \sum_{n_\ast}\mathcal{I}_{\text{in}_{n_\ast}}=N\, \mathcal{I}_{\text{in}_0}(p,q;v_{1,2})\qquad(\text{on the M wings}),\label{eq:NIin}
\end{equation}
with the factor of $N$ accounting for the images of in$_0$ under the $\mathbb{Z}_N$ center symmetry.

Putting the two findings in \eqref{eq:epsilonInftyM} and \eqref{eq:NIin} together and using the asymptotics \eqref{eq:N=4indexAsyCut}, we can simplify the asymptotic of the index on the M wing as
\begin{equation}
\begin{split}
    \mathcal{I}(p,q;v_{1,2})&\simeq N\exp\left[-\frac{i\pi}{\sigma\tau}(N^2-1)\left(\prod_{a=1}^3\big(\{\xi_a\}+\frac{\sigma+\tau}{3}-\frac{1+\eta}{2}\big)+\frac{\eta(\sigma^2+3\sigma\tau+\tau^2)}{12}\right)\right]\\
    &\quad\times Z_{3d}(\omega_1,\omega_2),\label{eq:N=4indexAsy}
    \end{split}
\end{equation}
where $Z_{3d}(\omega_1,\omega_2):=Z_{3d}(\omega_1,\omega_2;\infty)$.

To make contact with the result of Lezcano et al \cite{GonzalezLezcano:2020yeb}, we use
\begin{equation}
    \frac{1}{\Gamma_h(x;\omega_1,\omega_2)\Gamma_h(-x;\omega_1,\omega_2)}=4\sinh \big(\frac{\pi x}{-i\omega_1}\big)\sinh\big(\frac{\pi x}{-i\omega_2}\big)\label{eq:hGammaSinh}
\end{equation}
inside the integrand of $Z_{3d}(\omega_1,\omega_2)$, specialize to $\sigma=\tau$ (and hence $\omega_1=\omega_2=\omega$), define the re-scaled variable $\sigma'_\ell=\frac{\sigma_\ell}{-i\omega}$, and rotate the contour of $\sigma'_\ell$ back to the real axis to arrive at
\begin{equation}
    Z_{3d}(\omega,\omega)=\frac{1}{N!}\int_{-\infty}^{+\infty}\prod_{\ell=1}^{N-1}\mathrm{d}\sigma'_\ell\ e^{i\pi\eta N\sum_{j=1}^N \sigma_j'^2} \prod_{1\le i<j\le N} 4\sinh^2\big(\pi(\sigma'_i-\sigma'_j)\big).\label{eq:3dZS3}
\end{equation}
This coincides (up to a convention-dependent sign) with the $S^3$ partition function of the SU($N$)$_{-\eta N}$ Chern-Simons theory as encountered in \cite{GonzalezLezcano:2020yeb}. Incidentally, the nonzero value found for the latter partition function in \cite{GonzalezLezcano:2020yeb} demonstrates that $Z_{3d}(\omega_1,\omega_2)$ is not identically zero, and we expect it to remain nonzero for all $\omega_{1,2}$ in the upper half-plane.

We emphasize that while the expression \eqref{eq:3dZS3} is simpler looking, as we will see in the next section it is in fact \eqref{eq:3dZcut} that arises more naturally in the EFT perspective.

The index is \emph{fully deconfined} on the M wing, both in the sense that the $x_{ij}=0$ configurations maximally break the $\mathbb{Z}_N$ center symmetry, and in the sense that the maximal asymptotic growth is achieved on the M wing \cite{ArabiArdehali:2019orz}.

Note that the above derivation of the asymptotics is more rigorous than the one in \cite{GonzalezLezcano:2020yeb} because we use the real-analytic approach of Rains \cite{Rains:2006dfy} and avoid subtle steepest-descent contour deformation arguments.

Note also that in the preliminary analysis of the leading asymptotics of the 4d $\mathcal{N}=4$ index on the M wings in \cite{ArabiArdehali:2019tdm} some subtleties were not properly treated (see footnote~7 and the paragraph above Eq.~(2.6) in that work). In particular, the estimate \eqref{eq:outerEst} was used there, which is not uniformly valid for vector multiplet gamma functions near the dominant holonomy configurations $x_{ij}=0$. Furthermore, it was not proven there that phase oscillations in the integrand do not lead to cancellations (in other words, that completely destructive interference does not occur). The derivation above overcomes those subtleties because the estimates we use here are all-order exact and uniformly valid near $x_{ij}=0$. Our analysis thus fills the gaps in the derivation of the leading asymptotics of the index in \cite{ArabiArdehali:2019tdm}.

\subsubsection*{W wing}

Determination of the quantum moduli space on the W wing is difficult for general~$N$~\cite{ArabiArdehali:2019orz}. Here we focus on the simple case of SU(2).

As can be seen from Figure~\ref{fig:CatastSimp}, on the middle triangle of the W wing, the quantum moduli space $\mathfrak{h}_{\text{qu}}$ corresponds to $x_{12}=2x_1=\pm1/2$, and therefore consists of two isolated points: $x_1=\pm 1/4$. Strictly inside the middle triangle, these points are in the outer patch, and hence as discussed below \eqref{eq:3dZoutInfty} we have
\begin{equation}
    \mathcal{I}(p,q;v_{1,2})\simeq 2\times e^{\,\beta E_{\text{susy}}}\, e^{-V^{\text{out}}(\boldsymbol{x}^\ast)}\, Z^{\boldsymbol{x}^\ast_{}}_{3d}(\omega_1,\omega_2),\label{eq:I:W:middle}
\end{equation}
with $\boldsymbol{x}^\ast_{}=\pm1/4$. The exponential terms in \eqref{eq:I:W:middle} can be written more explicitly as
\begin{equation}
\begin{split}
    \beta E_\text{susy}-V^\text{out}_*&=-\fft{\pi i}{2\sigma\tau}\prod_{a=1}^3\left(\{2\xi_a\}+\fft{2(\sigma+\tau)}{3}-\fft{1-\eta}{2}\right)\\
    &\quad+\fft{\pi i}{\sigma\tau}\prod_{a=1}^3\left(\{\xi_a\}+\fft{\sigma+\tau}{3}-\fft{1+\eta}{2}\right)+\fft{(\sigma^2+3\sigma\tau+\tau^2)\eta\pi i}{4\sigma\tau}.
\end{split}
\end{equation}
This is obtained by substituting $x_{12}=\pm\fft12$ into \eqref{eq:VoutN=4} and then using \eqref{V:out} and \eqref{eq:EsusyN=4}. We have also used that 
\begin{equation}
    2\sum_{a=1}^3\{2\xi_a\}-3=-\eta
\end{equation}
in the middle triangle of the W wing.

Strictly inside the peripheral triangles of the W wing, Figure~\ref{fig:CatastSimp} implies that $V_h$ is minimized on an extended domain. More precisely, inside \emph{the triangle closest to the origin} the locus of minima of $V_h(x_{12})$ is
\begin{equation}
    |x_{12}\pm\frac{1}{2}|\le\frac{1}{2}+\xi_1+\xi_2,\label{W:extended:1}
\end{equation}
while inside \emph{the left-most triangle} the locus is
\begin{equation}
    |x_{12}\pm\frac{1}{2}|\le-\frac{1}{2}-\xi_1,\label{W:extended:2}
\end{equation}
and inside \emph{the lower-most triangle} it is
\begin{equation}
|x_{12}\pm\frac{1}{2}|\le-\frac{1}{2}-\xi_2.\label{W:extended:3}
\end{equation}
Combining (\ref{W:extended:1}), (\ref{W:extended:2}), and (\ref{W:extended:3}) together, the locus of minima of $V_h(x_{12})$ can be written compactly as
\begin{equation}
    |x_{12}\pm\fft12|\leq1/2-\{\xi_\text{min}\}=1-\{1/2+\xi_\text{min}\},\label{W:extended}
\end{equation}
where $\xi_3$ is determined by the constraint $\sum_{a=1}^3\xi_a\in\mathbb Z$ modulo an integer, and $\xi_\text{min}$ is defined through
\begin{equation}
    \{\xi_\text{min}\}:=\min[\{\xi_a\}\,|\,a=1,2,3].
\end{equation}
The quantum moduli space is the subset of this extended locus where $V^{\text{out}}_1$ is minimized. However, it turns out that $V^{\text{out}}_1$ is flat on the locus of minima of $V_h$ on all the three peripheral triangles. Therefore the quantum moduli space $\mathfrak{h}_{\text{qu}}$ coincides with the loci (\ref{W:extended}). The length of the quantum moduli space, denoted $|\mathfrak{h}_{\text{qu}}|$, is then given from (\ref{W:extended}) as
\begin{equation}
    |\mathfrak{h}_{\text{qu}}|=2(1-\{1/2+\xi_\text{min}\}).\label{eq:hquLengthWperipheral}
\end{equation}
We emphasize that $\mathfrak{h}_\text{qu}$ is parametrized by the first holonomy $x_1$ (recall that the second holonomy $x_2$ is determined modulo an integer by the SU(2) constraint $x_1+x_2\in\mathbb Z$).

Note that the quantum moduli space $\mathfrak{h}_{\text{qu}}$ given in \eqref{W:extended} intersects the outer patch as well as inner patches. Therefore we must sum the contributions from all those patches to derive the asymptotics of the index.

First, the contribution to the index from the intersection of the quantum moduli space and the outer patch, namely $\mathfrak{h}_{\text{qu}}\cap \mathcal{S}'_\epsilon$, reads
\begin{equation}
    \frac{1}{2!}\ \frac{2\pi}{\beta\sqrt{-\omega_1\omega_2}}(|\mathfrak{h}_{\text{qu}}|-2\epsilon)\ e^{\beta E_{\text{susy}}}\ e^{-V^{\text{out}}_\ast},\label{eq:IoutWp}
\end{equation}
similarly to \eqref{eq:IoutCutGen}. Note that because $V^{\text{out}}_{2,1}$ are flat on $\mathfrak{h}_{\text{qu}}\cap \mathcal{S}'_\epsilon$, the coefficients $k^\text{out}_{ij}$, $k^\text{out}_{jR}$ are zero and the partition function $Z_{n_\ast}^{3\mathrm{d}}(\omega_1,\omega_2;\frac{2\pi\epsilon}{\beta})$ is particularly simple here; compare with the discussion around \eqref{eq:flatContOut}. The $2\epsilon$ subtracted above from $|\mathfrak{h}_{\text{qu}}|$ arises because $\mathfrak{h}_{\text{qu}}$ has two connected components, and by intersecting it with $\mathcal{S}'_\epsilon$ we are excising two intervals of length $\epsilon/2$ from the ends of each component. The $1/2!$ is the $1/|W|$ factor in the present SU(2) case.

Next, the contributions from the four inner patches at the ends of the two components of $\mathfrak{h}_{\text{qu}}$ read
    \begin{equation}
    \begin{split}
     4\times &e^{\,\beta E_{\text{susy}}}\int_{\text{in}^{}_n}\frac{D\boldsymbol{x}}{(\sqrt{-\omega_1 \omega_2})^{r_G}}\, \big(\frac{2\pi}{\beta}\big)^{r_G}\times  \\
    &\qquad\qquad\left( e^{-V^{\text{in}_n}(\boldsymbol{x})}\, \frac{\prod_\chi\prod_{\rho^\chi\in L_n}\Gamma_h\big(r_\chi (\frac{\omega_1+\omega_2}{2})+\frac{2\pi}{\beta}(\rho^\chi\cdot \boldsymbol{x}+q^\chi\cdot\boldsymbol{\xi})^{}_\mathbb{Z}\big)}{\prod_{\alpha_+\in L_n}\Gamma_h\big(\frac{2\pi}{\beta}(\alpha_+\cdot\boldsymbol{x})^{}_\mathbb{Z}\big)\Gamma_h\big(\frac{2\pi}{\beta}(-\alpha_+\cdot\boldsymbol{x})^{}_\mathbb{Z}\big)}-e^{-V^{\text{out}}(\boldsymbol{x})}
    \right),\\
    +\,4\times &e^{\,\beta E_{\text{susy}}}\int_{\text{in}^{}_n}\frac{D\boldsymbol{x}}{(\sqrt{-\omega_1 \omega_2})^{r_G}}\, \big(\frac{2\pi}{\beta}\big)^{r_G}\, e^{-V^{\text{out}}(\boldsymbol{x})}, 
    \end{split}\label{eq:IinWp}
\end{equation}
where we have added and subtracted the last line similarly to \eqref{eq:out+(in-out)}. We have also taken advantage of the symmetries of the problem to take into account the contributions of the four patches by multiplying the contribution from one of them by four.

Let us now consider the expression inside the large parenthesis in \eqref{eq:IinWp}. According to \eqref{eq:GammaHasy} the first term approaches the second term exponentially fast as we move away from the end points of the two components of $\mathfrak{h}_{\text{qu}}$. Therefore we can take out the $e^{-V^{\text{out}}}$ factor and call the remaining function inside the parenthesis $g(\frac{2\pi(x-x^\ast)}{\beta})$, with $x^\ast$ the position of the end point. Since $g(\frac{2\pi(x-x^\ast)}{\beta})$ approaches zero exponentially fast as we move away from the end points, we can re-scale $\widetilde{x}=2\pi (x-x^\ast)/\beta$, and get for the asymptotics of the first integral in \eqref{eq:IinWp} the following expression:
\begin{equation}
    e^{\beta E_{\text{susy}}}\ e^{-V^{\text{out}}_\ast}\, D_1,\label{eq:IinWp1}
\end{equation}
with
\begin{equation}
    D_1=4\int_{-\infty}^{\infty} \frac{\mathrm{d}\widetilde{x}/2!}{\sqrt{-\omega_1\omega_2}}\, g(\widetilde{x})\, e^{-\Delta V^{\text{out}}(\widetilde{x})},
\end{equation}
where $\Delta V^{\text{out}}(\widetilde{x}):= V^{\text{out}}(\boldsymbol{x})-V^{\text{out}}_\ast.$

The second integral in \eqref{eq:IinWp} has contributions from inside $\mathfrak{h}_{\text{qu}}$, which cancel the $\epsilon$-dependent piece of \eqref{eq:IoutWp}. It also has contributions from outside of $\mathfrak{h}_{\text{qu}}$, which since $V^{\text{out}}$ decays away from $\mathfrak{h}_{\text{qu}}$ can be written as
\begin{equation}
    e^{\beta E_{\text{susy}}}\ e^{-V^{\text{out}}_\ast}\, D_2,\label{eq:IinWp2}
\end{equation}
with
\begin{equation}
    D_2=4\int_{0}^{\infty} \frac{\mathrm{d}\widetilde{x}/2!}{\sqrt{-\omega_1\omega_2}}\, e^{-\Delta V^{\text{out}}(\widetilde{x})}.
\end{equation}

Putting all the above contributions \eqref{eq:IoutWp}, \eqref{eq:IinWp1}, and \eqref{eq:IinWp2} together we get
\begin{equation}
    \mathcal{I}(p,q;v_{1,2})\simeq e^{\beta E_{\text{susy}}}\ e^{-V^{\text{out}}_\ast} (\frac{2\pi}{\beta\sqrt{-\omega_1\omega_2}}\frac{|\mathfrak{h}_{\text{qu}}|}{2!}+C_0),\label{eq:I:W:peripheral}
\end{equation}
with $C_0:=D_1+D_2$. This is of course a special case of \eqref{eq:ItotGenSimp}.

The exponential terms in \eqref{eq:I:W:peripheral} can be written more explicitly as
\begin{equation}
\begin{split}
    \beta E_\text{susy}-V^\text{out}_*&=-\fft{\pi i}{2\sigma\tau}\prod_{a=1}^3\left(\{2\xi_a\}+\fft{2(\sigma+\tau)}{3}-\fft{1+\eta}{2}\right)\\
    &\quad+\fft{\pi i}{\sigma\tau}\prod_{a=1}^3\left(\{\xi_a\}+\fft{\sigma+\tau}{3}-\fft{1+\eta}{2}\right)-\fft{(\sigma^2+3\sigma\tau+\tau^2)\eta\pi i}{12\sigma\tau}.
\end{split}\label{eq:I:W:peripheral:exp}
\end{equation}
This is obtained by substituting $x_{12}=\pm\fft12$ into \eqref{eq:VoutN=4} and then using \eqref{V:out} and \eqref{eq:EsusyN=4}. We have also used that 
\begin{equation}
    2\sum_{a=1}^3\{2\xi_a\}-3=\eta
\end{equation}
in the peripheral triangles of the W wing.

Here we compare the expression \eqref{eq:I:W:peripheral} for the Cardy-like asymptotics of the $\mathcal N=4$ SU(2) superconformal index in the peripheral triangles of the $W$-wing with the result from the Bethe-Ansatz approach \cite{Lezcano:2021qbj}. From (3.30) of \cite{Lezcano:2021qbj}, the Cardy-like asymptotics of $\mathcal N=4$ SU(2) index is given in the peripheral triangles of the $W$-wing as\footnote{In the convention of \cite{Lezcano:2021qbj}, assuming positive [negative] $k_{1,2}$ corresponds to $\eta_1=1\,[\eta_1=-1]$ in the $W$-wing, and the peripheral [middle] triangles correspond to the region with $\eta_1=\eta_2\,[\eta_1=-\eta_2]$. Also note $\eta_1^\text{there}=\eta^\text{here}$.} 
\begin{equation}
\begin{split}
	&\mathcal I(q,q;v_{1,2})\\
	&\simeq\bigg(\fft{i(1-\{1/2+\xi_\text{min}\})}{\tau}-\fft{i\eta}{6}+\fft{1}{2\pi}\bigg)e^{-\fft{\pi i}{2\tau^2}\prod_{a=1}^3\left(\{2\xi_a\}+\fft{4\tau}{3}-\fft{1+\eta}{2}\right)+\fft{\pi i}{\tau^2}\prod_{a=1}^3\left(\{\xi_a\}+\fft{2\tau}{3}-\fft{1+\eta}{2}\right)-\fft{5\eta\pi i}{12}},
\end{split}\label{eq:I:W:BA}%
\end{equation}
where $\Delta_a^\text{there}=\fft{\sigma+\tau}{3}+\xi_a^\text{here}$ and we take $p=q$ ($\sigma=\tau$, hence $\omega_1=\omega_2=\omega$) and $\xi_a\in\mathbb R$ for simplicity. The Cardy-like asymptotics from the Bethe-Ansatz approach (\ref{eq:I:W:BA}) is consistent with our result (\ref{eq:I:W:peripheral}) under the identification $\sigma=\tau=\fft{\beta\omega}{2\pi}$. First, it is straightforward to see that the exponential term in (\ref{eq:I:W:peripheral}), which is given as \eqref{eq:I:W:peripheral:exp}, reduces to the exponential term in the Bethe-Ansatz result (\ref{eq:I:W:BA}) under the identification $\sigma=\tau$. Next, note that the coefficients of the $\fft1\beta$-order in the polynomial parts of (\ref{eq:I:W:peripheral}) and (\ref{eq:I:W:BA}) match precisely since the length of the quantum moduli space is given as $|\mathfrak h_\text{qu}|=2(1-\{1/2+\xi_\text{min}\})$ as we saw in (\ref{eq:hquLengthWperipheral}). We leave the calculation of the constant $C_0$ in (\ref{eq:I:W:peripheral}) and its comparison with the counterpart $-\fft{i\eta}{6}+\fft{1}{2\pi}$ in (\ref{eq:I:W:BA}) for future research.

\subsubsection*{The bifurcation set}
The bifurcation set consists of the boundary of the butterfly in Figure~\ref{fig:CatastSimp}, as well as the boundaries of the middle triangles inside the two wings. Actually, the boundary of the middle triangle on the M wing is completely irrelevant for our purposes, as it is the structure of the \emph{minima} of $V_h$ that determines $\mathfrak{h}_{\text{qu}}$ and significantly affects the asymptotics of the index.

On the boundary of the butterfly $V^{\text{out}}_2$ vanishes. For simplicity we focus on the origin of the $\xi_1-\xi_2$ plane, but there is no particular difficulty in extending the analysis to the other parts of the boundary region.

At the origin of the $\xi_1-\xi_2$ plane not only $V^{\text{out}}_2$ but also $V^{\text{out}}_1$ vanishes. As a result we have $\mathfrak{h}_{\text{qu}}=\mathfrak{h}_{\text{cl}}$: the quantum moduli space coincides with the classical moduli space.

Focusing on the SU(2) case for simplicity again, there are two singular points at $x=0,1/2$, and hence two inner patches as can be seen from Figure~\ref{fig:su2decomp}. These are related to each other via the center symmetry, so we can consider only in$_0$ and multiply its contribution by two to account for in$_1$ as well.

The contribution of the outer patch is, as in \eqref{eq:IoutWp}, given by
\begin{equation}    \frac{1}{2!}\ \frac{2\pi}{\beta\sqrt{-\omega_1\omega_2}}(1-2\epsilon)\ e^{\beta E_{\text{susy}}}\ e^{-V_{0}^{\text{out}}},\label{eq:IoutOrig}
\end{equation}
because in this case $|\mathfrak{h}_{\text{qu}}|=|\mathfrak{h}_{\text{cl}}|=1$, while $V_{1,2}^{\text{out}}=0$, and $V_{0}^{\text{out}}$ is independent of $x$.

The contribution of the inner patches is, similarly to  \eqref{eq:IinWp}, given by
    \begin{equation}
    \begin{split}
     2\times &e^{\,\beta E_{\text{susy}}}\int_{\text{in}^{}_0}\frac{D\boldsymbol{x}}{(\sqrt{-\omega_1 \omega_2})^{r_G}}\, \big(\frac{2\pi}{\beta}\big)^{r_G}\times  \\
    &\qquad\qquad\left( e^{-V^{\text{in}_n}(\boldsymbol{x})}\, \frac{\prod_\chi\prod_{\rho^\chi\in L_n}\Gamma_h\big(r_\chi (\frac{\omega_1+\omega_2}{2})+\frac{2\pi}{\beta}(\rho^\chi\cdot \boldsymbol{x}+q^\chi\cdot\boldsymbol{\xi})^{}_\mathbb{Z}\big)}{\prod_{\alpha_+\in L_n}\Gamma_h\big(\frac{2\pi}{\beta}(\alpha_+\cdot\boldsymbol{x})^{}_\mathbb{Z}\big)\Gamma_h\big(\frac{2\pi}{\beta}(-\alpha_+\cdot\boldsymbol{x})^{}_\mathbb{Z}\big)}-e^{-V^{\text{out}}(\boldsymbol{x})}
    \right),\\
    +\,2\times &e^{\,\beta E_{\text{susy}}}\int_{\text{in}^{}_0}\frac{D\boldsymbol{x}}{(\sqrt{-\omega_1 \omega_2})^{r_G}}\, \big(\frac{2\pi}{\beta}\big)^{r_G}\, e^{-V^{\text{out}}(\boldsymbol{x})}. 
    \end{split}\label{eq:IinOrig}
\end{equation}

The first integral can be argued as in \eqref{eq:IinWp1} to be given by
\begin{equation}
    e^{\beta E_{\text{susy}}}\ e^{-V_0^{\text{out}}}\, C_0,\label{eq:IinOrig1}
\end{equation}
with
\begin{equation}
    C_0=2\int_{-\infty}^{\infty} \frac{\mathrm{d}\widetilde{x}/2!}{\sqrt{-\omega_1\omega_2}}\, h(\widetilde{x}).
\end{equation}
Here $h(\widetilde{x})$, with $\widetilde{x}=2\pi x/\beta$, is the exponentially decaying function arising from the second line of \eqref{eq:IinOrig} near $x=0$. The second integral in \eqref{eq:IinOrig} simply kills the $\epsilon$-dependent part of \eqref{eq:IoutOrig}. Putting everything together we hence get
\begin{equation}
    \mathcal{I}(p,q;v_{1,2})\simeq e^{\beta E_{\text{susy}}}\ e^{-V_0^{\text{out}}} (\frac{2\pi}{\beta\sqrt{-\omega_1\omega_2}}\,\frac{1}{2!}+C_0).
\end{equation}
Again, this is a special case of \eqref{eq:ItotGenSimp}.\\

\section{Effective field theory perspective}\label{sec:EFT}
\noindent The asymptotic formulas in the previous section were derived by direct asymptotic analysis of the integral expression (\ref{eq:EHIgen}), which in turn is obtained via Hamiltonian operator counting arguments \cite{Spiridonov:2010qv,Dolan:2008qi}. In this section we present a more physical derivation of the asymptotic expressions in a path-integral picture via the supersymmetric 3d EFT machinery pioneered by Di~Pietro-Honda-Komargoski \cite{DiPietro:2014bca,DiPietro:2016ond}, and developed to all orders in \cite{Cassani:2021fyv,ArabiArdehali:2021nsx}.

Here we refine the treatments in the latter two references by combining their respective advantages. In particular, we adopt the more general (and more EFT friendly) supersymmetric background of \cite{Cassani:2021fyv}, but follow the more systematic Wilsonian treatment of \cite{ArabiArdehali:2021nsx} (which unlike the approach in \cite{Cassani:2021fyv} can capture partially-deconfined or confined phases as well). We also add the novel ingredient of decomposition of the BPS moduli space (see Section~\ref{sec:lightModes}), which we believe completes the conceptual framework.

\subsection*{The 4d background}
The starting point of the EFT approach is the Lagrangian formulation of the index $\mathcal{I}(p,q;\boldsymbol{v})$ as a path-integral partition function $Z(p,q;\boldsymbol{v})$ on a Hopf surface that is topologically $S^3\times S^1$ \cite{Nawata:2011un,Assel:2014paa,Cassani:2021fyv}. The metric of the Hopf surface reads \cite{Cassani:2021fyv}
\begin{align}\label{eq:4dMetric}
\mathrm{d} s_4^2 \,&=\, \,\frac{1}{1+b^{-2}k_1^2\cos^2\theta+b^2k_2^2\sin^2\theta}\,\Big[\, \mathrm{d}t_E^2 + \left(b^2 \cos^2\theta + b^{-2}\sin^2\theta\right)\mathrm{d} \theta^2\nn\\[1mm]
&\qquad\qquad\qquad\qquad +  b^{-2}\cos^2\theta\left( \mathrm{d}\varphi_1 +k_1\mathrm{d}t_E\right)^2 + b^2\sin^2\theta \left(\mathrm{d}\varphi_2 +k_2\mathrm{d}t_E\right)^2 \Big]\,.
\end{align}
Here~$\theta \in [0, \pi/2]$, the angles~$\varphi_{1,2}$ are $2\pi$-periodic, and 
the Euclidean time coordinate has the independent periodicity
\begin{equation}
t_E \sim t_E + \beta  \,.
\end{equation}
The complex-structure moduli of the Hopf surface \eqref{eq:4dMetric} are
\begin{equation}
    \sigma=\frac{i\beta }{2\pi}(b+ik_1),\qquad \tau=\frac{i\beta }{2\pi}(b^{-1}+ik_2).\label{eq:csm}
\end{equation}
This is why the parameters $b,\beta,k_{1,2}$ were defined in the previous section via \eqref{eq:csmSec2}.

In the approach of \cite{Festuccia:2011ws}, to preserve supersymmetry on the Hopf surface one has to turn on various background supergravity fields, specifically the $V$ and $A$ fields of new minimal supergravity. We skip their detailed expressions here and refer the interested reader to \cite{Cassani:2021fyv}. On the other hand, the fugacities $v_{a}$ correspond to turning on background gauge fields
\begin{equation}
A^{(f)}_{a}=\frac{2\pi \xi_a}{\beta}\,\mathrm{d}t_E,\label{eq:flavorHol}  
\end{equation}
for the Cartan of the 
flavor symmetry.


\subsection*{Localization in 4d and the exact partition function}

On the 4d background described above, we place our dynamical 4d $\mathcal{N}=1$ vector and chiral multiplets. We then employ appropriate supersymmetric Lagrangians constructed as in \cite{Festuccia:2011ws}, and impose periodic boundary conditions along the $S^1$.\footnote{This corresponds to setting $n_0=0$ in \cite{Cassani:2021fyv}. More generally, one can consider twisted identifications of the form $\psi(t_E+\beta) \,=\, e^{\pi i n_0 ( r+F)}\psi(t_E)$, with $r$ the $U(1)_R$ charge and $F$ the fermion number of the field $\psi$.} The partition function $Z$ of the dynamical fields can then be computed \emph{exactly} via supersymmetric localization of the path-integral---see \cite{Nekrasov:2002qd,Pestun:2007rz,Nawata:2011un,Assel:2014paa}. Assuming $r_\chi>0$ as we have in this work, it turns out that on the localization (or BPS) locus all the dynamical chiral and vector multiplet fields (including the vector multiplet auxiliary field $D$) are zero, except the $t_E$ component of the vector field~\cite{Assel:2014paa}:
\begin{equation}
A_j\xrightarrow{\text{on BPS locus}}\ \frac{2\pi x_j}{\beta}\,\mathrm{d}t_E,\label{eq:gaugeHol}  
\end{equation}
with $x_j\in(-\frac{1}{2},\frac{1}{2}].$
Note that using the large gauge transformations (winding the $S^1$) the Cartan components are restricted to $(-\frac{\pi}{\beta},\frac{\pi}{\beta}]$, implying $x_j\in(-\frac{1}{2},\frac{1}{2}].$ A Weyl redundancy remains, which we do not fix at this stage, and instead incorporate into the measure of the localized path-integral. Thus
\begin{equation*}
    \mathfrak{h}_\text{cl}=(-\frac{1}{2},\frac{1}{2}]^{r_G}\text{\, \ is the moduli space of 4d BPS field configurations},
\end{equation*}
up to the Weyl redundancy.

For the case of our interest here, with $\xi_a$ real and hence $v_a$ on the unit circle, it was argued in \cite{ArabiArdehali:2019tdm} (based on \cite{Assel:2014paa,Ardehali:2015hya}) that the localized partition function $Z$ is given by\footnote{For general 4d $\mathcal{N}=1$ indices without flavor fugacities (\emph{i.e.} with $v_a=1$) the correct exponential factor was obtained in \cite{Ardehali:2015hya,Assel:2015nca}. For \emph{real-valued} flavor fugacities a generalization was given in \cite{Bobev:2015kza}. (That generalization is presumably valid only in a neighborhood of $v_a=1$, although the details are not properly understood.) In the present context, with flavor fugacities complex and \emph{on the unit circle}, it was argued in \cite{ArabiArdehali:2019tdm} that the result of \cite{Ardehali:2015hya,Assel:2015nca} applies.\label{ftnt:susyCasimir}}
\begin{equation}
    Z(p,q;\boldsymbol {v})=e^{-\beta E_{\mathrm{susy}}}\ \mathcal{I}(p,q;\boldsymbol {v}),\label{eq:ZvsI}
\end{equation}
with  $\mathcal{I}$ the matrix integral in \eqref{eq:EHIgen}, and $E_{\mathrm{susy}}$ as in \eqref{eq:Esusy}. In particular, the integral in \eqref{eq:EHIgen} is now seen to arise as the integral over the moduli space of BPS field configurations, and the division by $|W|$ in \eqref{eq:defDx} incorporates the Weyl redundancy. The rest of the matrix integral as well as the SUSY Casimir factor arise from one-loop contributions of the dynamical vector and chiral multiplets \cite{Nawata:2011un,Assel:2014paa,Ardehali:2015hya,ArabiArdehali:2019tdm}.

\subsection*{The goal of the present section}

Rather than going the exact 4d localization route of Eq.~\eqref{eq:ZvsI}, our aim in this section is to derive the asymptotics of $Z$ via 3d localization in the small-$\beta$ effective field theory. 

To be more concrete, recall that the central result of the previous section was the formula \eqref{eq:Iin} for the Cardy-like asymptotics of $\mathcal{I}_{\text{in}_n}$ (the contribution of the inner patch in$_n$ to the index); everything else followed from it via elementary manipulations. That asymptotic formula, according to Eq.~\eqref{eq:ZvsI}, is equivalent to
\begin{equation}
    \begin{split}
    Z_{\text{in}^{}_n}(p,q;\boldsymbol{v})&\simeq \int_{\text{in}^{}_n}\frac{D(\frac{2\pi\boldsymbol{x}}{\beta})}{(\sqrt{-\omega_1 \omega_2})^{r_G}} \, e^{-V^{\text{in}_n}(\boldsymbol{x})}\, \frac{\prod_\chi\prod_{\rho^\chi\in L_n}\Gamma_h\big(r_\chi (\frac{\omega_1+\omega_2}{2})+\frac{2\pi}{\beta}(\rho^\chi\cdot \boldsymbol{x}+q^\chi\cdot\boldsymbol{\xi})^{}_\mathbb{Z}\big)}{\prod_{\alpha_+\in L_n}\Gamma_h\big(\frac{2\pi}{\beta}(\alpha_+\cdot\boldsymbol{x})^{}_\mathbb{Z}\big)\Gamma_h\big(\frac{2\pi}{\beta}(-\alpha_+\cdot\boldsymbol{x})^{}_\mathbb{Z}\big)}
    .
    \end{split}\label{eq:ZinGeneralities}
\end{equation}
Reproducing this expression via localization in 3d EFT is the main goal of the present section.

\subsection{The 3d background}\label{sec:EFT:background}
Since we are interested in the Cardy-like limit where the circle size $\beta$ shrinks, we put the Hopf surface metric \eqref{eq:4dMetric} in a Kaluza-Klein (KK) form
\begin{equation}
    \mathrm{d}s_4^2=\mathrm{d}s_3^2+(\mathrm{d}t_E+c)^2.
\end{equation}
In the three-dimensional picture we then have a manifold $\mathcal{M}_3$ with metric
\begin{equation}
\begin{split}
    \mathrm{d} s^2_3 \,&=\, \frac{ \left(b^2 \cos^2\theta + b^{-2}\sin^2\theta\right)\mathrm{d} \theta^2 +  b^{-2}\cos^2\theta\,\mathrm{d}\varphi_1^2 + b^2\sin^2\theta\, \mathrm{d}\varphi_2^2}{1+b^{-2}k_1^2\cos^2\theta+b^2k_2^2\sin^2\theta}\\[1mm]
\,&\,\qquad - \left(\frac{ b^{-2}k_1 \cos^2\theta\,\mathrm{d}\varphi_1 + b^2k_2\sin^2\theta\,\mathrm{d}\varphi_2}{1+b^{-2}k_1^2\cos^2\theta+b^2k_2^2\sin^2\theta}\right)^2,\label{eq:M3metric}
\end{split}
\end{equation}
as well as a graviphoton
\begin{equation}
    c\,=\,\frac{ b^{-2}k_1 \cos^2\theta\,\mathrm{d}\varphi_1 + b^2k_2\sin^2\theta\,\mathrm{d}\varphi_2}{1+b^{-2}k_1^2\cos^2\theta+b^2k_2^2\sin^2\theta}.
\end{equation}

The KK reduction of the 4d background $V$ and $A$ fields  of new minimal supergravity leads to the 3d background fields $v$, $\mathcal{A}^{(R)}$, and $H$, which together with the 3d metric comprise the bosonic components of the 3d new minimal supergravity multiplet. We recite \cite{Cassani:2021fyv}:
\begin{equation}
\begin{split}
    H&=\frac{i+b^{-1}k_1+bk_2}{\sqrt{b^2\cos^2\theta+b^{-2}\sin^2\theta}},\\
    v&=-i\ast_3 \mathrm{d}c=\frac{2i\big( k_2\cos^2\theta\, \mathrm{d}\varphi_1 + k_1\sin^2\theta\, \mathrm{d}\varphi_2\big)}{\big(1+b^{-2}k_1^2\cos^2\theta+b^2k_2^2\sin^2\theta\big)\sqrt{b^2\cos^2\theta+b^{-2}\sin^2\theta}},
    \label{eq:H-v}
    \end{split}
\end{equation}
and refer the interested reader to \cite{Cassani:2021fyv} for the explicit expression of $\mathcal{A}^{(R)}$ (denoted $\mathcal{A}$ there).

The 4d background gauge field $A^{(f)\, a}$ sits in a 4d $\mathcal{N}=1$ vector multiplet with auxiliary field $D^{(f)\, a}=0$. Its KK reduction leads to a 3d background scalar $\sigma^{(f)\,a}=A_{t_E}^{(f)\,a}$ as well as a 3d background gauge field $\mathcal{A}_\mu^{(f)\, a}=-A_{t_E}^{(f)\,a}c_\mu$, which sit together with an auxiliary field $\mathcal{D}^{(f)\,a}=-A_{t_E}^{(f)\,a}H$ in a 3d $\mathcal{N}=2$ vector multiplet. See \cite{DiPietro:2016ond,ArabiArdehali:2021nsx}.

We emphasize that all 4d $\mathcal{N}=1$ multiplets---the background multiplets whose dimensional reduction was just described as well as the dynamical multiplets whose KK expansion will be described below---yield 3d $\mathcal{N}=2$ multiplets on $\mathcal{M}_3$.

\subsection{Light fields and cut-offs}\label{sec:lightModes}
After discussing the dimensional reduction of the background fields, we now consider the KK expansion of the dynamical fields.

The KK expansion leads to infinite towers of fields. This applies to all the 4d fields in the dynamical 4d $\mathcal{N}=1$ vector and chiral multiplets. If all the fields in these towers are kept, we are essentially back in the 4d realm; indeed the ``4d localization computation'' is actually done in this way \cite{Assel:2014paa}: by KK expanding the 4d fields, localizing the path-integrals of the 3d fields in the KK towers, and then summing over the towers.

Instead, we can integrate out the ``heavy'' 3d fields in the KK towers, and keep only the ``light'' fields. This leads to a Wilsonian approximation of the full partition function, which as we will see reproduces the asymptotics up to exponentially small corrections.

What makes the Wilsonian procedure particularly rich in the supersymmetric context is that often supersymmetric partition functions are integrals over a moduli space, and the separation of dynamical fields into ``light'' and ``heavy'' can not be done \emph{uniformly} over all of that moduli space. In the present setting, we will see that different parts of the BPS moduli space support different sets of light and heavy fields.

One of the main ideas of the present paper is that the EFT explanation of the asymptotics of the 4d index should proceed by decomposing the moduli space $\mathfrak{h}_\text{cl}$ of 4d BPS field configurations into various patches, and assigning to each patch its own 3d EFT according to the set of light fields the patch supports.

A clear sense of which 3d fields are light and which are heavy, is provided by the asymptotic hierarchy of scales emerging in the $\beta\to0$ limit.

Therefore two notions of ``cut-off'' arise in our analysis. The first one is the \emph{Wilsonian cut-off which sets the boundary between light and heavy} 3d fields. We denote this cut-off by $\Lambda_M$, and express it in terms of $\beta$ explicitly as
\begin{equation}
    \Lambda_{M}=\frac{2\pi\epsilon}{\beta}.\label{eq:EFTcut-off}
\end{equation}
Here $\epsilon$ is some $\beta$-independent, positive number. A specific choice of $\epsilon$ amounts to a precise scheme for separation of mass scales.

The second notion is that of the \emph{field-space cut-offs which set the boundaries between neighboring patches} of the moduli space. A specific choice of these cut-offs amounts to a precise scheme for decomposition of the moduli space.

The particular scheme that we adopt here for decomposition of $\mathfrak{h}_\text{cl}$ is based on the set of 3d fields that are light on each patch. Here by ``light'' we mean having mass less than the cut-off~$\Lambda_M$. In this way, we hence link our field-space decomposition scheme to our mass-separation scheme, so that our decomposition scheme is also fixed when we pick a specific value for $\epsilon.$ We emphasize though that this decision to link so closely the field-space decomposition and mass-separation schemes is motivated by the technical requirements of the present problem, and may not be as useful in other contexts.

Here we discuss the separation of 3d fields into light and heavy explicitly only for the fermions. The 3d $\mathcal{N}=2$ supersymmetry guarantees that each light fermion is accompanied by its scalar and/or vector super-partners, and similarly for the heavy fields.

The 3d fermion arising from the $n$th KK mode of the weight $\rho^{\chi}$ of a 4d chiral multiplet $\chi$ would have real-mass
(in the sign convention of \cite{DiPietro:2014bca})
\begin{equation}
    m^{[n]}_{\rho^{\chi}}=\frac{2\pi n}{\beta}-\frac{2\pi\, \rho^{\chi}\cdot \boldsymbol{x}}{\beta}-\frac{2\pi\, q^\chi\cdot\boldsymbol{\xi}}{\beta},\label{eq:chiRealMass}
\end{equation}
with the notation as in the previous section. This is light if
\begin{equation}
   |m^{[n]}_{\rho^{\chi}}|<\Lambda_M\  \Longleftrightarrow\quad |n- \rho^{\chi}\cdot \boldsymbol{x}- q^\chi\cdot\boldsymbol{\xi}|<\epsilon, \label{eq:lightChiFer}
\end{equation}
and heavy otherwise. Below we will denote the real-mass of the light 3d fermion arising from the weight $\rho^\chi$ by $m^{}_{\rho^\chi}$. Note that
\begin{equation}
m^{}_{\rho^\chi}=-\frac{2\pi}{\beta}(\rho^\chi\cdot \boldsymbol{x}+q^\chi\cdot\boldsymbol{\xi})^{}_\mathbb{Z}.\label{eq:lightRMchi}
\end{equation}

On the other hand, denoting the roots of the gauge group by $\alpha$, the 3d fermion arising from the $n$th KK mode of the $\alpha$ component of the vector multiplet would have real-mass
\begin{equation}
    m^{[n]}_{\alpha}=\frac{2\pi n}{\beta}-\frac{2\pi\, \alpha\cdot \boldsymbol{x}}{\beta}.\label{eq:vecRealMass}
\end{equation}
This is light if
\begin{equation}
   |m^{[n]}_{\alpha}|<\Lambda_M\  \Longleftrightarrow\quad |n- \alpha\cdot \boldsymbol{x}|<\epsilon, \label{eq:lightVecFer}
\end{equation}
and heavy otherwise. Below we will denote the real-mass of the light 3d fermion arising from the root $\alpha$ by $m^{}_{\alpha}$. Note that
\begin{equation}
m^{}_{\alpha}=-\frac{2\pi}{\beta}(\alpha^\chi\cdot \boldsymbol{x})^{}_\mathbb{Z}.\label{eq:lightRMvec}
\end{equation}

\subsubsection*{The outer patch}

Recall that in our convention $r_G$ of the roots are zero, and correspond to the Cartan of the gauge group. We can see from \eqref{eq:lightVecFer} that the $n=0$ mode of the Cartan fermions are light---indeed massless---for all $\boldsymbol{x}$.

If there are zero weights $\rho^\chi$ in the problem, we call them ``light'' only if $Q_\chi(\boldsymbol{\xi}):=q^\chi\cdot\boldsymbol{\xi}\in\mathbb{Z}$ (because otherwise we can take $\epsilon$ small enough so that \eqref{eq:lightChiFer} is not satisfied for any $n$). In other words, a weight $\rho^\chi$ is called light if it yields a light 3d fermion as in \eqref{eq:lightChiFer}. We can see from \eqref{eq:lightChiFer} that the $n=Q_\chi(\boldsymbol{\xi})$ mode of the fermions in these light zero weights are also light---massless indeed---for all $\boldsymbol{x}$.

The outer patch, from the EFT perspective, is the subset of the BPS moduli space $\mathfrak{h}_{\text{cl}}$ where there are no light 3d fermions besides the massless fermions of the previous two paragraphs. The massless fermions are of course accompanied by their 3d $\mathcal{N}=2$ super-partners. In particular, the massless fermions arising from the 4d $\mathcal{N}=1$ vector multiplet sit in massless 3d $\mathcal{N}=2$ gauge multiplets whose bosonic field content is the gauge field $\mathcal{A}_\mu$, the Coulomb branch scalar $\sigma$, and  
the auxiliary field~$\mathcal{D}$.

In the outer patch we thus have a 3d $\mathcal{N}=2$ gauge theory with $U(1)^{r_G}$ gauge group arising from the $n=0$ mode of the Cartan of the 4d vector multiplet, together with decoupled ($\rho^\chi=0$) massless 3d $\mathcal{N}=2$ chiral multiplets arising from the $n=Q_\chi(\boldsymbol{\xi})$ mode of the light zero weights of the 4d chiral multiplets.

Anticipating our 3d localization computation below, we note that (as emphasized in the present context in \cite{DiPietro:2016ond}) on the BPS locus of the 3d $\mathcal{N}=2$ $U(1)^{r_G}$ gauge theory we have
\begin{equation}
\begin{split}
\sigma^j = A^j_{t_E}\ \ \  &\xrightarrow{\text{on BPS locus}}\ \  \frac{2\pi x_j}{\beta},\\ 
\mathcal{A}^j_\mu =A^j_{\mu}-A^j_{t_E}c_\mu\,&\xrightarrow{\text{on BPS locus}}   \,- \frac{2\pi x_j}{\beta} c_\mu,\\
\mathcal{D}^j = D^j -A^j_{t_E} H \,&\xrightarrow{\text{on BPS locus}}\, -  \frac{2\pi x_j}{\beta}  H.\end{split}\label{eq:BPSlocus}
\end{equation} 

\subsubsection*{Inner patches}

Inner patches are the subsets of $\mathfrak{h}_{\text{cl}}$ where additional light 3d fermions arise. They can be distinguished from their neighboring patches by their differing set of light fermions. Note that in a patch in$_n$ the 4d fermion associated to a root $\alpha$ (or a weight $\rho^\chi$) yields a light 3d fermion only if $\alpha\in L_n$ (or $\rho^\chi\in L_n$). This explains our terminology in the previous section for $L_n$ as the set of ``light'' roots and weights. Of course 4d fermions associated to light roots and weights yield infinite KK towers of heavy 3d fields as well.

The extra light fermions of the inner patches also sit inside 3d $\mathcal{N}=2$ multiplets. In particular, if they arise from KK expansion of the 4d gauginos, they would be accompanied by light 3d vector fields that may combine with the massless Cartan gauge fields to enhance the gauge group of the low energy EFT from $U(1)^{r_G}$ to a non-abelian subgroup of $G$.

For example, in the inner patch in$_0$, not just the Cartan but all components of the 4d gauge field yield light 3d vector fields through their $n=0$ mode. This means that on in$_0$ we have a 3d $\mathcal{N}=2$ gauge theory with gauge group $G$. If there is a center symmetry (as in SU($N$) $\mathcal{N}=4$ theory) images of in$_0$ under the center symmetry will also have gauge group~$G$. These patches may or may not have additional light chiral multiplets as well.

There may also be patches where the gauge group is a proper non-abelian subgroup of~$G$. Those are the patches that dominate the index in partially-deconfined phases \cite{ArabiArdehali:2019orz,Hanada:2019kue,Cherman:2020zea}.

We emphasize that there could be inner patches where the EFT has the same $U(1)^{r_G}$ gauge group as the outer patch, but with additional light chiral multiplets arising from the 4d chiral multiplets. In other words, the gauge symmetry of the 3d EFT is not necessarily enhanced on inner patches.

On those inner patches where the gauge group is enhanced, we have to deal with non-abelian 3d vector multiplets. But anticipating our 3d localization computation below, we know that we need not worry about the non-Cartan components, because the BPS locus even in the non-abelian case is \eqref{eq:BPSlocus}.

Since in$_0$ is the most special inner patch, it is worth noting that in terms of the 3d Coulomb branch scalar we can characterize in$_0$ as (set $n=0$ in \eqref{eq:lightVecFer} and use \eqref{eq:BPSlocus})
\begin{equation}
  \text{in}_0:\qquad |\alpha\cdot \sigma|<\Lambda_{M}.  
\end{equation}
This is a sharp instance of the close relation between the mass-separation scheme and the field-space decomposition scheme.

\subsection{The Wilsonian effective action}\label{sec:EFT:action}
The tree-level action of the light 3d fields, denoted $S^L_\text{tree}$, comes from the UV: in the mode expansion of the 4d action we keep only the terms associated to the light fields. This is straightforward if one has access to the 4d Lagrangian on the Hopf surface, which can be found in  \cite{Assel:2014paa} (based on \cite{Festuccia:2011ws,Klare:2012gn,Dumitrescu:2012ha}).

The tree-level action is corrected at the one-loop level by the contribution arising from integrating out the heavy fields. The calculation needed for directly finding the complete one-loop correction is difficult (indeed daunting) because of the various kinds of fields present in our problem. The breakthrough realization of Di~Pietro and Komargodski \cite{DiPietro:2014bca} (refined further in \cite{DiPietro:2016ond,ArabiArdehali:2021nsx,Cassani:2021fyv}) was that one can recover the total result indirectly as follows. First, we compute certain one-loop generated Chern-Simons couplings that are easily under control and receive contributions only from the heavy fermions. Next, we supersymmetrize the resulting Chern-Simons actions. This accounts indirectly for the one-loop contribution of the heavy bosonic fields (including the KK modes of the dynamical gauge field and its ghosts) as well. Although a complete argument that the resulting supersymmetrized Chern-Simons actions exhaust the one-loop corrections is not currently available, as in \cite{ArabiArdehali:2021nsx,Cassani:2021fyv} we assume the exhaustion and find results that are in agreement with the previous section, save for a technical gap explained below (see also Problem~1 in Section~\ref{sec:discussion}).

The one-loop induced SUSY CS action is of the form (\emph{cf.} \cite{DiPietro:2016ond,ArabiArdehali:2021nsx})
\begin{equation}
\begin{split}
\delta S_\text{1-loop}&= \sum_\text{f}\big(\widetilde{S}^\text{f}_\text{g-g} + 2 \widetilde{S}^\text{f}_\text{g-R} + S^\text{f}_\text{R-R} + S^\text{f}_\text{grav}+S^\text{f}_v\big),
\label{eq:S1loop}
\end{split}
\end{equation}
where the sum is over all the heavy fermions in the theory. Here $\widetilde{S}^\text{f}_\text{g-g}$ stands for the supersymmetrized gauge-gauge CS action induced by integrating out a heavy fermion, together with the supersymmetrized gauge-KK and KK-KK actions that always accompany it. We also incorporate the background gauge fields $A^{(f)}_a$ into the same term following \cite{DiPietro:2016ond}, as elaborated on in Appendix~\ref{app:3d-CS}. The term $2 \widetilde{S}^\text{f}_\text{g-R}$ stands for the supersymmetrized gauge-R CS action, together with the supersymmetrized KK-R term that always accompanies it, while the factor of 2 accounts also for the supersymmetrized R-gauge and R-KK terms. The next two actions on the RHS of \eqref{eq:S1loop} are the supersymmetrized versions of RR and gravitational CS actions.

The last action on the RHS of \eqref{eq:S1loop} is new---it was not discussed in \cite{DiPietro:2016ond,ArabiArdehali:2021nsx,Cassani:2021fyv}. It stands for a yet-to-be-discovered supersymmetrized CS action in 3d new minimal supergravity, containing (mixed) CS terms involving $v$ (such as $v\wedge\mathrm{d}v$) but no RR or gravitational CS term. We have introduced it to resolve a difficulty (see the discussion below~\eqref{eq:Y:3d}) that was not addressed in \cite{DiPietro:2016ond,ArabiArdehali:2021nsx,Cassani:2021fyv}. It may also provide a resolution to a mismatch puzzle raised in \cite{Cassani:2021fyv} (see Eq.~\eqref{eq:modifiedG} in the appendix, and also Problem~1 in Section~\ref{sec:discussion}).

Assuming that the heavy fermion arises from the $n$th KK mode of a 4d chiral multiplet $\chi$, we have on the BPS locus \eqref{eq:BPSlocus} that (\emph{cf}. \cite{ArabiArdehali:2021nsx})
\begin{equation}\begin{split}
\widetilde{S}^\text{f}_\text{g-g} &\xrightarrow{\text{on BPS locus}} -\frac{i}{8 \pi}\, \mathrm{sgn}\big(m^{[n]}_{\rho^{\chi}}\big) \,  
\left(\frac{2\pi(n-\rho^{\chi}\cdot \boldsymbol{x}-q^\chi\cdot\boldsymbol{\xi})}{\beta} \right)^2\, A_{\mathcal{M}_3}, \\
 \widetilde{S}^\text{f}_\text{g-R} &\xrightarrow{\text{on BPS locus}}- \frac{i}{8\pi} \, \mathrm{sgn}\big(m^{[n])}_{\rho^{\chi}}\big)  \, \left(\frac{2\pi(n-\rho^{\chi}\cdot \boldsymbol{x}-q^\chi\cdot\boldsymbol{\xi})}{\beta} \right) \cdot  (r_\chi-1)\, \, \, L_{\mathcal{M}_3},\\
    S^\tf_\text{R-R}&=- \frac{i}{8\pi} \, \mathrm{sgn}\big(m^{[n]}_{\rho^{\chi}}\big)\, \bigl((r_\chi-1)^2-\frac{1}{6}\bigr) \, R_{\mathcal{M}_3}, \\
  S^\tf_\text{grav}&=- \frac{i}{192\pi} \, \mathrm{sgn}\big(m^{[n]}_{\rho^{\chi}}\big)\, G_{\mathcal{M}_3}.
    \end{split}\label{eq:S1loopSUSYCS}
\end{equation}
Here we have written the supersymmetrized CS actions as the one-loop CS coupling (see Appendix~\ref{app:3d-CS:chiral}) times the supersymmetrization of an $\int A\wedge\mathrm{d}A$ type term. That is except for the RR CS term whose coefficient is shifted by $-1/6$ to cancel the extra RR term arising from the supersymmetrized gravitational CS term. The precise form of $A_{\mathcal{M}_3}$, $L_{\mathcal{M}_3}$, $R_{\mathcal{M}_3}$, and $G_{\mathcal{M}_3}$ can be found in Appendix~\ref{app:3d-CS:susy}. We actually need to evaluate them on the 3d background of Section~\ref{sec:EFT:background}, and the resulting expressions can be found in Appendix~\ref{app:3d-CS:susy} as well.

We now perform the sum over the heavy 3d fermions. For the 3d fermions arising from a 4d chiral multiplet $\chi$ this involves a sum over the weights $\rho^\chi$ as
well as a sum over the KK numbers $n$. For the weights $\rho^\chi\in H_n$ we sum over all $n\in\mathbb{Z}$, and use
\begin{equation}
    \sum_{n\in\mathbb{Z}}\mathrm{sgn}(n+x)(n+x)^{j-1} = -\frac{2}{j} \, \overline{B}_{j}(x) \,,
\label{eq:KKsumBbar}
\end{equation}
to evaluate the KK sums. For the weights $\rho^\chi\in L_n$ we sum over all $n\in\mathbb{Z}$ except for the integer that yields a light 3d multiplet. Then use
\begin{equation}
    \sum'_{n\in\mathbb{Z}}\mathrm{sgn}(n+x)(n+x)^{j-1} = -\frac{2}{j} \, \overline{K}_{j}(x) \,,
\label{eq:KKsumKbar}
\end{equation} 
which follows from \eqref{eq:KKsumBbar} and \eqref{eq:KbarDef}, to simplify the result. The prime in the above sum means that the integer $n$ for which $|n+x|<\epsilon$ is excluded from the summation.

The inclusion of the heavy 3d fermions arising from the $n$th KK mode of a 4d vector multiplet leads to expressions similar to \eqref{eq:S1loopSUSYCS}:
\begin{equation}\begin{split}
\widetilde{S}^\text{f}_\text{g-g} &\xrightarrow{\text{on BPS locus}} -\frac{i }{8\pi}\, \mathrm{sgn}\big(m^{[n]}_{\alpha}\big) \,  
\left(\frac{2\pi(n-\alpha\cdot \boldsymbol{x})}{\beta} \right)^2\, A_{\mathcal{M}_3}, \\
 \widetilde{S}^\text{f}_\text{g-R} &\xrightarrow{\text{on BPS locus}}- \frac{i}{8\pi} \, \mathrm{sgn}\big(m^{[n]}_{\alpha}\big)  \, \left(\frac{2\pi(n-\alpha\cdot \boldsymbol{x})}{\beta} \right) \, \, L_{\mathcal{M}_3},\\
    S^\tf_\text{R-R}&=- \frac{i}{8\pi} \, \mathrm{sgn}\big(m^{[n]}_{\alpha}\big)\, \bigl(1-\frac{1}{6}\bigr) \, R_{\mathcal{M}_3}, \\
  S^\tf_\text{grav}&=- \frac{i}{192\pi} \, \mathrm{sgn}\big(m^{[n])}_{\alpha}\big)\, G_{\mathcal{M}_3}.
    \end{split}\label{eq:S1loopSUSYCSvec}
\end{equation}
Then the sum over 3d heavy fermions from the KK modes of a 4d vector multiplet can also be done using the formulas \eqref{eq:KKsumBbar} and \eqref{eq:KKsumKbar}.

Incorporating \eqref{eq:S1loopSUSYCS} and \eqref{eq:S1loopSUSYCSvec} and then performing the sums over the heavy fermions, we get
\begin{equation}
\begin{split}
\sum_\text{f}\widetilde{S}^\text{f}_\text{g-g} \xrightarrow{\text{on BPS locus}} &\ \frac{V^{\text{in}_n}_2(\boldsymbol{x})}{\beta^2},\\
    \sum_\text{f}2\widetilde{S}^\text{f}_\text{g-R} \xrightarrow{\text{on BPS locus}} &\ \frac{V^{\text{in}_n}_1(\boldsymbol{x})}{\beta},\\
    \sum_\text{f}\big({S}^\text{f}_\text{R-R}+{S}^\text{f}_\text{grav}+{S}^\text{f}_v\big) \overset{!}{=}&\  V^{\text{in}_n}_0(\boldsymbol{x}),    \end{split}\label{eq:S1loop&Vin}
\end{equation}
with $V^{\text{in}_n}_{2,1,0}(\boldsymbol{x})$ as in \eqref{eq:Vin}. Note that in absence of a precise expression for $S^\text{f}_v$, we are conjecturing here that its incorporation would lead to the last relation above. (An alternative form of this conjecture is \eqref{eq:modifiedG}). From \eqref{V:in} and \eqref{eq:S1loop}, we then have
\begin{equation}
    \delta S_\text{1-loop}\xrightarrow{\text{on BPS locus}}V^{\text{in}_n}(\boldsymbol{x}).\label{eq:deltaS=Vin}
\end{equation}

Some arguments in \cite{Cassani:2021fyv} suggest that there are no higher-loop corrections to the Wilsonian effective action. Although a complete non-renormalization argument directly in the 3d picture is not currently available, we assume it in the following and will find a perfect match with the results of the previous section. The effective action that we use for supersymmetric localization of the light 3d fields is hence
\begin{equation}
    S_{\text{eff}}=S^L_{\text{tree}}+\delta S_{\text{1-loop}}.\label{eq:Seff}
\end{equation}

In summary, the UV action is the tree-level 4d action, which can be KK expanded and expressed in terms of infinitely-many 3d fields. We integrate out the heavy 3d fields and obtain an IR effective action for the light 3d fields. Note that we do not integrate out the high-energy modes of the light 3d fields (with ``energy'' as measured via some time direction inside $\mathcal{M}_3$, and ``high'' meaning above the cut-off $\Lambda_M$); as the following localization calculation shows it is useful here not to further amputate the high-energy tail of the light 3d fields.

\subsection{Localization in the 3d EFT and the asymptotics of the partition function}\label{sec:EFT:local}

Supersymmetric localization of the 3d EFT on $\mathcal{M}_3$ gives
\begin{equation}
    \begin{split}
    Z^\text{EFT}_{\text{in}^{}_n}(p,q;\boldsymbol{v})=&\int e^{-S^L_{\text{tree}}-\delta S_{\text{1-loop}}}\\
    \xrightarrow{\text{localization}}&\int_{\frac{2\pi}{\beta}\text{in}^{}_n}\frac{D\boldsymbol{\sigma}}{(\sqrt{-\omega_1 \omega_2})^{r_G}} \, \frac{\prod_\chi\prod_{\rho^\chi\in L_n}\Gamma_h\big(r_\chi (\frac{\omega_1+\omega_2}{2})-m^{}_{\rho^\chi}\big)}{\prod_{\alpha_+\in L_n}\Gamma_h\big(-m^{}_{\alpha_+}\big)\Gamma_h\big(m^{}_{\alpha_+}\big)}
    \, e^{-V^{\text{in}_n}(\boldsymbol{x})}.
    \end{split}\label{eq:ZinLocalization}
\end{equation}
The $e^{-V^{\text{in}_n}(\boldsymbol{x})}$ factor arises from $\delta S_{\text{1-loop}}$ evaluated on the BPS locus. The rest is the familiar expression (see \emph{e.g.} \cite{Closset:2013vra} or Section~5 of \cite{Aharony:2013dha}) arising from localization of $S^L_{\text{tree}}$, assuming (based on expectations from holomorphy and dimensional reduction) that $\omega_1,\omega_2$ are the moduli of the transversely holomorphic foliation \cite{Closset:2013vra} of $\mathcal{M}_3$.

The above result \eqref{eq:ZinLocalization} matches our asymptotic expression \eqref{eq:Iin} upon using the explicit expressions for the light real-masses in \eqref{eq:lightRMchi} and \eqref{eq:lightRMvec}, as well as translating between $\boldsymbol{\sigma}$ and $\boldsymbol{x}$ via \eqref{eq:BPSlocus}. \\

\section{Discussion}\label{sec:discussion}
An interplay between asymptotic analysis and EFT description of exact results was explored in this work.

When the exact result is a localized supersymmetric partition function and the limit of interest introduces a scale hierarchy in the problem, an important aspect of the said interplay is that asymptotic estimates of the integrand \emph{lose uniform validity} precisely on those parts of the BPS moduli space \emph{where additional light fields emerge} that were not incorporated (or were treated as heavy) in the estimate. This calls for a decomposition of the moduli space into various patches according to the set of light fields they support, and using appropriate estimates with uniform validity over each patch.

In the previous sections we fleshed out the details of this procedure in the specific context of the Cardy-like limit of the 4d superconformal index. We gleaned in particular the following two lessons that have the promise of finding wider applicability in asymptotic analysis of supersymmetric partition functions:
\begin{itemize}
    \item that \emph{competition between patches} of the BPS moduli space can be a more powerful perspective than \emph{competition between saddles}, especially in presence of extended/degenerate saddles (this is the perspective that in Section~\ref{sec:asymptoticAnalysis} enabled us to establish the polynomial factor in the general formula \eqref{eq:ItotGenSimp}); 
    \item that not only the quantum moduli space as a whole, but also its intersections with the strata of the singular set of the classical moduli space (as a topologically stratified space) can be of significance for asymptotics of the partition function (see the discussion around \eqref{eq:out+(in-out)Refined}).
\end{itemize}

Below we highlight another lesson, as well as a few puzzles arising from our study.




\subsection{Non-renormalization: Wilsonian effective action versus $\log \mathcal{I}$}

We saw in the previous section that compatibility of the EFT arguments with the asymptotic analysis of Section~\ref{sec:asymptoticAnalysis} leads to the conclusion that the Wilsonian effective action \eqref{eq:Seff} does not receive higher-loop corrections; in other words, it is exact to all orders in $1/\Lambda_M$. (A direct EFT demonstration of this result is desirable; see Problem~2 below.)

When $\mathrm{dim}\mathfrak{h}_\text{qu}=0$, or more generally when the polynomial factor in \eqref{eq:ItotGenSimp} has only one non-zero term, the non-renormalization of $S_\text{eff}$ implies that the small-$\beta$ expansion of $\log Z(p,q;\boldsymbol{v})$ terminates at $\mathcal{O}(\beta^0)$---and consequently the small-$\beta$ expansion of $\log \mathcal{I}(p,q;\boldsymbol{v})$ terminates at $\mathcal{O}(\beta)$ with the linear term being $\beta E_\text{susy}$.

But when $\mathrm{dim}\mathfrak{h}_\text{qu}>0$ and the polynomial factor in \eqref{eq:ItotGenSimp} contains more than one term, it is easy to see that the small-$\beta$ expansion of $\log Z(p,q;\boldsymbol{v})$---or $\log \mathcal{I}(p,q;\boldsymbol{v})$---does not terminate. This is analogous to the standard statement in supersymmetric QFT that while the ($F$-term part of the) Wilsonian effective action is not renormalized beyond one loop, the ($F$-term part of the) 1PI action can receive corrections at higher orders.

A non-trivial aspect of our findings in this context, unexpected from standard supersymmetric QFT literature, is that an extended quantum moduli space---or presence of massless degrees of freedom---is \emph{not sufficient} for breakdown of the non-renormalization of $\log \mathcal{I}$ (\emph{i.e.}~termination of its small-$\beta$ expansion). The latter requires that the polynomial factor in \eqref{eq:ItotGenSimp} has at least two nonzero terms; \emph{i.e.} that $C_j=C^{\text{out}}_j(\epsilon)+\sum_{n_\ast}C^{\text{in}_{n_\ast}}_j(\epsilon)\neq 0$ for at least two different values of the non-negative integer $j$, with $C^{\text{out}}_j(\epsilon)$ and $C^{\text{in}_{n_\ast}}_j(\epsilon)$ introduced in \eqref{eq:IoutCutGenSimp} and \eqref{eq:IinCutGenSimp} respectively. Roughly speaking, the index should receive dominant contributions from regions of the moduli space that localize in the $\beta\to0$ limit to subsets with different dimensions.


This sheds new light on why inclusion of decoupled massless abelian vector multiplets---which give rise to an extended quantum moduli space---does not spoil the termination of the small-$\beta$ expansion of $\log\mathcal{I}$ \cite{Ardehali:2015hya}. That is because the index of decoupled abelian vector multiplets is a product of Pochhammer symbols, so they multiply the polynomial factor of \eqref{eq:ItotGenSimp} only by a monomial (see \eqref{eq:PochEst}). Therefore if that polynomial had only one nonzero term before their inclusion, it will have only one nonzero term after their inclusion as well, and the small-$\beta$ expansion of $\log\mathcal{I}$ still terminates.

\subsection{Open problems}

We have left a few technical issues in our analysis unaddressed. For example, in comparing the asymptotics of the $\mathcal{N}=4$ SU(2) index on the peripheral triangles of the W wing as follows from the Bethe-Ansatz approach and as follows from our integral based approach, we have not demonstrated that the coefficient $C_0$ in \eqref{eq:I:W:peripheral} equals $-\fft{i\eta}{6}+\fft{1}{2\pi}$ (see below \eqref{eq:I:W:BA}). Another technical issue is that we have not demonstrated that the $\mathcal{M}_3$ metric in \eqref{eq:M3metric} admits a transversely holomorphic foliation (THF) with moduli $\omega_{1,2}$ as in \eqref{eq:thfModuli}.

More importantly, our asymptotic analysis has led to formulation of sharp and conceptually interesting puzzles in Wilsonian EFT. We highlight three of them below, as examples of EFT projects that are motivated and informed by asymptotic analysis of exact results.
\begin{description}
  \item[Problem 1)] Demonstrate that the SUSY CS terms exhaust the one-loop corrections to the Wilsonian effective action of Section~\ref{sec:EFT}, at least on the BPS locus.
\end{description}
In other words, prove that no other SUSY action is generated at one loop besides the SUSY CS terms, or that if such extra actions are generated they vanish on the BPS locus so that the localization result \eqref{eq:ZinLocalization} is unaltered.

Note that the result  \eqref{eq:ZinLocalization} relied on the conjectural equality on the third line of \eqref{eq:S1loop&Vin}. So part of the above problem is to identify the SUSY CS action $S_v$ in \eqref{eq:S1loop}, and to show that once evaluated on the supersymmetric background of Section~\ref{sec:EFT} it leads to the equality in \eqref{eq:S1loop&Vin} (which otherwise would not be valid for general $\omega_{1,2}\in\mathbb{H}$ according to the discussion at the end of Appendix~\ref{app:3d-CS:susy}).

\begin{description}
  \item[Problem 2)] Prove a non-renormalization theorem in the EFT of Section~\ref{sec:EFT}, prohibiting higher-loop corrections to $S_\text{eff}$, at least on the BPS locus.
\end{description}
In other words, prove that the effective action \eqref{eq:Seff} is exact to all orders in $1/\Lambda_M$, at least on the BPS locus. Such a theorem would be a curved super-space counterpart of the standard $F$-term non-renormalization theorem on flat space \cite{Grisaru:1979wc}.

\begin{description}
  \item[Problem 3)] Why is zeta-function regularization appropriate for the KK sums featuring in the 3d effective action?
\end{description}
The matching between asymptotic analysis and EFT also relied on our use of the zeta-function regularization for the infinite KK sums arising in the 3d EFT---see~\eqref{eq:KKsumBbar}. It would be desirable to have a satisfactory argument directly in the EFT to justify 
this use. (See Appendix~D of \cite{ArabiArdehali:2016fjg} for a proof of appropriateness of the zeta-function regularization in a related context. There, non-commutativity of the small-$\beta$ limit and plethystic exponentiation necessitates regularization of the plethystic exponential of a small-$\beta$ expansion. Here, non-commutativity of the small-$\beta$ limit and KK expansion necessitates regularization of the KK sum over small-$\beta$ contributions.)




\begin{acknowledgments}

This paper is in many ways a follow-up to our earlier works \cite{GonzalezLezcano:2020yeb} and \cite{ArabiArdehali:2021nsx}. We would like to acknowledge the essential contribution of our collaborators in those projects, A.~Gonzalez~Lezcano, J.~Liu, S.~Murthy, and L.~Pando~Zayas, to the ideas that are developed further in this work. We are also grateful to D.~Cassani for sharing with us the computations leading to~\eqref{eq:bkgdBPSsusyCS}, and to A.~Bourget, A.~Cabo~Bizet, J.~Grimminger, A.~Hanany, and Z.~Zhong for related discussions.
AA began this project at KCL where he was supported by the ERC Consolidator Grant
N.~681908, “Quantum black holes: A macroscopic window into the microstructure of gravity”, and is now supported by the NSF grant PHY-1915093 and the Simons Foundation (Simons Collaboration on the Non-Perturbative Bootstrap, grants 397411 and 488647). JH is supported in part by an Odysseus grant G0F9516N from the FWO, by the KU Leuven C1 grant ZKD1118 C16/16/005, and by the Research Programme of the Research Foundation grant G.0926.17N from the FWO.

\end{acknowledgments}

\appendix

\section{Supersymmetric Chern-Simons actions}\label{app:3d-CS}
In this appendix, we consider 3d effective Chern-Simons (CS) terms generated by integrating out heavy fermions arising from the Kaluza-Klein (KK) expansion of a 4d $\mathcal N=1$ chiral multiplet. We also discuss supersymmetrization of the CS terms.

\subsection{Chern-Simons actions from integrating out heavy fermions}\label{app:3d-CS:chiral}
We start from the relevant part of the Euclidean 4d action of an $\mathcal N=1$ chiral multiplet, namely ($M,N\in\{x^1,x^2,x^3,t_E\}$ are 4d coordinate indices)
\begin{equation}
    S_\text{UV}\supseteq S_\text{UV}^\chi=\int \mathrm{d}^4x\sqrt{g}\,\left[i\tilde\chi\tilde\sigma^M D^{}_M\chi+\fft12V^M\tilde\chi\tilde\sigma^{}_M\chi\right]\label{S:chiral:1}
\end{equation}
where the covariant derivative is given as
\begin{equation}
    D_M=\nabla_M-i\rho^\chi\cdot A_M-iq^\chi\cdot A_M^{(f)}-i(r_\chi-1)A_M^{(R)},\label{cov:D}
\end{equation}
and $\chi$ stands for the fermion field in the chiral multiplet labeled by the same symbol. Here we have mainly followed the convention of \cite{Assel:2014paa}. To be specific, $\rho^\chi$ is the weight of the representation of a chiral multiplet with respect to the gauge symmetry and $r_\chi$ is the $R$-charge of the chiral multiplet (the $R$-charge of the fermion in the multiplet is therefore $r_\chi-1$). $q^\chi$ is the vector of charges under various flavor symmetries. $A_M$ stands for the Cartan of the dynamical gauge field, and $A_M^{(f)}$ is the background gauge field associated with flavor symmetries; the latter was not included in \cite{Assel:2014paa}, but it is necessary for capturing flavor fugacities. $A^{(R)}_M$ and $V_M$ are the $R$-symmetry background gauge field and the auxiliary 1-form in the gravity multiplet of the 4d new minimal supergravity \cite{Sohnius:1981tp,Sohnius:1982fw}. The sigma matrices and the connection $\nabla_M$ are given in our convention as ($A,B\in\{1,2,3,4\}$ are 4d frame indices)
\begin{equation}
	\sigma^M=(\vec\sigma,-iI_2),\qquad\tilde\sigma^M=(-\vec\sigma,-iI_2),\qquad\nabla^{}_M\chi=(\partial_M-\fft14\omega_M^{AB}\sigma_{[A}\tilde\sigma_{B]})\chi.\label{convention}
\end{equation}
Combining (\ref{cov:D}) and (\ref{convention}), we can rewrite (\ref{S:chiral:1}) explicitly as 
\begin{equation}
	S_\text{UV}^\chi=\int \mathrm{d}^4 x\, \sqrt{g}\left[ i\tilde\chi\tilde\sigma^M(\partial_M-\fft14\omega_M^{AB}\sigma_{[A}\tilde\sigma_{B]}-iY^{}_M)\chi\right],\label{S:chiral:2}
\end{equation}
where we have introduced a 4d 1-form $Y_M$ as
\begin{equation}
    Y_M\equiv\rho^\chi\cdot A_M+q^\chi\cdot A_M^{(f)}+(r_\chi-1)A_M^{(R)}+\fft12V_M.\label{eq:Y}
\end{equation}

Next we explore the KK compactification of the 4d action (\ref{S:chiral:2}) on $S^1$ of the 4d manifold that is topologically $\mathcal M_3\times S^1$, and whose metric is of the form ($\mu,\nu\in\{x^1,x^2,x^3\}$ are 3d coordinate indices)
\begin{equation}
	ds_4^2=g^{}_{MN}dx^Mdx^N=h_{\mu\nu}dx^\mu dx^\nu+(dt_E+c_\mu dx^\mu)^2,
\end{equation}
where $t_E\sim t_E+\beta$. Using ($a,b\in\{1,2,3\}$ are 3d frame indices)
\begin{equation}
	(e^{-1})^M{}_A=\begin{pmatrix}
		(e^{-1}_\text{(3d)})^\mu{}_a & \ \ 0\\
		-c_\mu(e^{-1}_\text{(3d)})^\mu{}_a & \ \ 1
		\end{pmatrix},\qquad\omega_4^{ab}=-\fft12 f^{ab},\qquad\omega_a^{4b}=-\fft12 f_a{}^b,
\end{equation}
in terms of the field strength of a KK photon, $f_{ab}\equiv\partial_ac_b-\partial_bc_a$, the 4d action (\ref{S:chiral:2}) can be rewritten as
\begin{equation}
\begin{split}
	S_\text{UV}^\chi&=\int \mathrm{d}t_E\, \mathrm{d}^3x\,\sqrt{h}\left[-\tilde\chi i\sigma^a\Big((e^{-1}_\text{(3d)})^\mu{}_a\partial_\mu-c_\mu(e^{-1}_\text{(3d)})^\mu{}_a\partial_{t_E}+\fft14\omega_a^{bc}\sigma_{[b}\sigma_{c]}+\fft{i}{4}f_a{}^b\sigma_b\right.\\
	&\kern11em\,-i(e^{-1}_\text{(3d)})^\mu{}_aY_\mu+ic_\mu(e^{-1}_{(3d)})^\mu{}_aY_{t_E}\Big)\chi\\
	&\kern8em\left.+\tilde\chi(\partial_{t_E}+\fft18f^{ab}\sigma_{[a}\sigma_{b]}-iY_{t_E})\chi\right].
\end{split}\label{S:chiral:3}
\end{equation}
Substituting the KK expansion of the 4d fermion
\begin{equation}
	\chi=\sum_{n\in\mathbb Z}\fft{1}{\sqrt{\beta}}e^{2\pi nit_E/\beta}\chi^{(n)},\qquad \tilde\chi=\sum_{n\in\mathbb Z}\fft{1}{\sqrt{\beta}}e^{-2\pi nit_E/\beta}\tilde\chi^{(n)},
\end{equation}
and integrating over the $x^4$ coordinate, the 4d action (\ref{S:chiral:3}) reads
\begin{equation}
\begin{split}
	S_\text{UV}^\chi&=\sum_{n\in\mathbb Z}\int \mathrm{d}^3x\,\sqrt{h}\bigg[\tilde\chi^{(n)}\Big(-i\sigma^a(e^{-1}_\text{(3d)})^\mu{}_a\partial_\mu-\fft{i}{4}\omega_a^{bc}\sigma^a\sigma_b\sigma_c+\fft18 f^{ab}\sigma_a\sigma_b+im^{[n]}_{\rho^\chi}\\
	&\kern10em-\sigma^a(e^{-1}_\text{(3d)})^\mu{}_a(\mathcal Y^{(0)}_\mu+\fft{2\pi n}{\beta}c_\mu)\Big)\chi^{(n)}\bigg]+\cdots ,
\end{split}\label{S:chiral:KK}
\end{equation}
with the dots arising from the non-zero KK modes of the dynamical gauge field.
Above we have defined a shifted fermion mass $m^{[n]}_{\rho^\chi}$ and a combined 3d gauge field $\mathcal Y^{(0)}_\mu$ as
\begin{equation}
    m^{[n]}_{\rho^\chi}=\fft{2\pi n}{\beta}-Y^{(0)}_{t_E},\qquad \mathcal Y_\mu^{(0)}=Y^{(0)}_\mu-Y^{(0)}_{t_E}c_\mu,\label{eq:mA}
\end{equation}
with the superscripts $(0)$ emphasizing that only the zeroth KK mode of the dynamical gauge field is included. Note that the superscript ``$[n]$'' of the mass $m^{[n]}_{\rho^\chi}$ in \eqref{eq:mA} does not mean that it is the $n$th mode of a KK expansion: it just represents that it is a function of $n$.

We would like to see how CS terms show up in the 3d effective action. For that purpose, first we write down the partition function of an $\mathcal N=1$ chiral multiplet in terms of the effective action $S_\text{eff}$ obtained by integrating out heavy fermions as
\begin{equation}
\begin{split}
    Z&=\int[\mathrm{d}\chi \mathrm{d}\tilde\chi \mathrm{d}\mathcal Y\cdots]\exp[-S_\text{UV}[\chi,\tilde\chi,\mathcal Y,\cdots]]\\
    &=\int[\mathrm{d}\mathcal Y\cdots]\exp[-S_\text{eff}[\mathcal Y,\cdots]],
\end{split}
\end{equation}
where `$\cdots$' represent other fields in the chiral multiplet. As mentioned above, we first consider the contribution from the $S^\chi$ part given in (\ref{S:chiral:KK}) to the effective action $S_\text{eff}$, namely $S^\chi_\text{eff}[\mathcal Y]\subseteq S_\text{eff}[\mathcal Y,\cdots]$. It is given as
\begin{equation}
    \exp[-S^\chi_\text{eff}[\mathcal Y]]=\int[\mathrm{d}\chi \mathrm{d}\tilde\chi]\exp[-S_\text{UV}^\chi[\chi,\tilde\chi,\mathcal Y]].\label{S:chi:eff}
\end{equation}
The contribution from other fields to the effective action $S_\text{eff}$ will then be obtained by supersymmetrizing $S^\chi_\text{eff}$. Note that, according to \eqref{S:chi:eff}, $S^\chi_\text{eff}[\mathcal Y]$ can be computed by evaluating connected Feynman diagrams as
\begin{equation}
    S^\chi_\text{eff}[\mathcal Y]=-\left\langle\exp[-S^\chi_\text{int}]\right\rangle^\text{connected},\label{I}
\end{equation}
where the interaction term $S^\chi_\text{int}$ is read off from (\ref{S:chiral:KK}) as
\begin{equation}
\begin{split}
	S^\chi_\text{int}&=\sum_{n\in\mathbb Z}\int \mathrm{d}^3x\,\sqrt{h}\left[\tilde\chi^{(n)}\bigg(-\fft{i}{4}\omega_a^{bc}\sigma^a\sigma_b\sigma_c+\fft18f^{ab}\sigma_a\sigma_b-\sigma^a(e^{-1}_\text{(3d)})^\mu{}_a(\mathcal Y^{(0)}_\mu+\fft{2\pi n}{\beta}c_\mu)\bigg)\chi^{(n)}\right].
\end{split}\label{S:chiral:KK:int}
\end{equation}
Evaluating the effective action (\ref{I}) at one-loop order, we will derive a CS term for the combined 3d gauge field $\mathcal Y^{(0)}_\mu$. The following calculation involves only elementary QFT but we provide details for completeness.

To compute the effective action (\ref{I}) at one-loop order explicitly, we consider the simplest case with flat $h_{\mu\nu}=\delta_{\mu\nu}$.\footnote{A more methodical approach would be to work on curved background and use the heat kernel.} Then we use the Pauli matrices as 3d Euclidean gamma matrices, namely $\gamma^\mu\in\{\sigma^1,\sigma^2,\sigma^3\}$, which satisfy the trace formulas
\begin{equation}
\begin{split}
	\tr(\gamma_\mu\gamma_\nu)&=2\delta_{\mu\nu},\\
	\tr(\gamma_\mu\gamma_\nu\gamma_\rho)&=2i\epsilon_{\mu\nu\rho},\\
	\tr(\gamma_\mu\gamma_\rho\gamma_\nu\gamma_\sigma)&=2(\delta_{\nu\sigma}\delta_{\mu\rho}-\delta_{\rho\sigma}\delta_{\mu\nu}+\delta_{\mu\sigma}\delta_{\nu\rho}).
	\end{split}\label{gamma:tr}
\end{equation}
Now the one-loop order effective action (\ref{I}) reads
\begin{equation}
\begin{split}
	S^\chi_\text{eff}[\mathcal Y]\big|_\text{1-loop}&=-\fft{1}{2!}\sum'_{m,n\in\mathbb Z}\int \mathrm{d}^3x\int \mathrm{d}^3y\,(\mathcal Y^{(0)\,\mu}(x)+\fft{2\pi m}{\beta}c^\mu(x))(\mathcal Y^{(0)\,\nu}(y)+\fft{2\pi n}{\beta}c^\nu(y))\\
	&\kern7em\times\left\langle\tilde\chi^{(m)}(x)\gamma_\mu\chi^{(m)}(x)\tilde\chi^{(n)}{}(y)\gamma_\nu\chi^{(n)}(y)\right\rangle,
\end{split}\label{I:A2:1}
\end{equation}
with the prime over the summation symbol excluding the light fermions. Using the Wick's theorem, as well as the Fourier transforms of the gauge field $\mathcal Y^{(0)}$ and the KK photon $c$, and also the fermion propagator read off from (\ref{S:chiral:KK}) as
\begin{equation}
	\left\langle\chi^{(m)}(x)\tilde\chi^{(n)}(y)\right\rangle=\delta^{mn}\int\fft{\mathrm{d}^3k}{(2\pi)^3}\fft{1}{\gamma_\rho k^\rho+im^{[n]}_{\rho^\chi}}e^{ik\cdot(x-y)},\label{psipsi}
\end{equation}
one can rewrite the one-loop order (\ref{I:A2:1}) as
\begin{equation}
\begin{split}
	&S^\chi_\text{eff}[\mathcal Y]\big|_\text{1-loop}\\
	&=\fft{1}{2!}\sum'_{n\in\mathbb Z}\int \mathrm{d}^3x\int \mathrm{d}^3y\int\fft{\mathrm{d}^3p}{(2\pi)^3}(\tilde{\mathcal Y}^{(0)\,\mu}(p)+\fft{2\pi n}{\beta}\tilde c^\mu(p))\,e^{ip\cdot x}\int\fft{\mathrm{d}^3q}{(2\pi)^3}(\tilde{\mathcal Y}^{(0)\,\nu}(q)+\fft{2\pi n}{\beta}\tilde c^\nu(q))\,e^{iq\cdot y}\\
	&\quad\times\int\fft{\mathrm{d}^3k_1}{(2\pi)^3}\int\fft{\mathrm{d}^3k_2}{(2\pi)^3}e^{ik_1\cdot(x-y)+ik_2\cdot(y-x)}\tr(\gamma_\mu\fft{1}{\gamma_\rho k_1^\rho+im^{[n]}_{\rho^\chi}}\gamma_\nu\fft{1}{\gamma_\rho k_2^\rho+im^{[n]}_{\rho^\chi}}).
\end{split}\label{I:A2:2}
\end{equation}
The overall minus sign comes from having a fermion loop. In (\ref{I:A2:2}), evaluating the $x,y$-integrals gives two Dirac-delta functions. Using one of them, one can evaluate the $k_2$-integral too and thereby simplify (\ref{I:A2:2}) as
\begin{subequations}
\begin{align}
	S^\chi_\text{eff}[\mathcal Y]\big|_\text{1-loop}&=\fft{1}{2!}\sum'_{n\in\mathbb Z}\int\fft{\mathrm{d}^3p}{(2\pi)^3}\int\fft{\mathrm{d}^3q}{(2\pi)^3}\,(\tilde{\mathcal Y}^{(0)\,\mu}(p)+\fft{2\pi n}{\beta}\tilde c^\mu(p))(\tilde{\mathcal Y}^{(0)\,\nu}(q)+\fft{2\pi n}{\beta}\tilde c^\nu(q))\nn\\
	&\kern12em\times(2\pi)^3\delta^3(p+q)\Pi^{[n]}_{\rho^\chi}{}_{\mu\nu}(p),\label{I:A2:3}\\
	\Pi^{[n]}_{\rho^\chi}{}_{\mu\nu}(p)&\equiv\int\fft{\mathrm{d}^3k}{(2\pi)^3}\tr(\gamma_\mu\fft{1}{\gamma_\rho k^\rho+im^{[n]}_{\rho^\chi}}\gamma_\nu\fft{1}{\gamma_\rho(p+k)^\rho+im^{[n]}_{\rho^\chi}}).\label{AA}
\end{align}
\end{subequations}

Here we compute $\Pi^{[n]}_{\rho^\chi}{}_{\mu\nu}(p)$, omitting the superscript ``$[n]$'' and the subscript ``$\rho^\chi$'' for notational convenience. Using the identity
\begin{equation}
	\int_0^1dx\fft{1}{((1-x)a+xb)^2}=\fft{1}{ab},\label{Feynman}
\end{equation}
and then changing the integration variable $k\to l=k+xp$, one can simplify (\ref{AA}) as
\begin{equation}
\begin{split}
	\Pi_{\mu\nu}(p)&=\int_0^1dx\int\fft{d^3l}{(2\pi)^3}\fft{N_{\mu\nu}}{(l^2+x(1-x)p^2+m^2)^2},\\
	N_{\mu\nu}&=4l_\mu l_\nu-2g_{\mu\nu}l^2-x(1-x)\left(4p_\mu p_\nu-2g_{\mu\nu}p^2\right)+2mp^\rho\epsilon_{\mu\nu\rho}-2m^2g_{\mu\nu}.\label{AA:1}
\end{split}
\end{equation}
Here we have also used the trace formulas (\ref{gamma:tr}). Using $l_\mu l_\nu\to\fft13g_{\mu\nu}l^2$ upon the $l$-integration and the identity
\begin{equation}
	\int d^nl\,\fft{(l^2)^\alpha}{(l^2+M^2)^\beta}=\fft{2\pi^\fft{n}{2}}{\Gamma(\fft{n}{2})}(M^2)^{\fft{n}{2}+\alpha-\beta}\fft{\Gamma(\alpha+\fft{n}{2})\Gamma(\beta-\alpha-\fft{n}{2})}{2\Gamma(\beta)},\label{integration}
\end{equation}
one can evaluate the $l$-integral in (\ref{AA:1}) that gives
\begin{equation}
\begin{split}
	\Pi_{\mu\nu}(p)&=\int_0^1dx\,\left[-\fft{2}{3(2\pi)^3}g_{\mu\nu}\fft{2\pi^\fft32}{\Gamma(\fft32)}(x(1-x)p^2+m^2)^\fft12\fft{\Gamma(\fft52)\Gamma(-\fft12)}{2\Gamma(2)}\right.\\
	&\kern5em+\fft{-x(1-x)(4p_\mu p_\nu-2g_{\mu\nu}p^2)+2mp^\rho\epsilon_{\mu\nu\rho}-2m^2g_{\mu\nu}}{(2\pi)^3}\\
	&\kern6em\left.\times\fft{2\pi^\fft32}{\Gamma(\fft32)}(x(1-x)p^2+m^2)^{-\fft12}\fft{\Gamma(\fft32)\Gamma(\fft12)}{2\Gamma(2)}\right].\label{AA:2}
\end{split}
\end{equation}
In the $m\to\infty$ limit, the $\mathcal{O}(|m|)$ terms from the first and the second lines of (\ref{AA:2}) cancel each other. Consequently, we obtain
\begin{equation}
	\lim_{m\to\infty}\Pi_{\mu\nu}(p)=\fft{m}{|m|}\fft{1}{4\pi}p^\rho\epsilon_{\mu\nu\rho}.\label{AA:m->infty}
\end{equation}

Substituting (\ref{AA:m->infty}) back into (\ref{I:A2:3}) then gives
\begin{equation}
\begin{split}
	\lim_{\beta\to0+}S^\chi_\text{eff}[\mathcal Y]\big|_\text{1-loop}&=\fft{1}{2!}\sum'_{n\in\mathbb Z}\int\fft{\mathrm{d}^3p}{(2\pi)^3}\int\fft{\mathrm{d}^3q}{(2\pi)^3}(\tilde{\mathcal Y}^{(0)\,\mu}(p)+\fft{2\pi n}{\beta}\tilde c^\mu(p))(\tilde{\mathbb A}^{(0)\,\nu}(q)+\fft{2\pi n}{\beta}\tilde c^\nu(q))\\
	&\kern12em\times(2\pi)^3\delta^3(p+q)\,\fft{\text{sgn}\big(m^{[n]}_{\rho^\chi}\big)}{4\pi}p^\rho\epsilon_{\mu\nu\rho}.
\end{split}\label{I:A2:4}
\end{equation}
Note that the large mass limit $m^{[n]}_{\rho^\chi}\to\infty$ for the heavy fermions is justified by the Cardy-like limit $\beta\to0^+$. Finally, using the inverse Fourier transforms of $\tilde{\mathcal Y}^{(0)}$ and $\tilde c$, and then integrating by parts we obtain
\begin{equation}
	\lim_{\beta\to0+}S^\chi_\text{eff}[\mathcal Y]\big|_\text{1-loop}=-\fft{i}{8\pi}\sum'_{n\in\mathbb Z}\text{sgn}\big(m^{[n]}_{\rho^\chi}\big)\int \mathrm{d}^3x\,\epsilon^{\mu\nu\rho}(\mathcal Y^{(0)}_\mu+\fft{2\pi n}{\beta}c_\mu)\partial_\nu(\mathcal Y^{(0)}_\rho+\fft{2\pi n}{\beta}c_\rho).\label{I:A2:5}
\end{equation}
The resulting CS term matches the one in the literature precisely. See \cite{DiPietro:2014bca,DiPietro:2016ond} for example. 

Adding a gravitational CS term (see Appendix A of \cite{Closset:2018ghr} for example, and refer to footnote~\ref{sign:susy} below for comments on the overall sign)
\begin{equation}
    -\fft{i}{192\pi}\sum'_{n\in\mathbb Z}\text{sgn}\big(m^{[n]}_{\rho^\chi}\big)\int \dd^3x\,\sqrt{g} \epsilon^{\mu\nu\rho}\Tr\bigl(\omega_\mu\partial_\nu\omega_\rho-\tfrac23 \omega_\mu\omega_\nu\omega_\rho\bigr)
\end{equation}
to (\ref{I:A2:5}), and then summing over all weights $\rho_\chi$ for the chiral multiplet $\chi$, we obtain the total effective CS terms. The result can be written as
\begin{equation}
\begin{split}
    S^\chi_\text{1-loop CS}&=-\fft{i}{8\pi}\sum_{\rho^\chi}\sum'_{n\in\mathbb Z}\text{sgn}\big(m^{[n]}_{\rho^\chi}\big)\int \mathrm{d}^3x\,\epsilon^{\mu\nu\rho}(\mathcal Y^{(0)}_\mu+\fft{2\pi n}{\beta}c_\mu)\partial_\nu(\mathcal Y^{(0)}_\rho+\fft{2\pi n}{\beta}c_\rho)\\
    &\quad-\fft{i}{192\pi}\sum_{\rho^\chi}\sum'_{n\in\mathbb Z}\text{sgn}\big(m^{[n]}_{\rho^\chi}\big)\int \dd^3x\,\sqrt{g} \epsilon^{\mu\nu\rho}\Tr\bigl(\omega_\mu\partial_\nu\omega_\rho-\tfrac23 \omega_\mu\omega_\nu\omega_\rho\bigr).
\end{split}\label{eq:S:eff}
\end{equation}
%

\subsection{Supersymmetrization of Chern-Simons terms}\label{app:3d-CS:susy}

One can attempt supersymmetrizing the effective CS terms in (\ref{eq:S:eff}) explicitly as (see \cite{ArabiArdehali:2021nsx})
\begin{equation}
\begin{split}
    S^\chi_\text{1-loop}&=-\fft{i}{8\pi}\sum_{\rho^\chi}\sum'_{n\in\mathbb Z}\text{sgn}\big(m^{[n]}_{\rho^\chi}\big)\left(\tilde{s}_\text{g-g}+2(r_\chi-1)\tilde{s}_\text{g-R}\right)\\
    &\quad-\fft{i}{8\pi}\left((r_\chi-1)^2-\fft16\right)\sum_{\rho^\chi}\sum'_{n\in\mathbb Z}\text{sgn}\big(m^{[n]}_{\rho^\chi}\big)s_\text{R-R}-\fft{i}{192\pi}\sum_{\rho^\chi}\sum'_{n\in\mathbb Z}\text{sgn}\big(m^{[n]}_{\rho^\chi}\big)s_\text{grav},
\end{split}\label{eq:S:eff:susy}
\end{equation}
in terms of the following 3d supersymmetric actions\footnote{Our $\mathcal{A}^{(R)}$ corresponds to  $\mathcal{A}_\text{there}^{(R)}-v/2$ in \cite{DiPietro:2016ond,ArabiArdehali:2021nsx}.}
\begin{subequations}
\begin{align}
	\tilde{s}_\text{g-g}&=\int_{\mathcal M_3}\left[\big(\rho^{\chi}\cdot\mathcal A+q^\chi\cdot\mathcal A^{(f)}+\fft{2\pi n}{\beta}c\big)\wedge d\big(\rho^{\chi}\cdot\mathcal A+q^\chi\cdot\mathcal A^{(f)}+\fft{2\pi n}{\beta}c\big)\right.\nn\\
	&\kern4em\left.-2i\big(\rho^\chi\cdot\mathcal D+q^\chi\cdot\mathcal D^{(f)}+\fft{2\pi n}{\beta}H\big)\big(\fft{2\pi n}{\beta}-\rho^{\chi}\cdot\sigma-q^\chi\cdot\sigma^{(f)}\big)(\ast_3\mathbbm1)\right],\label{S:susy:gg:fl}\\
	\tilde{s}_\text{g-R}&=\int_{\mathcal M_3}\left[\big(\rho^{\chi}\cdot\mathcal A+q^\chi\cdot\mathcal A^{(f)}+\fft{2\pi n}{\beta}c\big)\wedge d\mathcal A^{(R)}+iH\big(\rho^{\chi}\cdot\mathcal D+q^\chi\cdot\mathcal D^{(f)}+\fft{2\pi n}{\beta}H\big)(\ast_3\mathbbm1)\right.\nn\\
	&\kern4em\left.-\fft{i}{4}\big(\fft{2\pi n}{\beta}-\rho^{\chi}\cdot\sigma-q^\chi\cdot\sigma^{(f)}\big)\big(R^{(3)}+2v^\mu v_\mu+2H^2\big)(\ast_3\mathbbm1)\right],\label{S:susy:gR:fl}\\
	s_\text{R-R}&=\int_{\mathcal M_3}\left[\mathcal A^{(R)}\wedge d\mathcal A^{(R)}+i\fft{H}{2}\big(R^{(3)}+2v^\mu v_\mu+2H^2\big)(\ast_3\mathbbm1)\right],\label{S:susy:RR:fl}\\
	s_\text{grav}&=\int_{\mathcal M_3}\left[\Tr(\omega\wedge d\omega-\fft23\omega\wedge\omega\wedge\omega)+4\big(\mathcal A^{(R)}-v\big)\wedge d\big(\mathcal A^{(R)}-v\big)\right].\label{S:susy:grav:fl}
\end{align}\label{S:susy:fl}%
\end{subequations}
The first two actions (\ref{S:susy:gg:fl}) and (\ref{S:susy:gR:fl}) are from Appendix A of \cite{DiPietro:2016ond} with the inclusion of flavor symmetries\footnote{Note that $D$ in Appendix A of \cite{DiPietro:2016ond} should be replaced with $\mathcal D$.}
\begin{equation}
    \rho^\chi\cdot\mathcal A\to \rho^\chi\cdot\mathcal A+q^\chi\cdot\mathcal A^{(f)},\quad \rho^\chi\cdot\sigma\to \rho^\chi\cdot\sigma+q^\chi\cdot\sigma^{(f)},\quad \rho^\chi\cdot\mathcal D\to \rho^\chi\cdot\mathcal D+q^\chi\cdot\mathcal D^{(f)},
\end{equation}
and the last two actions (\ref{S:susy:RR:fl}) and (\ref{S:susy:grav:fl}) are from Appendix A of \cite{Closset:2018ghr}.\,\footnote{In the effective action (\ref{eq:S:eff:susy}), the signs of the CS terms are different from those in \cite{Closset:2018ghr}. This is due to the different sign conventions used for fermion real-masses: compare our Eq.~\eqref{eq:mA} with (A.1) and (A.11) of~\cite{Closset:2018ghr}.\label{sign:susy}}

To obtain the supersymmetric effective action (\ref{eq:S:eff:susy}) from (\ref{eq:S:eff}) in terms of (\ref{S:susy:fl}), we have relied on relations between 4d superfields and 3d ones, namely (see Appendix A of \cite{Cassani:2021fyv} for example)
\begin{equation}
\begin{alignedat}{3}
    \sigma&=A_{t_E},\qquad&\mathcal{A}_\mu &=A_{\mu}-A_{t_E}c_\mu,\qquad&\mathcal{D}&=D-A_{t_E}H,\\
    \sigma^{(f)}&=A^{(f)}_{t_E},\qquad&\mathcal{A}^{(f)}_\mu &=A^{(f)}_{\mu}-A^{(f)}_{t_E}c_\mu,\qquad&\mathcal{D}^{(f)}&=D^{(f)}-A^{(f)}_{t_E}H,\\
    H&=A^{(R)}_{t_E}=V_{t_E},\qquad&\mathcal A^{(R)}_\mu&=A^{(R)}_\mu-A^{(R)}_{t_E}c_\mu,\qquad &\fft12v_\mu&=V_\mu-V_{t_E}c_\mu,
\end{alignedat}\label{4d:3d}%
\end{equation}
with $v^\mu=-i\epsilon^{\mu\nu\rho}\partial_\nu c_\rho$. 
In particular, note that the combined one-form $\mathcal Y^{(0)}$ defined by (\ref{eq:Y}) and (\ref{eq:mA}) can be rewritten using (\ref{4d:3d}) as
\begin{equation}
    \mathcal Y^{(0)}=\rho^\chi\cdot\mathcal A+q^\chi\cdot\mathcal A^{(f)}+(r_\chi-1)\mathcal A^{(R)}+\fft14 v.\label{eq:Y:3d}
\end{equation}
It is this expression (\ref{eq:Y:3d}) that we use in (\ref{eq:S:eff}) to make contact with the supersymmetric actions (\ref{S:susy:fl}). 

We would like to emphasize that the combination of supersymmetric actions in (\ref{eq:S:eff:susy}) provides effective CS terms of the dynamical gauge field $\mathcal A$, the background gauge field for flavor symmetry $\mathcal A^{(f)}$, the background gauge field for $R$-symmetry $\mathcal A^{(R)}$, and the KK photon $c$ precisely, but does not yield correct (mixed) CS terms for the KK photon field strength $v$. In particular, from \eqref{eq:S:eff} and \eqref{eq:Y:3d} we get a coefficient $-\fft{i}{8\pi}\sum_{\rho^\chi}\sum'_{n\in\mathbb Z}\text{sgn}\big(m^{[n]}_{\rho^\chi}\big)\times (1/4)^2$ for $\int v\wedge \mathrm{d}v$, whereas the only $\int v\wedge \mathrm{d}v$ term in \eqref{eq:S:eff:susy} arises from $s_\text{grav}$ and has coefficient $-\fft{i}{192\pi}\sum_{\rho^\chi}\sum'_{n\in\mathbb Z}\text{sgn}\big(m^{[n]}_{\rho^\chi}\big)\times 4$; the two clearly do not match. The supersymmetric effective action (\ref{eq:S:eff:susy}) is therefore valid only if $v$ vanishes on the 3d manifold of interest, $\mathcal M_3$. This mismatch puzzle was not addressed in \cite{ArabiArdehali:2021nsx,Cassani:2021fyv}. To fix it for the general case with non-vanishing~$v$, we have introduced the extra (unspecified) SUSY CS action $S_v$ in \eqref{eq:S1loop}. We leave the precise determination of $S_v$ for future research.

On the BPS locus of the dynamical vector multiplet, and with the specific expression for the background vector multiplet for flavor symmetry as in
\begin{equation}
\begin{alignedat}{3}
    A_{t_E}&=\frac{2\pi\boldsymbol{x}}{\beta},\qquad &A_\mu&=0,\qquad &D&=0,\\
    A^{(f)}_{t_E}&=\frac{2\pi\boldsymbol{\xi}}{\beta},\qquad& A^{(f)}_{\mu}&=0,\qquad &D^{(f)}&=0,
\end{alignedat}\label{eq:BPSlocus:all}
\end{equation} 
the supersymmetric effective action (\ref{eq:S:eff:susy}), with $S_v$ included, reads
\begin{equation}
\begin{split}
    S^\chi_\text{1-loop}&=-\fft{i}{8\pi}\sum_{\rho^\chi}\sum'_{n\in\mathbb Z}\text{sgn}\big(m^{[n]}_{\rho^\chi}\big)\left(\left(\fft{2\pi(n-\rho^\chi\cdot\boldsymbol{x}-q^\chi\cdot\boldsymbol{\xi})}{\beta}\right)^2A_{\mathcal M_3}\right.\\
    &\kern12em\left.+2(r_\chi-1)\left(\fft{2\pi(n-\rho^\chi\cdot\boldsymbol{x}-q^\chi\cdot\boldsymbol{\xi})}{\beta}\right)L_{\mathcal M_3}\right)\\
    &\quad-\fft{i}{8\pi}\left((r_\chi-1)^2-\fft16\right)\sum_{\rho^\chi}\sum'_{n\in\mathbb Z}\text{sgn}\big(m^{[n]}_{\rho^\chi}\big)R_{\mathcal M_3}-\fft{i}{192\pi}\sum_{\rho^\chi}\sum'_{n\in\mathbb Z}\text{sgn}\big(m^{[n]}_{\rho^\chi}\big)G_{\mathcal M_3}+S_v.
\end{split}
\end{equation}
Definitions of $A_{\mathcal M_3},L_{\mathcal M_3},R_{\mathcal M_3},G_{\mathcal M_3}$ are as follows (these four supersymmetric CS actions involving background fields were studied in a series of papers by Closset and collaborators---see in particular \cite{Closset:2012vp,Closset:2018ghr})
\begin{equation}
\begin{split}
A_{\mathcal{M}_3}& \, =\,  \int_{\mathcal{M}_3} \dd^3x \, \sqrt{g} \, \bigl(i \,v^\mu \, c_\mu - 
    2 \, i \, H\bigr)\,, \\
    L_{\mathcal{M}_3}& \, =\, \int_{\mathcal{M}_3} \dd^3x \, \sqrt{g} \,
    \Bigl(i \, v^\mu\mathcal A^{(R)} - i\,\fft12 v^\mu \, v_\mu + i \, \fft12H^2 - i \, \fft14R^{(3)} \Bigr) \,,\\
R_{\mathcal{M}_3} & \, =\, \int_{\mathcal{M}_3} \dd^3x \, \sqrt{g}\, \Bigl(\epsilon^{\mu\nu\rho} \, 
\mathcal A^{(R)}_\mu \partial_\nu \mathcal A^{(R)}_\rho
+i\, \frac{H}{2}\left(R^{(3)}+2v_\mu v^\mu+2H^2\right) \Bigr) \,, \\
G_{\mathcal{M}_3}  & \, =\, \int_{\mathcal{M}_3} \dd^3x \, \sqrt{g}\, \Bigl( \epsilon^{\mu\nu\rho}
\Tr \bigl(\omega_\mu \, \partial_\nu \, \omega_\rho-\tfrac23 \omega_\mu \, \omega_\nu \, \omega_\rho \bigr)
	+4 \, \epsilon^{\mu\nu\rho} \bigl(\mathcal A^{{(R)}}_\mu- v_\mu \bigr)
	\partial_\nu \bigl(\mathcal A^{(R)}_\rho- v_\rho \bigr) \Bigr) \,.
\label{eq:CSactions2}
\end{split}
\end{equation}

The first two were discussed in \cite{DiPietro:2016ond}. To compare with \cite{DiPietro:2016ond} note that
\begin{equation}
 A_{\mathcal{M}_3}=-\pi^2 A_{\mathcal{M}_3}^{\text{there}}\qquad L_{\mathcal{M}_3}=-i\pi^2 L_{\mathcal{M}_3}^{\text{there}}.  
\end{equation}

All four have been discussed in \cite{ArabiArdehali:2021nsx,Cassani:2021fyv}. To compare with \cite{Cassani:2021fyv} note that
\begin{equation}
    A_{\mathcal{M}_3}=-\frac{i \beta^2}{\pi} I_1^{\text{there}},\qquad L_{\mathcal{M}_3}= -i \beta I_2^{\text{there}},\qquad R_{\mathcal{M}_3}=-4\pi i I_3^{\text{there}},\qquad G_{\mathcal{M}_3}=-192\pi i I_4^{\text{there}}.
\end{equation}
Another notational difference with \cite{Cassani:2021fyv} is that $\omega^\text{there}_{1,2}=-i\beta\omega_{1,2}$. We have chosen to work with the SUSY actions in \eqref{eq:CSactions2} (rather than $I_{1,2,3,4}$) because they have clear physical significance as the supersymmetrized KK-KK, R-KK, R-R, and gravitational CS terms, respectively. We have also chosen to keep the symbols $\omega_{1,2}$ for the THF moduli of $\mathcal{M}_3$, to be consistent with standard references such as \cite{Aharony:2013dha}.

To compare with \cite{ArabiArdehali:2021nsx} note that
\begin{equation}
    A_{\mathcal{M}_3}=\pi^2 A_{\mathcal{M}_3}^{\text{there}},\qquad L_{\mathcal{M}_3}=-\pi^2 L_{\mathcal{M}_3}^{\text{there}},\qquad R_{\mathcal{M}_3}=\pi^2 R_{\mathcal{M}_3}^{\text{there}},\qquad G_{\mathcal{M}_3}=\pi^2 G_{\mathcal{M}_3}^{\text{there}}.
\end{equation}

On the 3d background of Section~\ref{sec:EFT:background}, the actions in \eqref{eq:CSactions2} evaluate to \cite{Cassani:2021fyv}:
\begin{equation}
\begin{split}
    A_{\mathcal{M}_3}&=-\frac{4\pi^2}{\omega_1\omega_2},\\
    L_{\mathcal{M}_3}&= 2\pi^2\,\frac{\omega_1+\omega_2}{\omega_1\omega_2},\\
    R_{\mathcal{M}_3}&=-\pi^2\,\frac{(\omega_1+\omega_2)^2}{\omega_1\omega_2}.
    \end{split}\label{eq:bkgdBPSsusyCS}
\end{equation}
Moreover, we conjecture
\begin{equation}
    G_{\mathcal{M}_3}+\cdots\overset{!}{=}4\pi^2\,\frac{(\omega_1-\omega_2)^2}{\omega_1\omega_2},\label{eq:modifiedG}
\end{equation}
with the $\cdots$ standing for the modification that $G_{\mathcal{M}_3}$ would receive if the contribution from $S_v$ is incorporated into it. This conjecture is based on holomorphy, as \eqref{eq:modifiedG} is found to be valid for a limited range of $\omega_{1,2}$ in \cite{Cassani:2021fyv}. The work \cite{Cassani:2021fyv} reported however that $G_{\mathcal{M}_3}$ alone on the RHS of \eqref{eq:modifiedG} can not reproduce its LHS for general $\omega_{1,2}\in\mathbb{H}$, and raised this as a mismatch puzzle. We are suggesting that incorporating $S_v$ may resolve that mismatch. This would imply that with the inclusion of $S_v$ the third line in \eqref{eq:S1loop&Vin} is satisfied.

\bibliographystyle{JHEP}
\bibliography{indexrefs}

\providecommand{\href}[2]{#2}\begingroup\raggedright\begin{thebibliography}{10}

\bibitem{Nekrasov:2002qd}
N.~A. Nekrasov, \emph{{Seiberg-Witten prepotential from instanton counting}},
  \href{https://doi.org/10.4310/ATMP.2003.v7.n5.a4}{\emph{Adv. Theor. Math.
  Phys.} {\bfseries 7} (2003) 831}
  [\href{https://arxiv.org/abs/hep-th/0206161}{{\ttfamily hep-th/0206161}}].

\bibitem{Nekrasov:2003rj}
N.~Nekrasov and A.~Okounkov, \emph{{Seiberg-Witten theory and random
  partitions}}, \href{https://doi.org/10.1007/0-8176-4467-9_15}{\emph{Prog.
  Math.} {\bfseries 244} (2006) 525}
  [\href{https://arxiv.org/abs/hep-th/0306238}{{\ttfamily hep-th/0306238}}].

\bibitem{Seiberg:1994rs}
N.~Seiberg and E.~Witten, \emph{{Electric - magnetic duality, monopole
  condensation, and confinement in N=2 supersymmetric Yang-Mills theory}},
  \href{https://doi.org/10.1016/0550-3213(94)90124-4}{\emph{Nucl. Phys. B}
  {\bfseries 426} (1994) 19}
  [\href{https://arxiv.org/abs/hep-th/9407087}{{\ttfamily hep-th/9407087}}].

\bibitem{Seiberg:1994aj}
N.~Seiberg and E.~Witten, \emph{{Monopoles, duality and chiral symmetry
  breaking in N=2 supersymmetric QCD}},
  \href{https://doi.org/10.1016/0550-3213(94)90214-3}{\emph{Nucl. Phys. B}
  {\bfseries 431} (1994) 484}
  [\href{https://arxiv.org/abs/hep-th/9408099}{{\ttfamily hep-th/9408099}}].

\bibitem{DiPietro:2014bca}
L.~Di~Pietro and Z.~Komargodski, \emph{{Cardy formulae for SUSY theories in $d
  =$ 4 and $d =$ 6}},
  \href{https://doi.org/10.1007/JHEP12(2014)031}{\emph{JHEP} {\bfseries 12}
  (2014) 031} [\href{https://arxiv.org/abs/1407.6061}{{\ttfamily 1407.6061}}].

\bibitem{DiPietro:2016ond}
L.~Di~Pietro and M.~Honda, \emph{{Cardy Formula for 4d SUSY Theories and
  Localization}}, \href{https://doi.org/10.1007/JHEP04(2017)055}{\emph{JHEP}
  {\bfseries 04} (2017) 055}
  [\href{https://arxiv.org/abs/1611.00380}{{\ttfamily 1611.00380}}].

\bibitem{Romelsberger:2005eg}
C.~Romelsberger, \emph{{Counting chiral primaries in $N = 1$, $d=4$
  superconformal field theories}},
  \href{https://doi.org/10.1016/j.nuclphysb.2006.03.037}{\emph{Nucl. Phys.}
  {\bfseries B747} (2006) 329}
  [\href{https://arxiv.org/abs/hep-th/0510060}{{\ttfamily hep-th/0510060}}].

\bibitem{Kinney:2005ej}
J.~Kinney, J.~M. Maldacena, S.~Minwalla and S.~Raju, \emph{{An Index for 4
  dimensional super conformal theories}},
  \href{https://doi.org/10.1007/s00220-007-0258-7}{\emph{Commun. Math. Phys.}
  {\bfseries 275} (2007) 209}
  [\href{https://arxiv.org/abs/hep-th/0510251}{{\ttfamily hep-th/0510251}}].

\bibitem{Cardy:1986ie}
J.~L. Cardy, \emph{{Operator Content of Two-Dimensional Conformally Invariant
  Theories}}, \href{https://doi.org/10.1016/0550-3213(86)90552-3}{\emph{Nucl.
  Phys.} {\bfseries B270} (1986) 186}.

\bibitem{Choi:2018hmj}
S.~Choi, J.~Kim, S.~Kim and J.~Nahmgoong, \emph{{Large AdS black holes from
  QFT}},  \href{https://arxiv.org/abs/1810.12067}{{\ttfamily 1810.12067}}.

\bibitem{ArabiArdehali:2019zac}
A.~Arabi~Ardehali, L.~Cassia and Y.~L\"u, \emph{{From Exact Results to Gauge
  Dynamics on $\mathbb{R}^3\times S^1$}},
  \href{https://doi.org/10.1007/JHEP08(2020)053}{\emph{JHEP} {\bfseries 08}
  (2020) 053} [\href{https://arxiv.org/abs/1912.02732}{{\ttfamily
  1912.02732}}].

\bibitem{Dolan:2008qi}
F.~A. Dolan and H.~Osborn, \emph{{Applications of the Superconformal Index for
  Protected Operators and q-Hypergeometric Identities to N=1 Dual Theories}},
  \href{https://doi.org/10.1016/j.nuclphysb.2009.01.028}{\emph{Nucl. Phys. B}
  {\bfseries 818} (2009) 137}
  [\href{https://arxiv.org/abs/0801.4947}{{\ttfamily 0801.4947}}].

\bibitem{Spiridonov:2009za}
V.~P. Spiridonov and G.~S. Vartanov, \emph{{Elliptic Hypergeometry of
  Supersymmetric Dualities}},
  \href{https://doi.org/10.1007/s00220-011-1218-9}{\emph{Commun. Math. Phys.}
  {\bfseries 304} (2011) 797}
  [\href{https://arxiv.org/abs/0910.5944}{{\ttfamily 0910.5944}}].

\bibitem{Spiridonov:2011hf}
V.~P. Spiridonov and G.~S. Vartanov, \emph{{Elliptic hypergeometry of
  supersymmetric dualities II. Orthogonal groups, knots, and vortices}},
  \href{https://doi.org/10.1007/s00220-013-1861-4}{\emph{Commun. Math. Phys.}
  {\bfseries 325} (2014) 421}
  [\href{https://arxiv.org/abs/1107.5788}{{\ttfamily 1107.5788}}].

\bibitem{Gaiotto:2012xa}
D.~Gaiotto, L.~Rastelli and S.~S. Razamat, \emph{{Bootstrapping the
  superconformal index with surface defects}},
  \href{https://doi.org/10.1007/JHEP01(2013)022}{\emph{JHEP} {\bfseries 01}
  (2013) 022} [\href{https://arxiv.org/abs/1207.3577}{{\ttfamily 1207.3577}}].

\bibitem{Ardehali:2015bla}
A.~Arabi~Ardehali, \emph{{High-temperature asymptotics of supersymmetric
  partition functions}},
  \href{https://doi.org/10.1007/JHEP07(2016)025}{\emph{JHEP} {\bfseries 07}
  (2016) 025} [\href{https://arxiv.org/abs/1512.03376}{{\ttfamily
  1512.03376}}].

\bibitem{Honda:2019cio}
M.~Honda, \emph{{Quantum Black Hole Entropy from 4d Supersymmetric Cardy
  formula}}, \href{https://doi.org/10.1103/PhysRevD.100.026008}{\emph{Phys.
  Rev.} {\bfseries D100} (2019) 026008}
  [\href{https://arxiv.org/abs/1901.08091}{{\ttfamily 1901.08091}}].

\bibitem{ArabiArdehali:2019tdm}
A.~Arabi~Ardehali, \emph{{Cardy-like asymptotics of the 4d $ \mathcal{N}=4 $
  index and AdS$_{5}$ blackholes}},
  \href{https://doi.org/10.1007/JHEP06(2019)134}{\emph{JHEP} {\bfseries 06}
  (2019) 134} [\href{https://arxiv.org/abs/1902.06619}{{\ttfamily
  1902.06619}}].

\bibitem{Cabo-Bizet:2019osg}
A.~Cabo-Bizet, D.~Cassani, D.~Martelli and S.~Murthy, \emph{{The asymptotic
  growth of states of the 4d $ \mathcal{N}=1 $ superconformal index}},
  \href{https://doi.org/10.1007/JHEP08(2019)120}{\emph{JHEP} {\bfseries 08}
  (2019) 120} [\href{https://arxiv.org/abs/1904.05865}{{\ttfamily
  1904.05865}}].

\bibitem{Kim:2019yrz}
J.~Kim, S.~Kim and J.~Song, \emph{{A 4d $N=1$ Cardy Formula}},
  \href{https://arxiv.org/abs/1904.03455}{{\ttfamily 1904.03455}}.

\bibitem{ArabiArdehali:2019orz}
A.~Arabi~Ardehali, J.~Hong and J.~T. Liu, \emph{{Asymptotic growth of the 4d $
  \mathcal{N} $ = 4 index and partially deconfined phases}},
  \href{https://doi.org/10.1007/JHEP07(2020)073}{\emph{JHEP} {\bfseries 07}
  (2020) 073} [\href{https://arxiv.org/abs/1912.04169}{{\ttfamily
  1912.04169}}].

\bibitem{ArabiArdehali:2021nsx}
A.~Arabi~Ardehali and S.~Murthy, \emph{{The 4d superconformal index near roots
  of unity and 3d Chern-Simons theory}},
  \href{https://arxiv.org/abs/2104.02051}{{\ttfamily 2104.02051}}.

\bibitem{GonzalezLezcano:2020yeb}
A.~Gonz\'alez~Lezcano, J.~Hong, J.~T. Liu and L.~A. Pando~Zayas,
  \emph{{Sub-leading Structures in Superconformal Indices: Subdominant Saddles
  and Logarithmic Contributions}},
  \href{https://arxiv.org/abs/2007.12604}{{\ttfamily 2007.12604}}.

\bibitem{Amariti:2020jyx}
A.~Amariti, M.~Fazzi and A.~Segati, \emph{{The superconformal index of
  $\mathcal{N}=4$ $USp(2N_c)$ and $SO(N_c)$ SYM as a matrix integral}},
  \href{https://arxiv.org/abs/2012.15208}{{\ttfamily 2012.15208}}.

\bibitem{Amariti:2021ubd}
A.~Amariti, M.~Fazzi and A.~Segati, \emph{{Expanding on the Cardy-like limit of
  the superconformal index of the SCI of 4d $\mathcal{N}=1$ ABCD SCFTs}},
  \href{https://arxiv.org/abs/2103.15853}{{\ttfamily 2103.15853}}.

\bibitem{Lezcano:2021qbj}
A.~G. Lezcano, J.~Hong, J.~T. Liu and L.~A.~P. Zayas, \emph{{The Bethe-Ansatz
  approach to the $\mathcal N=4$ superconformal index at finite rank}},
  \href{https://arxiv.org/abs/2101.12233}{{\ttfamily 2101.12233}}.

\bibitem{Festuccia:2011ws}
G.~Festuccia and N.~Seiberg, \emph{{Rigid Supersymmetric Theories in Curved
  Superspace}}, \href{https://doi.org/10.1007/JHEP06(2011)114}{\emph{JHEP}
  {\bfseries 06} (2011) 114} [\href{https://arxiv.org/abs/1105.0689}{{\ttfamily
  1105.0689}}].

\bibitem{Pestun:2007rz}
V.~Pestun, \emph{{Localization of gauge theory on a four-sphere and
  supersymmetric Wilson loops}},
  \href{https://doi.org/10.1007/s00220-012-1485-0}{\emph{Commun. Math. Phys.}
  {\bfseries 313} (2012) 71} [\href{https://arxiv.org/abs/0712.2824}{{\ttfamily
  0712.2824}}].

\bibitem{Assel:2014paa}
B.~Assel, D.~Cassani and D.~Martelli, \emph{{Localization on Hopf surfaces}},
  \href{https://doi.org/10.1007/JHEP08(2014)123}{\emph{JHEP} {\bfseries 08}
  (2014) 123} [\href{https://arxiv.org/abs/1405.5144}{{\ttfamily 1405.5144}}].

\bibitem{Cassani:2021fyv}
D.~Cassani and Z.~Komargodski, \emph{{EFT and the SUSY Index on the 2nd
  Sheet}},  \href{https://arxiv.org/abs/2104.01464}{{\ttfamily 2104.01464}}.

\bibitem{Hwang:2018riu}
C.~Hwang, S.~Lee and P.~Yi, \emph{{Holonomy Saddles and Supersymmetry}},
  \href{https://doi.org/10.1103/PhysRevD.97.125013}{\emph{Phys. Rev. D}
  {\bfseries 97} (2018) 125013}
  [\href{https://arxiv.org/abs/1801.05460}{{\ttfamily 1801.05460}}].

\bibitem{Cabo-Bizet:2021plf}
A.~Cabo-Bizet, \emph{{On the 4d superconformal index near roots of unity: Bulk
  and Localized contributions}},
  \href{https://arxiv.org/abs/2111.14941}{{\ttfamily 2111.14941}}.

\bibitem{Ruijsenaars:1997}
S.~N.~M. Ruijsenaars, \emph{First order analytic difference equations and
  integrable quantum systems}, {\emph{Journal of Mathematical Physics}
  {\bfseries 38} (1997) 1069}.

\bibitem{Ardehali:2015hya}
A.~Arabi~Ardehali, J.~T. Liu and P.~Szepietowski, \emph{{High-Temperature
  Expansion of Supersymmetric Partition Functions}},
  \href{https://doi.org/10.1007/JHEP07(2015)113}{\emph{JHEP} {\bfseries 07}
  (2015) 113} [\href{https://arxiv.org/abs/1502.07737}{{\ttfamily
  1502.07737}}].

\bibitem{Goldstein:2020yvj}
K.~Goldstein, V.~Jejjala, Y.~Lei, S.~van Leuven and W.~Li, \emph{{Residues,
  modularity, and the Cardy limit of the 4d $\mathcal{N}=4$ superconformal
  index}},  \href{https://arxiv.org/abs/2011.06605}{{\ttfamily 2011.06605}}.

\bibitem{Jejjala:2021hlt}
V.~Jejjala, Y.~Lei, S.~Van~Leuven and W.~Li, \emph{{$SL(3,\mathbb{Z})$
  Modularity and New Cardy Limits of the $\mathcal{N}=4$ Superconformal
  Index}},  \href{https://arxiv.org/abs/2104.07030}{{\ttfamily 2104.07030}}.

\bibitem{Rains:2006dfy}
E.~M. Rains, \emph{{Limits of elliptic hypergeometric integrals}},
  \href{https://doi.org/10.1007/s11139-007-9055-3}{\emph{Ramanujan J.}
  {\bfseries 18} (2007) 257}
  [\href{https://arxiv.org/abs/math/0607093}{{\ttfamily math/0607093}}].

\bibitem{Assel:2015nca}
B.~Assel, D.~Cassani, L.~Di~Pietro, Z.~Komargodski, J.~Lorenzen and
  D.~Martelli, \emph{{The Casimir Energy in Curved Space and its Supersymmetric
  Counterpart}}, \href{https://doi.org/10.1007/JHEP07(2015)043}{\emph{JHEP}
  {\bfseries 07} (2015) 043}
  [\href{https://arxiv.org/abs/1503.05537}{{\ttfamily 1503.05537}}].

\bibitem{Gadde:2011ik}
A.~Gadde, L.~Rastelli, S.~S. Razamat and W.~Yan, \emph{{The 4d Superconformal
  Index from q-deformed 2d Yang-Mills}},
  \href{https://doi.org/10.1103/PhysRevLett.106.241602}{\emph{Phys. Rev. Lett.}
  {\bfseries 106} (2011) 241602}
  [\href{https://arxiv.org/abs/1104.3850}{{\ttfamily 1104.3850}}].

\bibitem{Bourdier:2015wda}
J.~Bourdier, N.~Drukker and J.~Felix, \emph{{The exact Schur index of
  $\mathcal{N}=4$ SYM}},
  \href{https://doi.org/10.1007/JHEP11(2015)210}{\emph{JHEP} {\bfseries 11}
  (2015) 210} [\href{https://arxiv.org/abs/1507.08659}{{\ttfamily
  1507.08659}}].

\bibitem{Bourdier:2015sga}
J.~Bourdier, N.~Drukker and J.~Felix, \emph{{The $\mathcal{N}=2$ Schur index
  from free fermions}},
  \href{https://doi.org/10.1007/JHEP01(2016)167}{\emph{JHEP} {\bfseries 01}
  (2016) 167} [\href{https://arxiv.org/abs/1510.07041}{{\ttfamily
  1510.07041}}].

\bibitem{Closset:2017bse}
C.~Closset, H.~Kim and B.~Willett, \emph{{$ \mathcal{N} $ = 1 supersymmetric
  indices and the four-dimensional A-model}},
  \href{https://doi.org/10.1007/JHEP08(2017)090}{\emph{JHEP} {\bfseries 08}
  (2017) 090} [\href{https://arxiv.org/abs/1707.05774}{{\ttfamily
  1707.05774}}].

\bibitem{Benini:2018mlo}
F.~Benini and P.~Milan, \emph{{A Bethe Ansatz type formula for the
  superconformal index}},  \href{https://arxiv.org/abs/1811.04107}{{\ttfamily
  1811.04107}}.

\bibitem{Spiridonov:2010qv}
V.~P. Spiridonov and G.~S. Vartanov, \emph{{Superconformal indices of
  ${\mathcal N}=4$ SYM field theories}},
  \href{https://doi.org/10.1007/s11005-011-0537-2}{\emph{Lett. Math. Phys.}
  {\bfseries 100} (2012) 97} [\href{https://arxiv.org/abs/1005.4196}{{\ttfamily
  1005.4196}}].

\bibitem{poston2014catastrophe}
T.~Poston and I.~Stewart, \emph{Catastrophe theory and its applications}.
  Courier Corporation, 2014.

\bibitem{Aharony:2013dha}
O.~Aharony, S.~S. Razamat, N.~Seiberg and B.~Willett, \emph{{3d dualities from
  4d dualities}}, \href{https://doi.org/10.1007/JHEP07(2013)149}{\emph{JHEP}
  {\bfseries 07} (2013) 149} [\href{https://arxiv.org/abs/1305.3924}{{\ttfamily
  1305.3924}}].

\bibitem{Closset:2013vra}
C.~Closset, T.~T. Dumitrescu, G.~Festuccia and Z.~Komargodski, \emph{{The
  Geometry of Supersymmetric Partition Functions}},
  \href{https://doi.org/10.1007/JHEP01(2014)124}{\emph{JHEP} {\bfseries 01}
  (2014) 124} [\href{https://arxiv.org/abs/1309.5876}{{\ttfamily 1309.5876}}].

\bibitem{Nawata:2011un}
S.~Nawata, \emph{{Localization of $N$=4 Superconformal Field Theory on $S^1
  \times S^3$ and Index}},
  \href{https://doi.org/10.1007/JHEP11(2011)144}{\emph{JHEP} {\bfseries 11}
  (2011) 144} [\href{https://arxiv.org/abs/1104.4470}{{\ttfamily 1104.4470}}].

\bibitem{Bobev:2015kza}
N.~Bobev, M.~Bullimore and H.-C. Kim, \emph{{Supersymmetric Casimir Energy and
  the Anomaly Polynomial}},
  \href{https://doi.org/10.1007/JHEP09(2015)142}{\emph{JHEP} {\bfseries 09}
  (2015) 142} [\href{https://arxiv.org/abs/1507.08553}{{\ttfamily
  1507.08553}}].

\bibitem{Hanada:2019kue}
M.~Hanada and B.~Robinson, \emph{{Partial-Symmetry-Breaking Phase
  Transitions}}, \href{https://doi.org/10.1103/PhysRevD.102.096013}{\emph{Phys.
  Rev. D} {\bfseries 102} (2020) 096013}
  [\href{https://arxiv.org/abs/1911.06223}{{\ttfamily 1911.06223}}].

\bibitem{Cherman:2020zea}
A.~Cherman and A.~Dhumuntarao, \emph{{Confinement and graded partition
  functions for $\mathcal{N}=4$ SYM}},
  \href{https://doi.org/10.1103/PhysRevD.103.066013}{\emph{Phys. Rev. D}
  {\bfseries 103} (2021) 066013}
  [\href{https://arxiv.org/abs/2012.12341}{{\ttfamily 2012.12341}}].

\bibitem{Klare:2012gn}
C.~Klare, A.~Tomasiello and A.~Zaffaroni, \emph{{Supersymmetry on Curved Spaces
  and Holography}}, \href{https://doi.org/10.1007/JHEP08(2012)061}{\emph{JHEP}
  {\bfseries 08} (2012) 061} [\href{https://arxiv.org/abs/1205.1062}{{\ttfamily
  1205.1062}}].

\bibitem{Dumitrescu:2012ha}
T.~T. Dumitrescu, G.~Festuccia and N.~Seiberg, \emph{{Exploring Curved
  Superspace}}, \href{https://doi.org/10.1007/JHEP08(2012)141}{\emph{JHEP}
  {\bfseries 08} (2012) 141} [\href{https://arxiv.org/abs/1205.1115}{{\ttfamily
  1205.1115}}].

\bibitem{Grisaru:1979wc}
M.~T. Grisaru, W.~Siegel and M.~Rocek, \emph{{Improved Methods for
  Supergraphs}},
  \href{https://doi.org/10.1016/0550-3213(79)90344-4}{\emph{Nucl. Phys. B}
  {\bfseries 159} (1979) 429}.

\bibitem{ArabiArdehali:2016fjg}
A.~Arabi~Ardehali, \emph{{High-temperature asymptotics of the 4d superconformal
  index}}, Ph.D. thesis, Michigan U., 2016.
\newblock \href{https://arxiv.org/abs/1605.06100}{{\ttfamily 1605.06100}}.

\bibitem{Sohnius:1981tp}
M.~F. Sohnius and P.~C. West, \emph{{An Alternative Minimal Off-Shell Version
  of N=1 Supergravity}},
  \href{https://doi.org/10.1016/0370-2693(81)90778-4}{\emph{Phys. Lett. B}
  {\bfseries 105} (1981) 353}.

\bibitem{Sohnius:1982fw}
M.~Sohnius and P.~C. West, \emph{{The Tensor Calculus and Matter Coupling of
  the Alternative Minimal Auxiliary Field Formulation of $N=1$ Supergravity}},
  \href{https://doi.org/10.1016/0550-3213(82)90337-6}{\emph{Nucl. Phys. B}
  {\bfseries 198} (1982) 493}.

\bibitem{Closset:2018ghr}
C.~Closset, H.~Kim and B.~Willett, \emph{{Seifert fibering operators in 3d
  $\mathcal{N}=2$ theories}},
  \href{https://doi.org/10.1007/JHEP11(2018)004}{\emph{JHEP} {\bfseries 11}
  (2018) 004} [\href{https://arxiv.org/abs/1807.02328}{{\ttfamily
  1807.02328}}].

\bibitem{Closset:2012vp}
C.~Closset, T.~T. Dumitrescu, G.~Festuccia, Z.~Komargodski and N.~Seiberg,
  \emph{{Comments on Chern-Simons Contact Terms in Three Dimensions}},
  \href{https://doi.org/10.1007/JHEP09(2012)091}{\emph{JHEP} {\bfseries 09}
  (2012) 091} [\href{https://arxiv.org/abs/1206.5218}{{\ttfamily 1206.5218}}].

\end{thebibliography}\endgroup

\end{document}